\documentclass[%
a4,
 preprint,%
 amssymb, amsmath,nofootinbib,%
 aip,cha,%
]{revtex4-1}
\usepackage{epsfig}
\usepackage{epsf}
\usepackage{bm}                 
\usepackage{amsmath}
\usepackage{amssymb}
\usepackage{dcolumn}
\usepackage{graphicx}
\usepackage{psfrag}
\usepackage{latexsym}
\usepackage{color}
\usepackage{caption}  
\usepackage{subcaption}
\usepackage{bm}%
\usepackage[utf8]{inputenc}  
\usepackage[T1]{fontenc}
\usepackage[colorlinks=true,linkcolor=blue]{hyperref}%
\expandafter\ifx\csname package@font\endcsname\relax\else
 \expandafter\expandafter
 \expandafter\usepackage
 \expandafter\expandafter
 \expandafter{\csname package@font\endcsname}%
\fi
\newcommand{\rw}[1]{\textcolor{black}{#1}}
\newcommand{\raw}[1]{\textcolor{black}{#1}}

\newcommand{\sam}[1]{\textcolor{green}{#1}}

\def\mean#1{\left< #1 \right>}

\newcommand{\pdbb}{\mathcal{P}}
\begin{document}

\title{Bouncing oil droplets, de Broglie's quantum thermostat and convergence to equilibrium}%
\author{Mohamed Hatifi}%
\affiliation{Aix  Marseille  Universit\'e,  CNRS,
Centrale  Marseille, Institut Fresnel UMR 7249,13013 Marseille, France. email: hatifi.mohamed@gmail.com}%

\author{Ralph Willox}%
\affiliation{Graduate School of Mathematical Sciences, the University of Tokyo,\\3-8-1 Komaba, Meguro-ku, 153-8914 Tokyo, Japan}
\author{Samuel Colin}%
\affiliation{Centro Brasileiro de Pesquisas F\'{\i}sicas,
Rua Dr.\ Xavier Sigaud 150,
\mbox{22290-180, Rio de Janeiro -- RJ, Brasil.}}%

\author{Thomas Durt}%

\affiliation{Aix  Marseille  Universit\'e,  CNRS,
Centrale  Marseille, Institut Fresnel UMR 7249,13013 Marseille, France. email: thomas.durt@centrale-marseille.fr}%

\date{2 April 2018}%
\revised{10 August 2018}%

\maketitle

\section{Abstract}{Recently, the properties of bouncing oil droplets, also known as `walkers', have attracted much attention because they are thought to offer a gateway to  a  better  understanding  of  quantum  behaviour. They indeed constitute a macroscopic realization of wave-particle duality, in the sense that their trajectories are guided by a self-generated surrounding wave. 
The aim of this paper is to try to describe walker phenomenology in terms of de Broglie-Bohm dynamics and of a stochastic version thereof. In particular, we first study how a stochastic modification of  the de Broglie pilot-wave theory, {\it \`a la} Nelson, affects  the  process  of  relaxation to quantum equilibrium, and we prove an H-theorem for the relaxation to {quantum} equilibrium under Nelson-type dynamics. We then compare the onset of equilibrium in the stochastic and the de Broglie-Bohm approaches and we propose some simple experiments by which one can test the applicability of our theory to the context of bouncing oil droplets. Finally, we compare our theory to actual observations of walker behavior in a 2D harmonic potential well. }

\section{Introduction}
`Walkers' are realized as oil droplets generated at the surface of a vibrating oil bath. As shown by Couder and Fort  \cite{couder1,couder2,couder3}, the vibration of the bath prevents the coalescence of the droplets with the surface, allowing them to remain stable for very long times. Moreover, the trajectories of  the walkers are guided by an external wave \cite{Harris2013,Bush2015} that they themselves generate at the surface of the oil bath. From this point of view, walkers are reminiscent of wave-particle duality \cite{couder2,Couder2012} and they seem to offer deep analogies with de Broglie-Bohm particles \cite{Bush2015a}.
Up until now, different aspects of walker dynamics have been studied in a purely classical framework, typically in a hydrodynamical approach \cite{couder3,Bush2015}. For instance, certain models address their deformations due to their bouncing off the surface of the bath, in function of the density and viscosity of the oil and other parameters \cite{Bush2015}. Other studies describe the dynamics of the surface waves that the walkers generate during the bouncing process, and how those waves in turn guide their trajectories. In these models, this complex behaviour is characterized by a memory time which relates the dynamics of the walker bouncing at time $t$, to its successive bouncing positions in the past \cite{eddi2011,Perrard2014}. The presence of such a memory effect establishes a first difference with quantum mechanics. Normally, in quantum mechanics, it is assumed that all results of any possible future measurements to be performed on a quantum system, are encapsulated in its present quantum state \cite{Durt1999}: its wave function at the present time $t$. 

Droplets also transcend the most common interpretations of quantum theory  which prohibit any description of the system in terms of instantaneous, classical-like, trajectories. Droplets and their trajectories are visible with the naked eye at any time and standard interpretations of quantum mechanics do not apply. This is why we believe that it is necessary and worthwhile to adapt \rw{realist (causal)} formalisms such as de Broglie-Bohm (dBB) dynamics\cite{bohm521,bohm522} \rw{or a stochastic version thereof \`a la Nelson}\cite{Nelson1967}, to explore the analogy with quantum systems. This is the main motivation of the present paper.

Another difference between walker trajectories and quantum trajectories is that the quantum description is intrinsically probabilistic and non-classical, while there exist regimes in which the trajectory of the walkers is indeed deterministic and classical (for example, when they bounce exactly in phase with the bath, they can be shown to follow straight lines at constant velocity \cite{Labousse2014,Fort2010,Dubertrand2016,Tadrist2017}). {However, there also exist  regimes in which a Brownian motion is superimposed on their flow lines (e.g. above the Faraday threshold), and other regimes where the trajectories appear to be chaotic\cite{Bush2015}. {In fact, in} several regimes} droplets appear to exhibit ergodic behaviour. In practice, ergodicity has been established on the basis of the following observations: if we prepare a walker at the surface of the liquid bath (a corral for instance), it will progressively explore each part of the surface, following an apparently random motion \cite{Harris2013}. If one then visualizes the statistics of the sojourn time of the walker in each of these regions, a striking pattern emerges, bearing more than a simple resemblance to an interference pattern.\cite{Harris2013,Bush2015a} It is this, again remarkable, manifestation of wave-particle duality that first attracted our attention and which lies at the origin of this paper. {Actually, the onset of quantum equilibrium in the framework of \rw{dBB dynamics and in stochastic versions thereof} is an important foundational issue in itself, which motivated numerous studies {(see e.g. \cite{bohm-vigier,Nelson1967,bohmh,kyprianidis,valentini91a,Petroni1994,Petroni1995,guerra1995} and also \cite{efthymiopoulos4} and references therein).} Several authors in the past {have indeed tried to explain how the Born rule emerges from individual trajectories, which is a highly non-trivial problem.} In the case of dBB dynamics it is easy to show that in simple situations the relaxation to the Born statistical distribution does not occur at all, but recent studies{\cite{valentini042,cost10,toruva,colin2012,efthymiopoulos3,abcova}} show that in sufficiently complex situations (several modes of different energies for instance) {the system might exhibit mixing}, which explains the onset of quantum equilibrium in such cases. As we shall show in the present paper, in the case of \rw{Nelson-type dynamics} the quantum Brownian motion {imposed in such a model} accelerates the relaxation to Born's distribution,  {and in fact ensures that relaxation to the Born rule will \rw{almost} always occur (as we shall also show)}. {In our view, for the above reasons, de Broglie-Bohm and \rw{Nelson-type} dynamics are good candidates} for explaining how wavelike statistics emerges after averaging a set of apparently chaotic and/or stochastic trajectories.}

Briefly summarized, our main goal is to explain the emergence of {aforementioned} interference patterns in the framework of the dynamical models of de Broglie-Bohm \rw{and of a stochastic version thereof which is  based on the models of Bohm-Vigier\cite{bohm-vigier} and Bohm-Hiley\cite{bohmh} but which is formally close to Nelson\cite{Nelson1967}. Both models are introduced in section \ref{s1}.}
Here, it is worth noting that thus far there is no experimental evidence that droplets indeed follow de Broglie-Bohm and/or Nelson trajectories.
\rw{Our approach therefore differs radically from previous studies on droplets, in the sense that we impose a quantum dynamics by brute force whereas, until now, the attempt to illustrate how chaos may underlie quantum stochasticity has been a pillar of the research on walkers/droplets. In fact, Nelson's original goal, in proposing his dynamics, was to derive an effective wave equation from the properties of an underlying Brownian motion, as in classical statistical mechanics where a diffusion equation is derived from microscopic properties of the atoms. \raw{There actually exists an impressive number of attempts in that direction, as e.g. stochastic electro-dynamics \cite{Bush2015,DelaPena2013,DelaPena2015}. However,  there exists (as far as we know) no way to derive an effective Schr\"odinger equation from hydrodynamical models of droplets.}}

By choosing exactly the opposite approach, i.e. by imposing quantum-like dynamics on the droplets, we pursue \rw{three} goals. The first one is to \rw{describe} the onset of quantum equilibrium (and ergodicity). A second objective is to formulate precise quantitative predictions regarding this relaxation process, which can possibly be validated by future experiments. A third objective is to show, for the first time, that certain dBB trajectories present a deep structural resemblance with certain trajectories that have been reported in the literature about droplets trapped in a harmonic potential.

A short discussion of the onset of equilibrium in de Broglie-Bohm dynamics and the importance of coarse-graining is given in section  \ref{s2}. In the case of \rw{our stochastic, Nelson-type} dynamics, we derive in section \ref{Nelsonrelax} a new H-theorem showing the relaxation to quantum equilibrium,  which does not rely on coarse-graining and is valid at all scales.  We pay particular attention to the ergodicity  of trajectories in the case of our \rw{stochastic} dynamics (which mix properties of the de Broglie-Bohm  dynamics with Brownian motion).  We apply these ideas to discuss ergodicity in the case of  \rw{the stochastic treatment of a} particle trapped in a harmonic potential (section \ref{s5}), and next, to describe the dynamics of a droplet trapped in a harmonic potential, 
 in section \ref{s6}. In that section (in \ref{s6bis}) we also propose some simple experiments by which one can test the applicability of \rw{a Nelson-type dynamics} to the context of bouncing oil droplets, and we briefly discuss the problems caused by the presence of zeros in the interference pattern that is encoded in the statistics of the trajectories. In section \ref{s7}, we study a situation during which the attractor of the probability distribution is no longer a static eigenstate of the (static) Hamiltonian and we compare the onset of equilibrium in the dBB and \rw{stochastic} formalisms in that special framework. \rw{In section \ref{s9}, we tackle the dynamics of droplets in a 2D harmonic potential through a simple model where the pilot wave is treated as a dynamical object. This constitutes a preliminary attempt, ultimately aimed at establishing a dynamics that would combine stochastic and/or dBB dynamics with a feedback of the trajectory on the wave, a problem which has never been addressed in the framework of dBB or Nelson dynamics,} but which is a fundamental feature of droplet phenomenology. The last section is devoted to conclusions and open questions.
A short overview of the numerical methods used and supplementary technical details of the calculations are given in appendix.

\section{dBB and Nelson dynamics \label{s1}}
\subsection{The dBB theory}\label{sec-dBB}

In the following {quick} overview of the dBB theory 
we shall limit ourselves to the case of a single particle.
In the dBB theory, 
particle positions exist at all times and they are merely revealed by position measurements{, instead of `originating' with the measurement as the standard interpretation of quantum mechanics would have it}.
{The dynamics is described by a wave function which obeys the Schr\"odinger equation:
\begin{equation}\label{se}
i\hbar\frac{\partial \Psi({\bf x},t)}{\partial t}= -\frac{\hbar^{2}}{2m}\Delta\Psi({\bf x},t)+V({\bf x},t)\Psi({\bf x},t),
\end{equation}
where $V({\bf x},t)$ is an external potential and $m$ the mass of the particle, {\em and} by a position {${\bf x}$}.
In order to reproduce the predictions of standard quantum mechanics one must have that the positions are distributed according to
\begin{equation}\label{equivariance}
\pdbb({\bf x},t)=|\Psi({\bf x},t)|^2, 
\end{equation}
where $\pdbb({\bf x},t)$ is the distribution of particle positions over an ensemble of trajectories. An ensemble satisfying condition \eqref{equivariance} is said to be in {\em quantum equilibrium}.}

{It is also commonly assumed that (\ref{equivariance}) is satisfied at some initial time. 
Therefore, {in order to be at (quantum) equilibrium for all $t$,} the condition to enforce is 
\begin{equation}\label{equivariance0}
\frac{\partial \pdbb({\bf x},t)}{\partial t}=\frac{\partial|\Psi({\bf x},t)|^2}{\partial t}.
\end{equation}
As is well-known, the probability density  $\vert\Psi({\bf x},t)\vert^{2}$ satisfies the continuity equation
\begin{equation}
\frac{\partial \vert\Psi({\bf x},t)\vert^{2}}{\partial t}+ \boldsymbol{\nabla}\cdot\boldsymbol{j}({\bf x},t) =0,
\label{cont}
\end{equation}
where 
\begin{equation}
\boldsymbol{j}=\frac{\hbar}{m}\,\mathfrak{Im}\left(\,{\Psi}^*\,\boldsymbol{\nabla}\,\Psi\,\right),\label{j1}
\end{equation}
is the (probability) current describing the flow of the probability {due to \eqref{se}}.}

The probability density $\pdbb$, on the other hand, will satisfy a continuity equation
\begin{equation}
\frac{\partial \pdbb}{\partial t}+ \boldsymbol{\nabla}\cdot\left(\pdbb\,\textbf{v}\right)=0,
\label{cont2}
\end{equation}
where $\textbf{v}$ is the velocity field {for the particle}.
Therefore (\ref{equivariance0}) will be satisfied if\footnote{{The expression \eqref{velfield} for the velocity field is of course not the only possible one: any solution of the form
\begin{equation*}
\textbf{v}'({\bf x},t)=\textbf{v}({\bf x},t)+\frac{\boldsymbol{\nabla} \times f({\bf x},t)}{\vert\Psi({\bf x},t)\vert^{2}},
\end{equation*}
where $f$ is a scalar function, will also give rise to (\ref{equivariance0}) {(see Ref. \cite{deotto} for more details)}}. 
} 
\begin{equation}\label{velfield}
\textbf{v}({\bf x},t)=\frac{\textbf{j}({\bf x},t)}{\vert\Psi({\bf x},t)\vert^{2}}.
\end{equation}
Secondly, if one expresses the wave function in terms of its phase $S({\bf x},t)$ and modulus $R({\bf x},t)=\sqrt{|\Psi({\bf x},t)|^2\,}$,
\begin{equation}
\Psi({\bf x},t)=R({\bf x},t)e^{i\,S({\bf x},t)/\hbar},
\label{bohm}
\end{equation}
one finds that 
\begin{equation}
\boldsymbol{j} = \frac{|\Psi({\bf x},t)|^2}{m} \boldsymbol{\nabla} S
,\label{guid2}
\end{equation}
and that the velocity of the particle at time $t$ is given by:
\begin{equation}
\frac{d{\bf x}(t)}{dt}={\bf v}({\bf x},t)=
\frac{1}{m}\boldsymbol{\nabla}S({\bf x},t)\bigg{|}_{{\bf x}={\bf x}(t)}.
\label{guid3}
\end{equation}
Integrating the system \eqref{guid3} we recover the dBB trajectory. 
From the above it should be clear that the dBB theory is deterministic. {Any} stochastic element only comes from our lack of knowledge of the initial positions.

{In the context of bouncing droplets}, we shall view the external wave generated by the droplet as being  in one-to-one correspondence with the `pilot wave' $\Psi$ which guides the position of the dBB particle.

\subsection{\rw{A simple realization of de Broglie's quantum thermostat -- Nelson dynamics}}
  
As mentioned in the introduction, the trajectories of walkers are often characterized by a non-negligible stochastic (Brownian) component which sets them apart from the smooth dBB trajectories. From this point of view, it seems worthwhile to {try to} model walkers dynamics in terms of stochastic generalisations of dBB dynamics.

de Broglie himself\footnote{Quoting de Broglie: {\em ``...Finally, the particle's motion is the combination of a regular motion defined by the guidance formula, with a random motion of Brownian character... any particle, even isolated, has to be imagined as in continuous ``energetic contact'' with a hidden medium, which constitutes a concealed thermostat. This hypothesis was brought forward some fifteen years ago by Bohm and Vigier \cite{bohm-vigier}, who named this invisible thermostat the ``subquantum medium''...  If a hidden sub-quantum medium is assumed, knowledge of its nature would seem desirable...''} (In\cite{deBroglieend} Ch.XI: On the necessary introduction of a random element in the double solution theory. The hidden thermostat and the Brownian motion of the particle in its wave.) }, in fact, considered such generalizations of the deterministic dBB dynamics (which he called the ``quantum thermostat hypothesis'') to be highly welcome because they might provide a physically sound picture of the hidden dynamics of static quantum states. For instance, if we consider the position of an electron prepared in the ground state of a hydrogen atom, the dBB dynamics predicts that its position will remain frozen at the same place throughout time, which is counterintuitive {to say the least}. Adding a stochastic component to its velocity could, in principle, explain why averaging the position of the electron over time is characterized by an exponentially decreasing probability density function, in agreement with the Born rule (provided, of course, that ergodicity is present in the problem in exactly the right proportion). A first proposal in this sense was formulated by Bohm and Vigier in 1954 \cite{bohm-vigier} \rw{which, later on, was made more precise by Bohm and Hiley\cite{bohmh}, but stochastic derivations of Schr\"odinger's equation by Nelson\cite{Nelson1967} (and others \cite{DelaPena2013,DelaPena2015} in the framework of stochastic electrodynamics) can also be considered to provide models of the quantum thermostat.}  In this paper we shall consider a particular model of the quantum thermostat \rw{ in which, as in the Bohm-Vigier model,  a single spinless particle suspended in a Madelung fluid moves with the local velocity of the resulting field, given by $\eqref{guid3}$, and is subjected to fluctuations coming from the latter (cf. Figure 1). However, following Nelson, we shall model these fluctuations by means of a particular stochastic process.}\footnote{\rw{To be precise: our model is formally the same as Nelson's in that it relies on the same stochastic process. However, in spirit, it is closer to the Bohm-Hiley model\cite{bohmh} in that we do not assume to be at quantum equilibrium (an assumption which is fundamental to Nelson's theory, as was already pointed out by Bohm and Hiley\cite{bohmh} ; see also Ref. \cite{kyprianidis} for a detailed presentation and a comparison of both approaches).}}

\begin{figure}
\begin{center}
\includegraphics[scale=0.40]{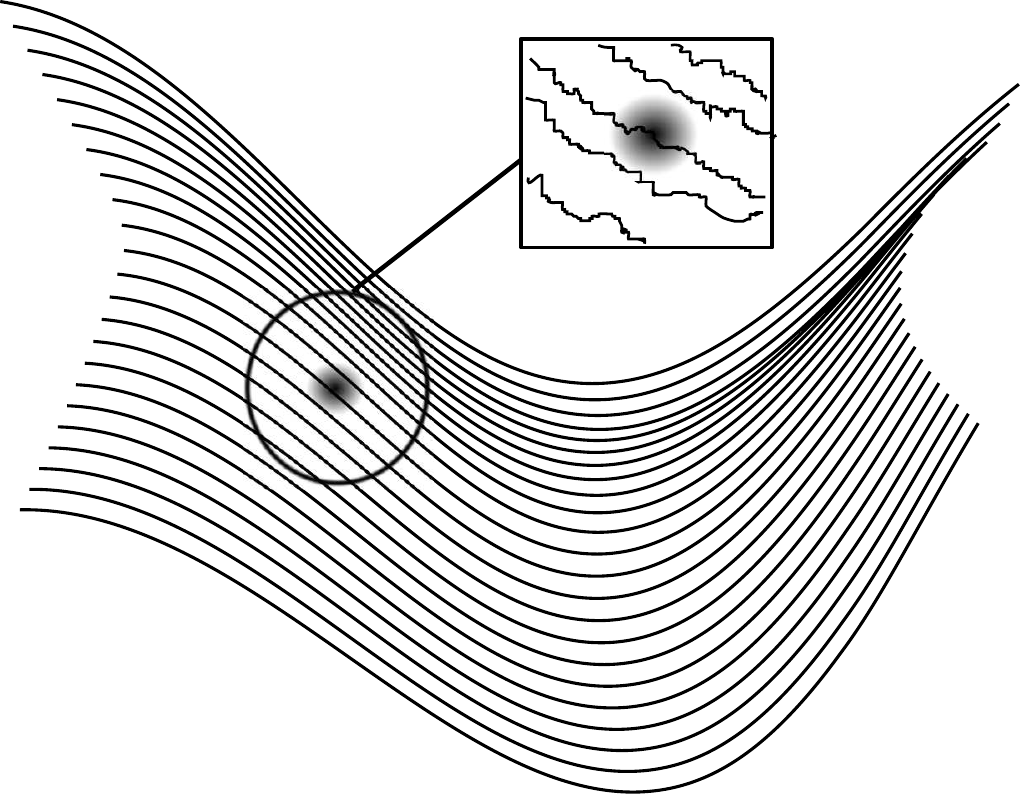}  
\captionof{figure}{\small A particle suspended in a Madelung fluid and subject to local fluctuations.}
\label{fig1}
\end{center}
\end{figure}

This process is defined on a probabilistic space $\Omega$, characterized by a probability distribution $P(\textbf{x},t)$ and obeying an Ito stochastic differential equation of the general form:
\begin{equation}
d\boldsymbol{x} = \left.\left[\,\frac{1}{m}\boldsymbol{\nabla} S + \boldsymbol{\gamma}\,\right]\right\vert_{x=x(t)}dt + \sqrt{\alpha}\,d\boldsymbol{W}(t),
\label{ito}
\end{equation}
where $\alpha$ is the (constant) diffusion coefficient that characterizes the strength of the random part and $d\boldsymbol{W}(t)$ is a Wiener process in three dimensions. The function $\boldsymbol{\gamma}(\textbf{x},t)$ in \eqref{ito} is a systematic drift, the so-called {\em osmotic velocity}, which we shall fix in the following way.

The conservation equation of the probability distribution (which we denote by $P$, in order to stress the difference with the probability in the dBB dynamics which is denoted by $\mathcal{P}$), can be written as the Fokker-Planck equation:
\begin{equation}
\frac{\partial P}{\partial t}= - \boldsymbol{\nabla}\cdot\left(\frac{P}{m}\boldsymbol{\nabla} S +\boldsymbol{\gamma}\,P\right)+\frac{\alpha}{2}\,\Delta P.\label{FPeq}
\end{equation}
If we now require that the quantum equilibrium $P(q,t)=\vert\Psi(q,t)\vert^{2}$  be a solution of this Fokker-Planck equation, we obtain from \eqref{cont}, \eqref{guid2} and  \eqref{FPeq} that
\begin{equation}
  \boldsymbol{\nabla}\cdot\left(\boldsymbol{\gamma}\,\vert\Psi\vert^{2}-\frac{\alpha}{2}\,\boldsymbol{\nabla}\vert\Psi\vert^{2}\right)=0,
\end{equation}
which is a constraint on the osmotic velocity.
The simplest solution of this constraint is \cite{Bacciagaluppi1999}
\begin{equation}
\boldsymbol{\gamma}(\textbf{x},t)= \frac{\alpha}{2}\,\frac{\boldsymbol{\nabla}\vert\Psi\vert^{2} }{\vert\Psi\vert^{2}}.
\label{vos}
\end{equation}
{In the rest of the paper we choose} the osmotic drift velocity to be \eqref{vos}, {with $\alpha$ an a priori free parameter, Nelson's choice for $\alpha$ ($\alpha=\hbar/m$) being irrelevant if we apply this formalism to droplets.}

In summary, \rw{our} Nelson dynamics is fully defined by the following Ito equation,
\begin{equation}
d\textbf{x}(t)= \left.\left[\,\frac{1}{m}\boldsymbol{\nabla} S + \frac{{\alpha}}{2} \frac{\boldsymbol{\nabla}\vert\Psi\vert^{2}}{\vert\Psi\vert^{2}}\,\right]\right\vert_{x=x(t)}\,dt + \sqrt{\alpha}d\boldsymbol{W}(t),\label{Nelsondyn1}
\end{equation}
where $dW_{i}(t)$ represents a Wiener process with 
\begin{equation}\label{Wienerprocess}
<dW_{i}(t)>=0\qquad \text{and}\qquad <dW_{i}(t)dW_{j}(t')>=\frac{1}{2}\,\delta_{ij}\,\delta(\,t-t'\,),
\end{equation}
the Fokker-Planck equation for the associated probability distribution $P({\bf x},t)$,
\begin{equation}
\frac{\partial P}{\partial t}=\frac{{\alpha}}{2}\,\Delta P - \boldsymbol{\nabla}\cdot\left(\frac{P}{m}\boldsymbol{\nabla} S +\frac{{\alpha}}{2} \frac{P}{\vert\Psi\vert^{2}}\boldsymbol{\nabla}\vert\Psi\vert^{2}\right),
\label{fp}
\end{equation}
where $\Psi(\textbf{x},t)$ satisfies the Schrodinger equation:
\begin{equation}
i\hbar\,\frac{\partial \Psi}{\partial t}=-\frac{\hbar^{2}}{2m}\Delta \Psi + V \Psi .\label{Nelsondyn4}
\end{equation}
 \\
At quantum equilibrium, i.e. when $P(\textbf{x},t)=\vert\Psi(\textbf{x},t)\vert^{2}$, the diffusion velocity is balanced by the osmotic term and the Bohm velocity is recovered, on average.

We shall now discuss the details of the relaxation towards quantum equilibrium, in the dBB and \rw{stochastic} formalisms. 
  
\section{Relaxation to quantum equilibrium in the de Broglie-Bohm theory \label{s2}}
In our presentation of the dBB theory for a single particle, in section \ref{sec-dBB}, we assumed that the particle positions are initially distributed according to Born's law
\begin{equation}\label{equ0}
\pdbb({\bf x},t_i)=|\Psi({\bf x},t_i)|^2,
\end{equation}
over an ensemble. Then the dynamics ensure that the same relation will hold for any later time. 
This is the assumption de Broglie and Bohm made in their original papers \cite{deBroglie27,bohm521,bohm522}.
{Although Bohm tried, already in the 1950s (first on his own -- see, e.g., sec. 9 in\cite{bohm521} -- and later with Vigier \cite{bohm-vigier}), to relax this assumption by modifying the dynamics, to many authors working today on the dBB theory} it is still an assumption which has to be made\footnote{\rw{The final objective of de Broglie, Bohm, Vigier, Nelson (and many other contributors to various realistic hidden variable interpretations in which quantum systems are assumed to be localized in space at any time) was to rationalize wave-like statistics in terms of individual trajectories. The same problem occurs in droplets phenomenology and according to us, it admits no fully satisfying solution yet.}}.

According to Valentini \cite{valentini-phd,valentini91a,valentini91b}, however, there is no need to assume that the {particle} positions are initially distributed according to Born's law or to modify the dynamics. 
His claim is that an ensemble in which Born's law is not satisfied (so-called quantum non-equilibrium) will evolve naturally towards quantum equilibrium, provided that the wave function 
leads to sufficiently complex dynamics. 
This relaxation process has to take place on a coarse-grained level and can only occur if the initial distributions do not display any fine-grained micro structure.

Let us first explain the need for coarse-graining.
Let us introduce the function $f=\pdbb/\vert\Psi\vert^{2}$, as in \cite{valentini042}. An important implication of $\eqref{cont2}$ is that the function $f$ is conserved along the dBB trajectories:
\begin{equation}
\frac{d f({\bf x},t)}{dt}\equiv\frac{\partial f({\bf x},t)}{\partial t}+\dot{{\bf x}} \cdot\boldsymbol{\nabla}f({\bf x},t)=0 .\label{contf}
\end{equation}
Hence we have that
\begin{equation}
\pdbb({\bf x},t)=\frac{\pdbb({\bf x}_i,t_i)}{|\Psi({\bf x}_i,t_i)|^2}|\Psi({\bf x},t)|^2,
\end{equation}
where ${\bf x}_i$ is the initial position of the particle which leads to ${\bf x}$, when evolving from $t_i$ to $t$ according to the dBB dynamics. 
{If one assumes that $\pdbb({\bf x}_i,t_i)/|\Psi({\bf x}_i,t_i)|^2\neq1$, relaxation to quantum equilibrium is clearly impossible, at least at the microscopic level. However, as argued by Valentini\cite{valentini91a}, relaxation {\em is} possible at the coarse-grained level, provided the initial distribution does not display any fine-grained microstructure.}

The operational definition of the coarse-graining is as follows. 
We divide the domain of interest $A\subset\Omega$ into small cubes of equal edge length $\epsilon$ (we call them coarse-graining cells, or CG cells for short). 
These CG cells do not overlap and their union is equal to $A$. The coarse-grained densities, which we denote by 
$\overline{\pdbb}({\bf x},t)$ and $\overline{|\Psi({\bf x},t)|^2}$, are then defined as
\begin{gather}
\overline{\pdbb}({\bf x},t)=\frac{1}{\epsilon^3}\int_{\textrm{CG cell}\ni{\bf x}} d^3 x  \pdbb({\bf x},t),\label{cgquantities1}\\[-2mm]\nonumber\\
\overline{|\Psi({\bf x},t)|^2}=\frac{1}{\epsilon^3}\int_{\textrm{CG cell}\ni{\bf x}} d^3 x  |\Psi({\bf x},t)|^2,\label{cgquantities2}
\end{gather}
where the domain of integration is the CG cell containing ${\bf x}$.

We can now discuss the second assumption: the lack of fine-grained micro structure in the initial distribution. Let us assume we have a non-equilibrium distribution $\pdbb({\bf x},t_i)$ which relaxes to quantum equilibrium 
at the coarse-grained level, under the dynamics generated by the wave function $\Psi({\bf x},t)$.
As the dBB theory is time-reversal invariant, in the time-reversed situation, under the dynamics generated by $\Psi^\ast({\bf x},-t)$ we would have a distribution that moves away 
from quantum equilibrium. Thus it would seem that time-reversal invariance {contradicts the possibility of} relaxation to quantum equilibrium. {This conclusion is unwarranted however: as the initial distribution $\pdbb({\bf x},t_i)$ relaxes to quantum equilibrium,} it retains information on the original values of $f$ (which are constant in time) and thereby acquires a fine-grained microstructure, which means that at the final time $t_f$, 
$\pdbb({\bf x},t_f)$ will differ significantly from $\overline{\pdbb}({\bf x},t_f)$. Therefore, in the time-reversed situation, the initial distribution would exhibit a fine-grained micro-structure, {which is prohibited under our assumption}, thereby breaking the time-reversal invariance.

In order to quantify the difference between the distribution $\pdbb({\bf x},t)$ and the quantum equilibrium condition $\vert\Psi({\bf x},t)\vert^{2}$ at the coarse-grained level, 
Valentini \cite{valentini-phd,valentini91a,valentini91b} has introduced the entropy-like function
\begin{equation}
\overline{H}(t)=\int_\Omega\,d^3 x ~\overline{\pdbb}\,\ln\left(\overline{\pdbb}/\overline{\vert\Psi\vert^{2}}\right),
\label{hfun}
\end{equation}
where $\overline{\pdbb}$ and $\overline{\vert\Psi\vert^{2}}$ as in \eqref{cgquantities1} and \eqref{cgquantities2}, for which he has shown the {(quantum)} H-theorem,
\begin{equation}
\overline{H}(t)\leq \overline{H}(t_i),
\end{equation}
under the assumption of no fine-grained micro-structure. It should be stressed however that this is not necessarily a monotonic decay and therefore does not prove that quantum equilibrium will always be reached. 
It merely indicates a tendency towards relaxation.
The strongest support for the idea of relaxation to quantum equilibrium comes from numerical simulations of the evolution of non-equilibrium distributions 
for various quantum systems \cite{valentini042,cost10,toruva,colin2012,efthymiopoulos3,abcova} (\raw{see Ref. \cite{efthymiopoulos4} and references therein for a review}). 
The first numerical simulations were performed by Valentini and Westman\cite{valentini042} who showed, in the case of a 2D box, 
that relaxation quickly takes place for a wave function which is a superposition of the first 16 modes of energy (the superposition being equally weighted). 
It was also hinted that the nodes of the wave function, with their associated vorticity, play a crucial role in the relaxation process, as purveyors of chaos {(or mixing)} in the dynamics.
This later claim was properly understood in\cite{efthymiopoulos}.
The dependence of the relaxation timescale on the coarse-graining length $\epsilon$ and on the number of energy modes was studied in\cite{toruva}. 
In\cite{abcova}, it was shown that quantum systems with a low number of modes are likely to never fully relax, in which case $\overline{H}$
reaches a {non-zero residue value}. However, such a scenario becomes unlikely as the number of modes increases.

According to the quantum non-equilibrium hypothesis, standard quantum mechanics is only one facet of the pilot-wave theory, that of quantum equilibrium, leaving the possibility for {possible} new physics: that of quantum non-equilibrium. {{One should assume of course, that during our time we have only had (or can only have) access to systems for which quantum equilibrium has already been reached.}}
But that does not mean that quantum non-equilibrium never existed in the early universe {(which could {possibly} be inferred from the observation of the remnants of the early fractions of seconds of the universe, just after the big bang \sam{\cite{valentini2010}})}, or that some{, yet undetetected,} exotic quantum systems cannot still be in quantum non-equilibrium today \sam{\cite{underwood}}. {{This is why droplets are appealing, because if their dynamics do present analogies with dBB dynamics, their study will allow us to observe relaxation to (quantum) equilibrium `in real time' in the lab., with our naked eyes,} which is not possible with quantum systems for which we have no direct access to individual trajectories.}

\section{An H-theorem for Nelson dynamics}\label{Nelsonrelax}
Let us start by introducing an analog of Valentini's entropy\footnote{\rw{It should be noted} that the entropy \eqref{hv} or the functional \eqref{Lf} we shall use to quantify the relaxation to quantum equilibrium, are very different from the entropies usually considered in the context of classical H-theorems (like e.g. the Boltzmann entropy). One should bear in mind however that \rw{\it quantum equilibrium} is radically different from classical equilibrium \cite{bricmont} and has no connection whatsoever with relaxation to quantum \rw{\it thermal} equilibrium, for the simple reason that the Born distribution of positions reached by an ensemble of trajectories {\it  \`a la} Nelson or dBB is not a thermal distribution.} \eqref{hfun} for the probability distribution $P({\bf x},t)$ associated with \rw{our} Nelson dynamics, as defined by (\ref{Nelsondyn1}--\ref{Nelsondyn4}),
\begin{equation}
H_V(t)=\int_\Omega\,d^3 x~ P\,\ln\left(\frac{P}{\vert\Psi\vert^{2}}\right),
\label{hv}
\end{equation}
\rw{which is a special instance of a relative entropy known as the Kullback-Leibler divergence\cite{jungel}. We also define a} second 
non-negative functional,
\begin{equation}
L_f(t)=\int_\Omega\,d^3 x~ f (P-{\vert\Psi\vert^{2}}),
\label{Lf}
\end{equation}
where 
\begin{equation}
f({\bf x},t) = \frac{P({\bf x},t)}{{\vert\Psi({\bf x},t)\vert^{2}}}.\label{fdef}
\end{equation}
Note that we always impose the boundary conditions $\vert\Psi\vert^{2}\big|_{\partial\Omega}=P\big|_{\partial\Omega}=0$ and $f\big|_{\partial\Omega}=1$ so as to avoid divergence of these integrals on the boundary of $\Omega$.

To understand why functionals \eqref{hv} and \eqref{Lf} are non-negative and why they are zero if and only if (quantum) equilibrium is reached (that is to say when  $f=1$ everywhere in space), it is important to note that the integrands of $H_V$ and $L_f$ satisfy the inequalities\footnote{This is immediate from the trivial inequality: $\forall x>0, \quad (1-1/x) \leq \ln x \leq x-1$.}
\begin{equation}
(P - \vert\Psi\vert^{2}) \leq\, P \,\ln \frac{P}{\vert\Psi\vert^{2}}\,  \leq\, \frac{P}{\vert\Psi\vert^{2}} (P - \vert\Psi\vert^{2}),\label{ineq1}
\end{equation}
for which any of the possible equalities only hold when $P=\vert\Psi\vert^{2}$.
Now, since both  $P({\bf x},t)$ and $\vert\Psi({\bf x},t)\vert^{2}$ are probability distributions, i.e. \rw{since we have $\int_\Omega P dx = \int_\Omega \vert\Psi\vert^{2} dx=1$,} it follows from \eqref{ineq1} that whenever $H_V(t)$ and $L_f(t)$ are well-defined, they satisfy the inequalities:
\begin{equation}
0 \leq H_V(t) \leq L_f(t). \label{ineq2}
\end{equation}

Moreover,  \rw{for the same reason}, $L_f$ can be re-expressed as $\int_\Omega\,d^3 x~ \big[f (P-{\vert\Psi\vert^{2}}) - (P-{\vert\Psi\vert^{2}})\big]$, the integrand in which is non-negative due to \eqref{ineq1}. Therefore, $L_f$ can only be zero if its integrand  is zero, i.e.: if $P={\vert\Psi\vert^{2}}$ (if $P$, ${\vert\Psi\vert^{2}}$ and $f$ are sufficiently smooth, which is something we shall always assume unless otherwise stated). Similarly\cite{valentini91a} one also has that $H_V$ can only be zero when $P={\vert\Psi\vert^{2}}$ everywhere in $\Omega$.

{Let us now prove the relaxation to quantum equilibrium.} Substituting $P= f\, {\vert\Psi\vert^{2}}$ in the Fokker-Planck equation \eqref{fp}, and using the continuity equation \eqref{cont} and relation \eqref{guid2}, it is easily verified that
\begin{equation}
{\vert\Psi\vert^{2}} \frac{\partial f}{\partial t} = \frac{\alpha}{2} \nabla \cdot ({\vert\Psi\vert^{2}}\, \nabla f) - \frac{{\vert\Psi\vert^{2}}}{m} (\nabla f) (\nabla S).\label{eqpourf}
\end{equation}
Rewriting $L_f$ as
\begin{equation}
L_f = \int_\Omega\,d^3 x~ f (f-1) {\vert\Psi\vert^{2}},
\end{equation} 
its behaviour in time can be calculated using \eqref{eqpourf}, \eqref{cont} and \eqref{guid2}:
\begin{align}
\frac{d L_f}{d t} =& \int_\Omega\,d^3 x~ \left[ - \nabla\cdot \left( \frac{{\vert\Psi\vert^{2}}}{m} (f^2-f) \nabla S \right) + \frac{\alpha}{2} (2 f -1) \nabla \cdot ({\vert\Psi\vert^{2}}\, \nabla f)\right]\\
=& \, \frac{\alpha}{2}\,\int_\Omega\,d^3 x~  \left[ \nabla\cdot \left[ (2f-1) {\vert\Psi\vert^{2}}\, \nabla f \right]  - 2 \big(\nabla f\big)^2 {\vert\Psi\vert^{2}}\right]\\
=& \, -\alpha \,\int_\Omega\,d^3 x~  \big(\nabla f\big)^2\,  {\vert\Psi\vert^{2}},
\end{align}
which is of course strictly negative, for all $t$, as long as $\nabla f$ \rw{and $\vert\Psi\vert^{2}$ are} not identically zero. Hence, \rw{if $\vert\Psi\vert^{2}$ is not zero throughout $\Omega$}, $L_f$ will decrease monotonically for as long as $f$ is not (identically) equal to 1 on $\Omega$, and therefore necessarily converges to 0, a value it can only attain when $f\equiv1$ or, equivalently, when $P\equiv{\vert\Psi\vert^{2}}$. We have thus established a strong H-theorem showing that, in the case of Nelson dynamics, any probability distribution $P$ necessarily converges to $\vert\Psi\vert^{2}$\rw{, if the latter does not become identically zero. Note that this excludes the case of a free particle for which $\lim_{t\to+\infty} \vert\Psi(x,t)\vert^{2}=0$, for all $x$, which means that $\frac{dL_f}{dt}$ tends to zero even when $f$ does not converge to 1.}

A result, similar to the above, is also easily established for $H_V$ since $L_f$ dominates the latter, or alternatively from the formula
\begin{equation}
\frac{d H_V}{d t} = - \frac{\alpha}{2}\, \int_\Omega\,d^3 x~ \big(\nabla f\big)^2\,\frac{\vert\Psi\vert^{2}}{f}.
\end{equation}

The above results show that \rw{(excluding the case of the free particle)} Nelson dynamics, naturally, exhibits relaxation towards quantum equilibrium, and this for general initial probability distributions (at least, as long as the initial distribution is smooth enough). In this stochastic setting there is therefore no need for any assumptions on the microstructure of the initial distributions, nor is there any need for the coarse-grained hypothesis when deriving an H-theorem. 

\rw{Note that these results also show that we have, in fact, convergence of the distribution $P$ to the quantum equilibrium distribution $\vert\Psi\vert^{2}$ in $L^1$ norm. This is a consequence of the so-called Csisz\'ar-Kullback-Pinsker inequality \cite{jungel}},
\begin{equation}
L_1 \leq \sqrt{2 H_V},
\end{equation}
where
\begin{equation}
L_1=\int_\Omega\,d^3 x~ \big\vert P-\vert\Psi_{}\vert^{2} \big\vert.\label{L1}
\end{equation}
\rw{This generalizes the results by Petroni and Guerra \cite{Petroni1994,Petroni1995} obtained in their study of the relaxation towards quantum equilibrium in the framework of the Nelson dynamics of a single particle in a harmonic potential.} The $L^1$ norm is also used by Efthymiopoulos et al \cite{efthymiopoulos4} in the context of the dBB theory.

We shall illustrate these results by means of numerical simulations for the case of a ground state for the 1D-harmonic oscillator in section \ref{ergodicity}, for the case of the 2D-harmonic oscillator in section \ref{corralsec}, and in the case of a coherent state in section \ref{CSrelax}.

A last important remark concerns the influence of possible zeros in the equilibrium distribution $\Psi({\bf x},t)$, which would give rise to singularities in the osmotic velocity terms in the Ito equation \eqref{Nelsondyn1} or the Fokker-Planck equation \eqref{fp} (or equivalently in equation \eqref{eqpourf}) and might make the functions $H_V$ and $L_f$ ill-defined.
In section \ref{corralsec} we discuss the case of the first excited state of the 1D-harmonic oscillator, for which $\Psi({\bf x},t)$ has a node at $x=0$, and one could in fact imagine studying higher excited states for which one has a finite number of nodes. In that case, the osmotic velocity \eqref{vos} will have simple poles at a finite number of positions in $x$. At the level of the Ito equation one would not expect a finite set of poles to cause any particular problems, not only because the probability of hitting a pole exactly in the stochastic evolution is zero but also because the osmotic term tends to move the particle away from the pole very quickly. Similarly, a finite number of simple poles in the convection-diffusion equation \eqref{eqpourf} for $f$ only influence the velocity field in the convection term in a finite number of distinct places and it is to be expected that this would have the effect of actually enhancing the mixing of information in the system.

Moreover, it is also clear that simple nodes in $\Psi({\bf x},t)$ only give rise to (a finite number of) logarithmic singularities in the integrand of $H_V$ and that the integral  \eqref{hv} therefore {still} converges. The H-theorem for $H_V$ derived above is thus still valid and an arbitrary distribution $P$ (sufficiently smooth) will still converge to quantum equilibrium, even in the presence of nodes for $\Psi({\bf x},t)$. The same cannot be said however of the function $L_f$ as simple zeros in $\Psi({\bf x},t)$ give rise to double poles in the integrand and a possible divergence of the integral \eqref{Lf}. Hence, at the beginning of the evolution, for arbitary $P$, the function $L_f$ might take an infinitely large value\footnote{The integrand only diverges when $\vert\Psi\vert^{2}\ll P$, i.e. when it is positive.}, but as soon as convergence sets in (which is guaranteed by the H-theorem for $H_V$), the divergent parts in its integrand will be smoothed out and the function $L_f$ will take finite values that converge to zero as time goes on.\footnote{Of course, when calculating these quantities for the results of numerical simulations, there is always some amount of coarse-graining going on and genuine infinities never occur.}

\section{Relaxation to quantum equilibrium and Nelson dynamics: static case}\label{s5}
In this section, in order to simplify the discussion, we will only consider the case of stationary states $\Psi_{st}(x)$ for the one dimensional Schr\"odinger equation, i.e. energy levels for which $S=-E\,t$ and which therefore have zero Bohm velocity \eqref{guid3}: $\boldsymbol{\nabla}S \equiv S_x =0$. 

\subsection{Fokker-Planck operator and a formal connection to a Schr\"odinger equation}\label{ss51} 
There exists a wide literature \cite{Gardiner1985,Risken1996} concerning {a {particular} method for studying} the convergence of  solutions of the Fokker-Planck equation to a stationary one, which is only sporadically mentioned in the literature devoted to Nelson dynamics \cite{Petroni1998}. This approach makes it possible to quantify very precisely the speed of convergence to equilibrium, in terms of (negative) eigenvalues of the Fokker-Planck operator. In order to show this, let us rewrite the Fokker-Planck equation \eqref{fp} in terms of the Fokker-Planck operator $\widehat{\mathcal{L}}$ :
\begin{equation}
\frac{\partial P}{\partial t}=\widehat{\mathcal{L}}P=\left[\, {-\frac{\partial \gamma}{\partial x}} -\gamma(x)\frac{\partial }{\partial x} +\frac{\alpha}{2} \frac{\partial^{2} }{\partial x^{2}}\,\right]P,
\label{l1}
\end{equation} {where \eqref{vos}:
\begin{equation}\boldsymbol{\gamma}(\textbf{x})= 
\alpha\frac{\big(\vert\Psi_{st}\vert\big)_x}{\vert\Psi_{st}\vert}.\label{osmo}\end{equation}}
Note that, due to the presence of the first derivative $\frac{\partial }{\partial x}$, the $\widehat{\mathcal{L}}$ operator is not Hermitian. 

Now, in order to {establish} the H-theorem, we must prove that in the long-time limit this equation tends to a stationary solution $P_{st}=\vert\Psi_{st}\vert^{2}$. 
The key idea here is to \raw{transform the Fokker-Planck equation to a simple diffusion equation} through the transformation
\begin{equation}
P(x,t)=\sqrt{P_{st}(x)}\,\,g(x,t),
\label{p}
\end{equation}
under which the r.h.s. of equation $\eqref{l1}$ reduces to
\begin{equation}
\widehat{\mathcal{L}}\,P=\sqrt{P_{st}(x)}\,\,\widehat{\mathcal{H}}_{st}\,\,g(x,t),
\label{l2}
\end{equation}
 \\
where $\widehat{\mathcal{H}}_{st}$ is now a Hermitian operator:  
\begin{equation}
\widehat{\mathcal{H}}_{st}=\frac{\alpha}{2} \frac{\partial^{2} }{\partial x^{2}} - \frac{1}{2} \left(\,\frac{\partial \gamma}{\partial x}+\frac{\gamma^{2}}{\alpha}\,\right).
\end{equation}

The function $g(x,t)$ thus obeys \raw{a `Schr\"odinger-like' equation (though with imaginary time)} with an effective potential \raw{($\widehat{\mathcal{H}}_{st}$)} which depends on $\gamma(x)$:
\begin{equation}
\frac{\partial g(x,t)}{\partial t}=\widehat{\mathcal{H}}_{st}\,\,g(x,t).
\label{sch}
\end{equation}
\\
{Note that} the effective potential is exactly the Bohm-quantum potential defined by
\begin{equation}
Q_{\Psi}=-\frac{\hbar^{2}}{2\,m}\,\frac{1}{\vert\Psi_{st}\vert} \frac{\partial^{2} {\vert\Psi_{st}\vert}}{\partial x^{2}},
\end{equation}
which can be expressed in terms of the osmotic velocity \eqref{osmo} as:
\begin{equation}
\frac{Q_{\Psi}}{{m\,\alpha}}=-\frac{1}{2} \left(\,\frac{\partial \gamma}{\partial x}+\frac{\gamma^{2}}{\alpha}\,\right).
\end{equation}

\subsection{Superposition ansatz}\label{ss52}
\label{subsec:sup}W can now represent the solution of \eqref{sch} as a superposition of discrete eigenvectors (all orthogonal, as the operator {$\widehat{\mathcal{H}}_{st}$} is Hermitian) and impose the superposition ansatz \cite{brics2013solve}: 
\begin{equation}
g(x,t) = \sum_{k=0}^{\infty}\,a_{k}(t)\,g_{k}(x).
\label{g}
\end{equation}
Equation $\eqref{sch}$ is separable and gives rise to the eigenvalue problem:
\begin{equation}
\frac{1}{a_{k}(t)}\frac{d a_{k}(t)}{d t}=\frac{1}{g_{k}(x)}\widehat{\mathcal{H}}_{st}\,g_{k}(x)=-\lambda_{k}.
\label{schgk}
\end{equation}
As a result we have 
\begin{equation}
g(x,t) = \sum_{k=0}^{\infty}\,a_{k}e^{-\lambda_{k}t}\,g_{k}(x),
\label{gsg}
\end{equation}
for a set of constants $a_k$ and where all the $\lambda_{k}$ are real (as ${\mathcal{H}}$ is Hermitian), for eigenfunctions $g_{k}(x)$ that satisfy the orthonormality conditions:
\begin{equation}
\int_{-\infty}^{\infty}\,dx\,g_{k}(x)g_{l}(x)=\delta_{k,l}.
\label{orth}
\end{equation}
Thus, we have the expression
\begin{equation}
P(x,t) = \sum_{k=0}^{\infty}\,a_{k}e^{-\lambda_{k}t}\,\sqrt{P_{st}(x)}\,g_{k}(x).
\label{pn}
\end{equation}
 By construction, the function $\sqrt{P_{st}(x)}$ is an eigenstate of the effective Hamiltonian with energy 0. We shall associate the label $\lambda_{0}$ with this energy level. 
 
In order to have a well defined probability distribution and to avoid any divergence in time, it is clear that all eigenvalues $-\lambda_k$ have to be negative, which requires $\Psi_{st}$ to be the ground state of the effective Hamiltonian ${\mathcal{H}}_{st}$\raw{: just as in the case of the usual Schr\"odinger equation, the eigenvalues $-\lambda_k$ in equation \eqref{schgk} are all negative only if $\Psi_{st}(x)$ has no zeros\footnote{\raw{See also Ref. \cite{v3} (appendix 2) for an elementary proof that all $\lambda_{k}$ are indeed positive if $\Psi_{st}(x)$ has no zeros.}}.}

If $\Psi_{st}(x)$ does have zeros the osmotic velocity will have singularities. In \raw{Ref. \cite{v3} (appendix 2)}, we consider what happens in the case when $\Psi_{st}(x)$ is an excited state of the harmonic oscillator and we derive a formal solution in terms of the eigenvalues $-\lambda_{k}$ which are now not all negative, {thus revealing the appearance of instabilities for cases where the above formalism would still be valid}.


\subsection{One dimensional oscillator and the evolution of Gaussian distributions for the ground state}\label{ss54}
In \raw{Ref. \cite{v3}(appendix 2)} we discuss the application of  the method of the effective Hamiltonian outlined in section \ref{ss51} to this particular problem, and we derive a Green function for the associated Fokker-Planck equation when $\Psi_{st}$ is the ground state of the one dimensional oscillator. \raw{This Green function $K_{P}(x,x',t)$ is defined through}

\begin{align}
P(x,t) &=\int_{-\infty}^{\infty}\,dx'\,P(x',0)\,K_{P}(x,x',t),
\label{pgreen}
\end{align}

where the kernel $K_{P}$ is given by

\begin{align}\label{pgreen2}
K_{P}(x,x',t)= \left(\frac{a}{\pi\,sinh(\omega\,t)} \right)^{\frac{1}{2}}\,&e^{\omega\left(n+\frac{1}{2}\right) t}\,\nonumber\\
\times &e^{\frac{-a}{sinh(\omega\,t)}\left[(x^{2}+x'^{2})\,cosh(\omega\,t)+(x^{2}-x'^{2})\,sinh(\omega\,t)-2\,xx'\right]}.
\end{align}

An important property of the Green function \eqref{pgreen2} for this case is that if $\vert\Psi(x)\vert^{2}$ and $P(x,0)$ are Gaussian, then $P(x,t)$ will still be Gaussian \eqref{pgreen}. Let us define the ground state as
\begin{equation}
{\vert\Psi_{st}\vert^{2} \equiv\,}\vert\Psi(x)\vert^{2}=\sqrt{\frac{2a}{\pi}}e^{-2\,a\,x^{2}},
\label{gs}
\end{equation}for which we can then write:
\begin{equation}
P(x,t)=\sqrt{\frac{2b(t)}{\pi}}e^{-2\,b(t)\,\left(x-\mean{x(t)}\,\right)^{2}}.
\label{pg}
\end{equation}

Injecting $\eqref{pg}$ in the Fokker-planck equation {\eqref{l1}} gives a differential equation for $\mean{x(t)}$,
\begin{equation}
\frac{d\mean{x(t)}}{dt}=-2a\alpha\,\mean{x(t)},
\end{equation}
which is readily solved:
\begin{equation}
\mean{x(t)}=\mean{x_{0}}e^{-2a\alpha\,t},\label{meanNels}
\end{equation}
as well as an equation for $b(t)$
\begin{equation}
\frac{1}{2b
(t)}\,\frac{db(t)}{dt}+2\alpha\left(b(t)-a\right)=0,
\end{equation}
with solution:
\begin{equation}
b(t)=\frac{a}{1-\left(1-\frac{a}{b_{0}}\right)e^{-4a\alpha\,t}}.\label{bsol}
\end{equation}
From \eqref{pg} and \eqref{bsol} we can then calculate the width of the non-equilibrium Gaussian as: 
\begin{align}
\sigma_{x}^{2} (t)\equiv\frac{1}{4b(t)}&=\frac{1}{4a}\left[\left(1-e^{-4a\alpha\,t}\right)+ \frac{a}{b_0} e^{-4a\alpha\,t}\right]\nonumber\\[-3mm]\nonumber\\
&=\sigma_{eq}^{2}\left(1-e^{-4a\alpha\,t}\right)+\sigma_x^2(0)\, e^{-4a\alpha\,t},
\label{sig}
\end{align}
where $\sigma_{eq}^{2}$ represents the width $1/(4a)$ of the equilibrium distribution \eqref{gs}.

\begin{figure}[!t]
\begin{center}
\includegraphics[scale=0.33]{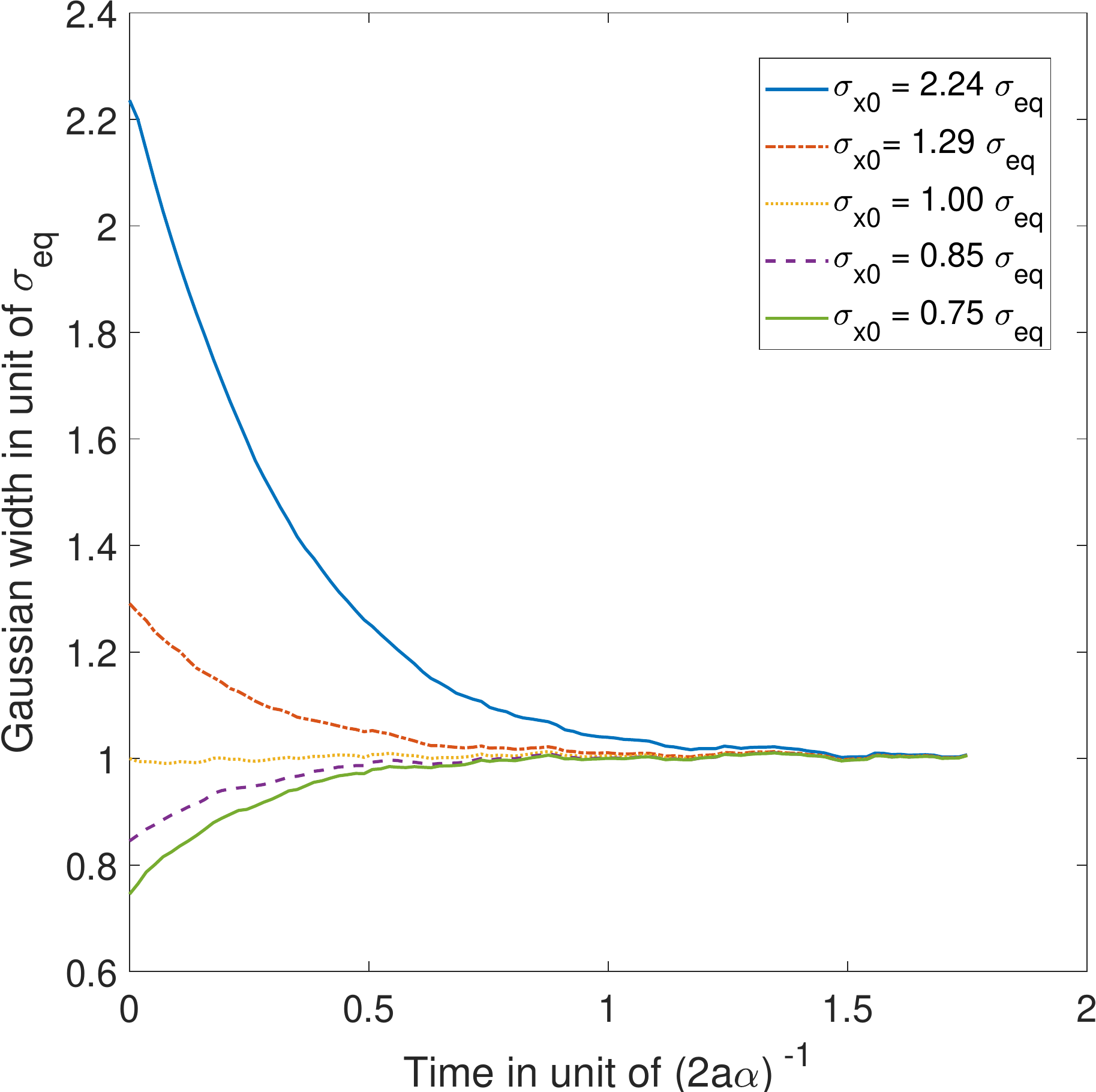}  
\captionof{figure}{\small Simulations of 10 000 trajectories {(calculated from the Ito equation \eqref{Nelsondyn1} for the ground state (\ref{gs}) of the 1D harmonic oscillator)}, whose initial positions are normally distributed, for 5 different choices of distribution width (for $a=0.5$ and $\alpha=1$). We observe, in each case, convergence to the equilibrium (\ref{gs}) as predicted by the theory. }
\label{fig2}
\end{center}
\end{figure}

Clearly, $\mean{x}\mathop{=}\limits^{t\rightarrow\infty}\mean{x}_{eq}=0$ with a characteristic relaxation time inversely proportional to the diffusion coefficient $\alpha$. Moreover, 
\begin{equation}
\frac{d\sigma_x(t)}{dt} \propto 4 a \alpha \big( \sigma_{eq}^2 - \sigma_x^2(0) \big)\, e^{-4a\alpha\,t},
\end{equation}
which has the same sign as that of the difference $( \sigma_{eq} - \sigma_x(0) )$. Hence, $\sigma_x(t)$ converges monotonically  to the equilibrium value $\sigma_{eq}$, with a characteristic time inversely proportional to the diffusion coefficient $\alpha$, as can be seen in Figure \ref{fig2}.

\subsection{Ergodicity in the relaxation to quantum equilibrium for the ground state of the harmonic oscillator}\label{ergodicity}
\begin{figure}[!t]
\begin{center}
\includegraphics[scale=0.35]{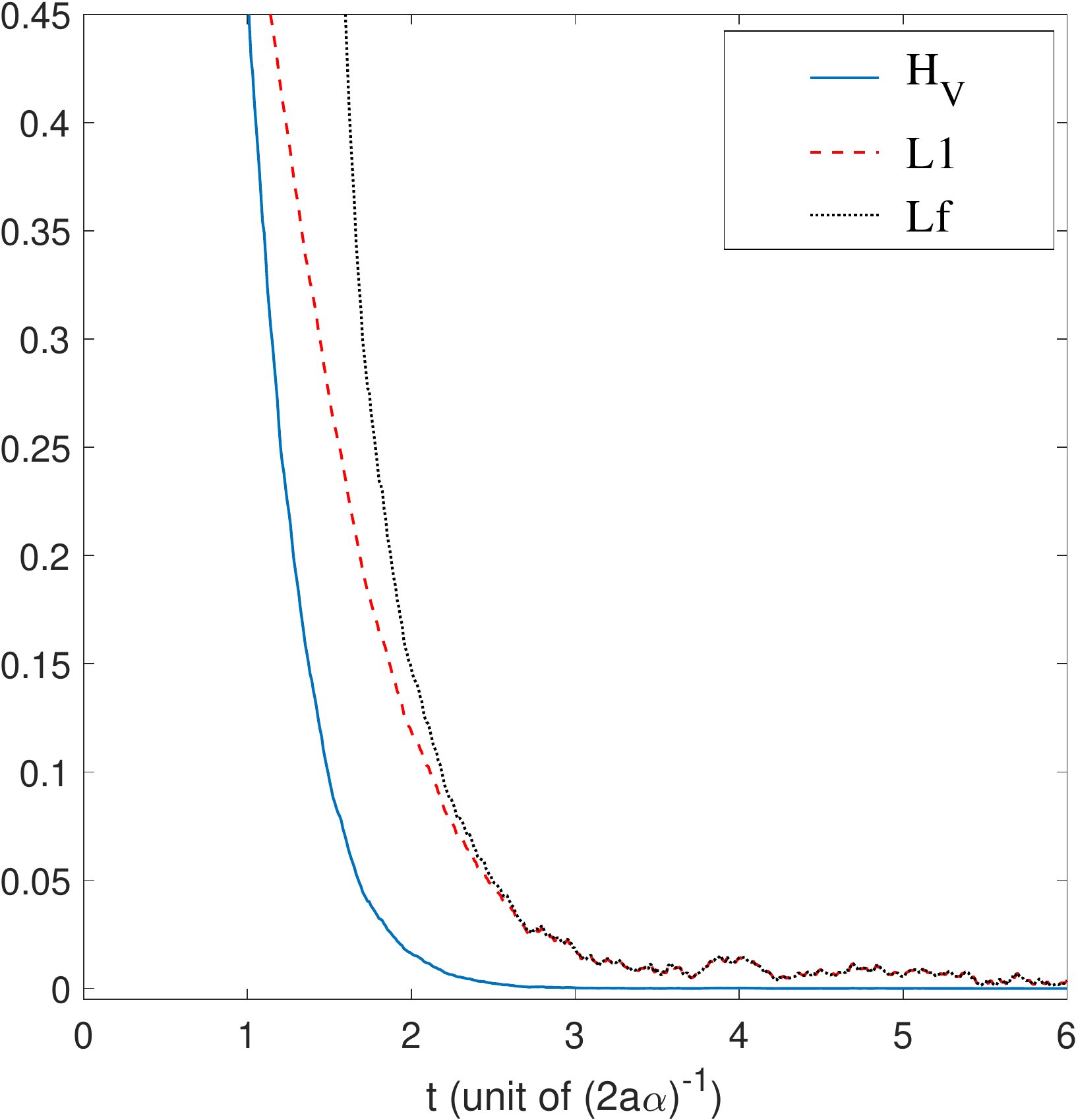}  
\captionof{figure}{\small{Time evolution of $H_V$ \eqref{hv}, $L_f$ \eqref{Lf} and the $L_1$ norm \eqref{L1} , for a uniform initial probability distribution, calculated from the Ito equation \eqref{Nelsondyn1} for the ground state of the 1D harmonic oscillator. Relaxation towards the distribution of the ground state $\vert\Psi_{st}\vert^{2}$ \eqref{gs} is clearly visible. The simulation is performed for $\alpha=1$, $a=0.5$, $\Delta t=0.01$, for $20\,000$ uniformly distributed initial conditions.}}
\label{fig-entropiesGS}
\end{center}
\end{figure}

We have just shown how Gaussian initial distributions converge towards quantum equilibrium, but one could also ask the same question for non-Gaussian initial distributions. Convergence is guaranteed by the H-theorem of section \ref{Nelsonrelax}, but contrary to the Gaussian case, we have no clear measure for the rate of convergence, except for the entropy-like functions $H_V$ \eqref{hv} and $L_f$ \eqref{Lf}, or the $L_1$ norm \eqref{L1}, defined in section \ref{Nelsonrelax}. The evolution, in time, of these three quantities is shown in Figure \ref{fig-entropiesGS}, for the stochastic trajectories obtained from 20000 uniformly distributed initial conditions.
The relaxation towards quantum equilibrium is clearly visible in all three quantities. As expected, the convergence of $H_V$ is extremely fast. {Note that, although initially very large, $L_f$ quickly matches $L_1$, up to numerical fluctuations.}

One important question concerning this relaxation process is of course that of possible ergodicity. Since we want to study the ergodic properties of Nelson dynamics in a numerical way, we choose the definition of ergodicity that is, in our approach, the easiest to test. \raw{Defining the time average $\widehat h$ of a function $h$ on $\Omega$ by the limit (if it exists)
\begin{equation}
\widehat{h} = \lim_{t\to+\infty} \frac{1}{t}\, \int_0^t h\big({\bf x}_{t'}\big)\,dt',
\end{equation}
where ${\bf x}_{t'}$ represents the position of a particle at time $t'$, as obtained form the Ito stochastic differential equation \eqref{Nelsondyn1} for an initial condition ${\bf x}$, we say\cite{ergogray} that the corresponding stochastic process is ergodic if the time average of any bounded function $h$ on $\Omega$ is always independent of ${\bf x}$.} Since for bounded $h$ the time average is also invariant under shifts in time, we can say that we have ergodicity if all time averages of such functions are in fact constants. The main reason for choosing this particular definition is that it is well-suited to empirical testing, since it is of course sufficient to establish constancy of the time averages for all indicator functions $\chi_A$ of arbitrary (measurable) sets $A\subset\Omega$,  for the analogous property to ensue automatically for all bounded functions on $\Omega$.\footnote{\raw{Another reason for choosing this particular definition is that it is also applicable to non-stationary stochastic processes, as in the case of the coherent state of section \ref{s7}. Note also that for stationary processes, this definition is of course equivalent to the usual one, asking that the time average of any bounded function on $\Omega$ is equal to its space average: $\widehat{h} =  \int_\Omega d\mu\, h({\bf x})$.}} More precisely, we {need} to verify that 
\begin{equation}
\widehat{\chi}_A = \lim_{t\to+\infty} \frac{1}{t}\, \int_0^t \chi_A\big({\bf x}_{t'}\big)\,dt',\label{chihat}
\end{equation}
is independent of both $t$ and ${\bf x}$, for any measurable $A\subset\Omega$. Remember that one has of course that $\chi_A\big({\bf x}_{t}) = \chi_{\phi_t^{-1}A}({\bf x})$, where $\phi_t^{-1}A = \left\{ {\bf x}\in\Omega\,\vert\, {\bf x}_t \in A\right\}$.

In the present case, i.e. that of the Nelson dynamics defined by the stationary (ground) state of the 1D harmonic oscillator, it is clear that the distribution $\vert\Psi_{st}\vert^2$ obtained from the ground state eigenfunction $\Psi_{st}$ is a stationary solution to the associated Fokker-Planck equation \eqref{fp}. This distribution provides a natural invariant measure $\mu$ on $\Omega$: $d\mu = \vert\Psi_{st}\vert^2 dx$, for which $ \int_\Omega d\mu=1$ and
\begin{equation}
\mu(A) = \int_A \vert\Psi_{st}\vert^2 dx = \mu(\phi_t^{-1}A),\quad ^\forall t>0,\, ^\forall\!A\in\Omega.
\end{equation}

If a stationary stochastic process is ergodic, i.e. if for such a process all $\widehat{\chi}_A$ are indeed constants, the values of these constants \raw{ are simply the measures of the subsets $A$\cite{arnavez}.} Therefore, when one needs to decide whether or not a stationary stochastic process is ergodic, it suffices to establish that $\widehat{\chi}_A=\mu(A)$, for any $A\in\Omega$.

The usual way to check this condition is to consider sampling time averages for a sufficiently refined `binning' of $\Omega$. Starting from a particular initial particle position $\bf x$, we calculate the trajectory ${\bf x}_t$ that follows from the Ito stochastic equation \eqref{Nelsondyn1}, for a sufficiently long time $t$. As was explained for the coarse-graining in section \ref{s2}, the configuration space $\Omega$ is subdivided into a large number of non-overlapping cells or `bins' $A_k$ ($k=1, \hdots, N_b$), each with the same volume $\Delta{\bf x}$. The trajectory ${\bf x}_{t'}~ (t'\in[0,t])$ is then sampled at regular intervals $\Delta t$, yielding $N+1$ sample positions ${\bf x}_{n \Delta t}~ (n=0, \hdots, N)$, for $N=t/\Delta t$. We then define the sampling function $\varphi_{N,k}$
\begin{equation}
\varphi_{N,k} = \frac{1}{N} \sum_{n=0}^N \chi_{A_k}({\bf x}_{n \Delta t}),
\end{equation}
which is a discretization of $\frac{1}{t}\int_0^t \chi_A\big({\bf x}_{t'}\big)\,dt'$ in \eqref{chihat} and which gives the frequency with which the (sample of the) orbit visited the bin $A_k$. Hence, if in the limit $N\to+\infty$, for diminishing bin sizes $\Delta{\bf x}$ and sampling steps $\Delta t$, the normalized distribution obtained from $\varphi_{N,k}/\Delta{\bf x}$ tends to a constant distribution (and, in particular, does not depend on the initial positions $\bf x$) then the stochastic process is ergodic according to the above definition. 

\begin{figure}[!t]
\centering{
\begin{subfigure}[b]{0.33\textwidth}
\textbf{a)}
\includegraphics[width=\textwidth]{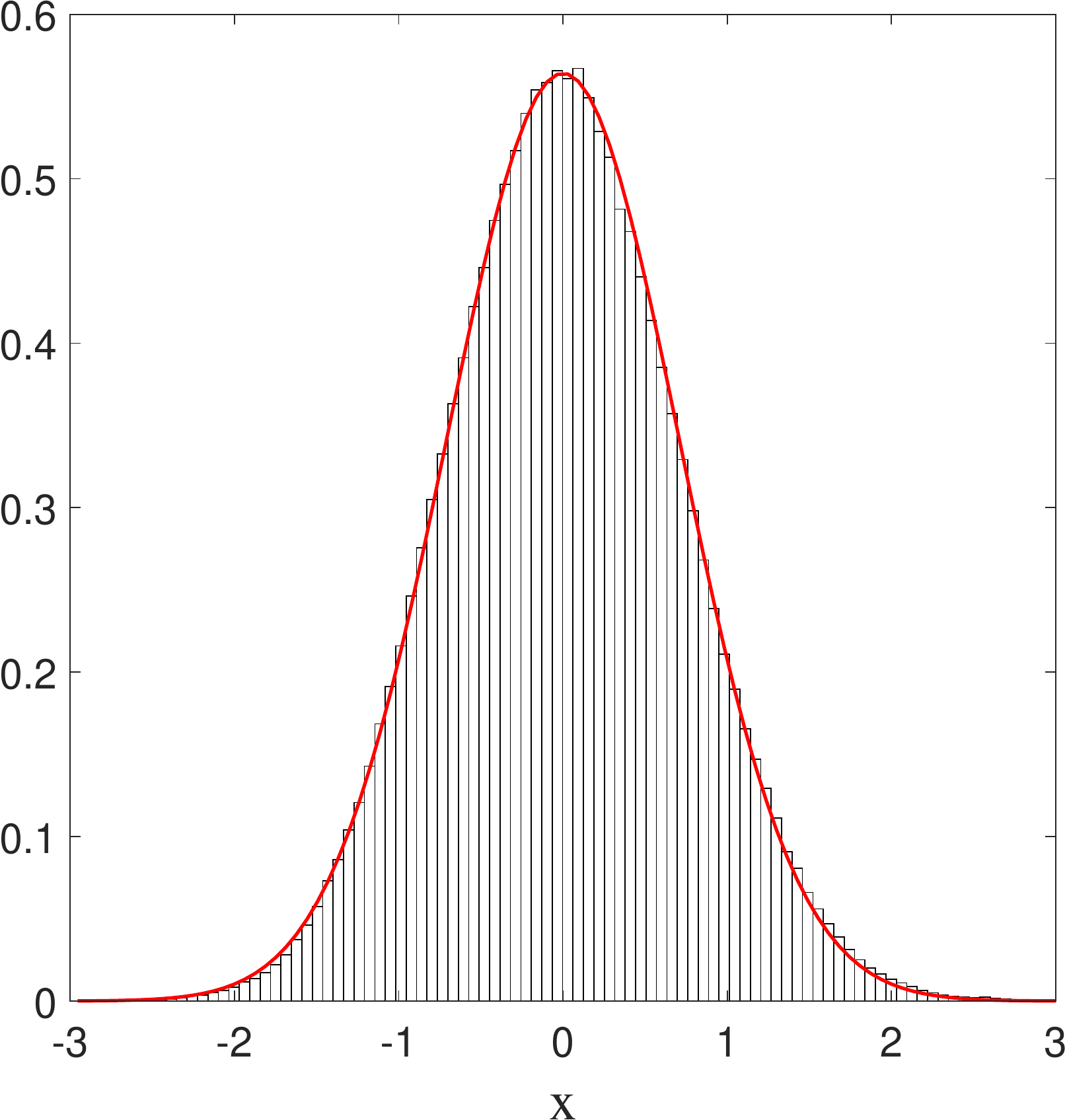}
\end{subfigure}
\begin{subfigure}[b]{0.33\textwidth}
\textbf{b)}
\includegraphics[width=\textwidth]{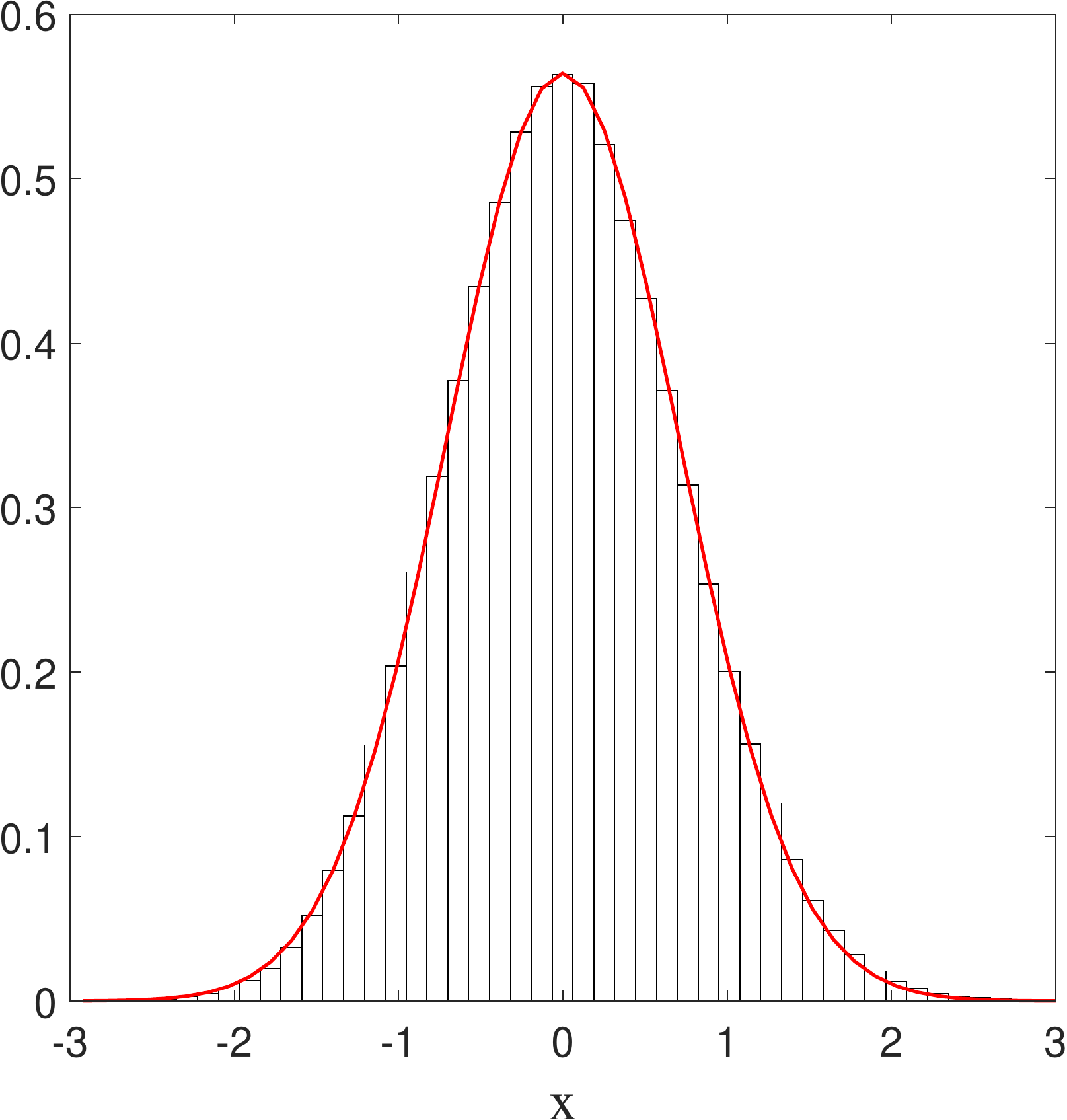}
\end{subfigure}
\begin{subfigure}[b]{0.33\textwidth}
\textbf{c)}
\includegraphics[width=\textwidth]{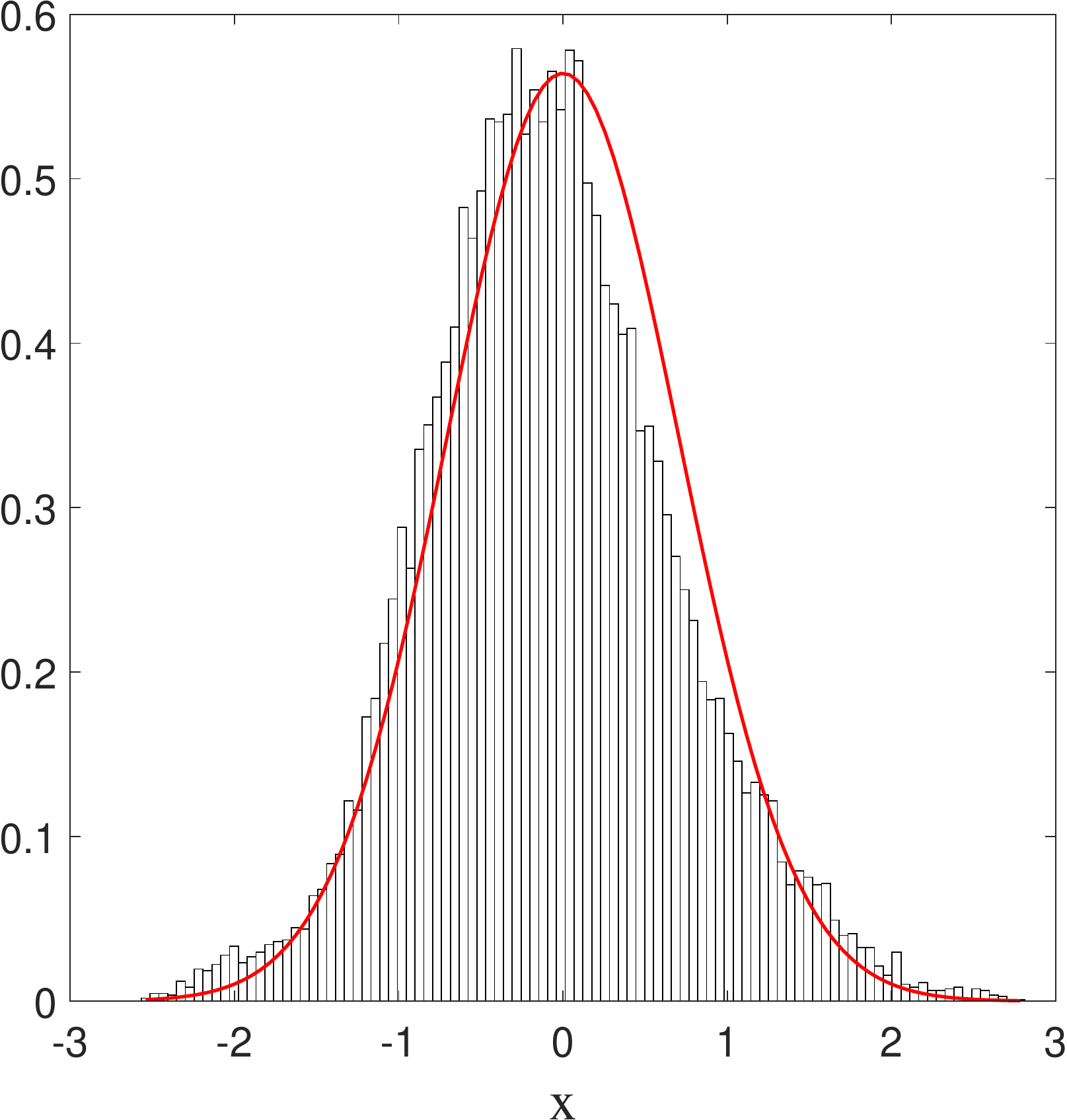}
\end{subfigure}
\begin{subfigure}[b]{0.33\textwidth}
\textbf{d)}
\includegraphics[width=\textwidth]{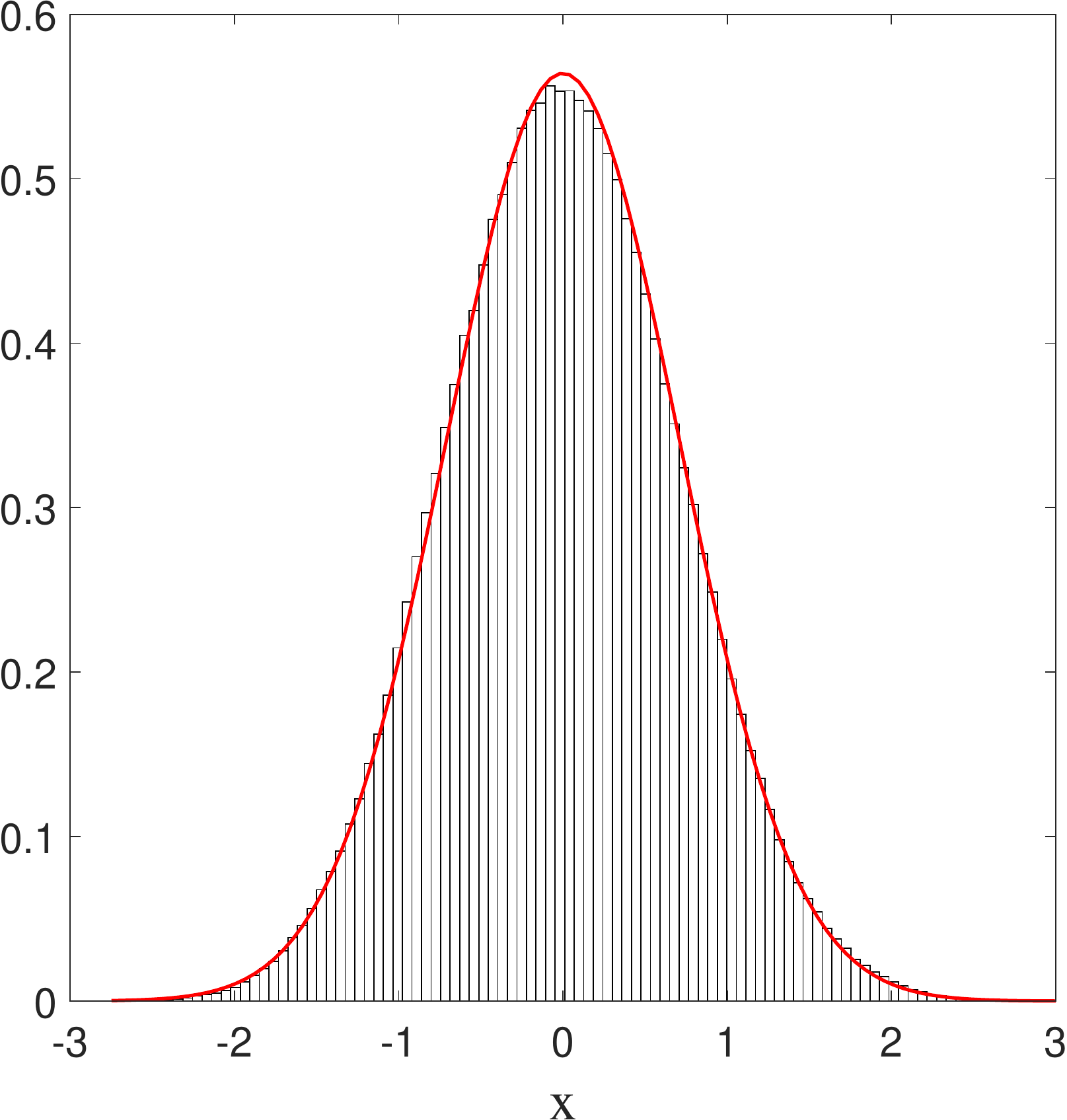}
\end{subfigure}}
\caption{\small Histograms of the positions of a single particle, subject to Nelson dynamics for the ground state of the 1D harmonic oscillator. The full (red) curve corresponds to the quantum probability $\vert\Psi_{st}\vert^{2}$. Here $a=0.5$, $\alpha=1$ and the total simulation time (t=10000) is sampled with $\Delta t=0.01$. \\ $\textbf{a)}$ The initial particle position is $x_{0}=2.5$ and the number of bins $N_{b}=100$ (each with spatial size $\Delta x=  0.0635$). $\textbf{b)}$ Same as $\textit{a)}$ but with $N_{b}=50$ and  $\Delta x=  0.1270$. $\textbf{c)}$ Same as $\textit{a)}$ but with $t=200$. $\textbf{d)}$ Same as $\textit{a)}$ but for $x_{0}=-0.85$.}
\label{gaussg}
\end{figure}

Moreover, since in that case $\widehat{\chi}_{A_k}= \mu(A_k) $, this normalized distribution must in fact coincide with that for the invariant measure for the stationary process. For example, in the case at hand, if the normalized distribution we obtain is indeed independent of the initial positions, then since $\mu(A_k) = \vert\Psi_{st}(x)\vert^2\big|_{x=\xi}\, \Delta x$ for some point $\xi\in A_k$, we must have that for sufficiently large $N$
\begin{equation}
\frac{\varphi_{N,k}}{\Delta x} \approx \frac{ \mu(A_k)}{\Delta x} = \vert\Psi_{st}(x)\vert^2\big|_{x=\xi},
\end{equation}
i.e.: the empirical distribution obtained from this sampling time average must coincide with the stationary quantum probability $\vert\Psi_{st}\vert^2$.
This is exactly what we obtain from our numerical simulations, as can be seen from the histograms depicted in Figure \ref{gaussg}. 
After a certain amount of time, the histograms we obtain indeed converge to the equilibrium distribution, and this for arbitrary initial positions. The convergence clearly improves if we increase the integration time, or if we diminish the spatial size of the bins (while diminishing the sampling time step in order to keep the occupancy rate of each bin high enough).  Although purely numerical, we believe this offers conclusive proof for the ergodicity of the Nelson dynamics associated with the ground state of the harmonic oscillator in one dimension.

\begin{center}
\begin{figure}
 \centering
\includegraphics[scale=0.33]{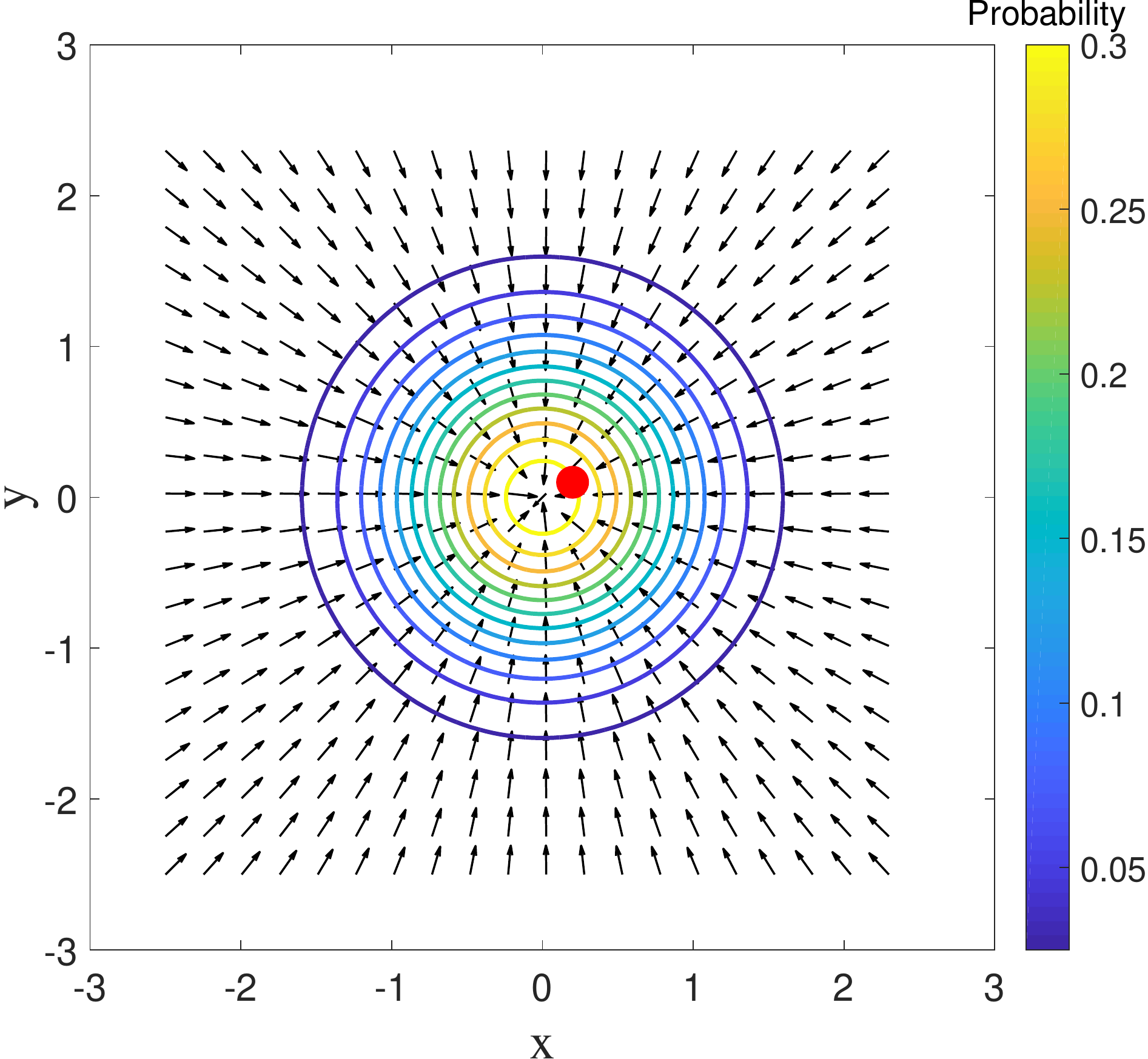}  
\includegraphics[scale=0.33]{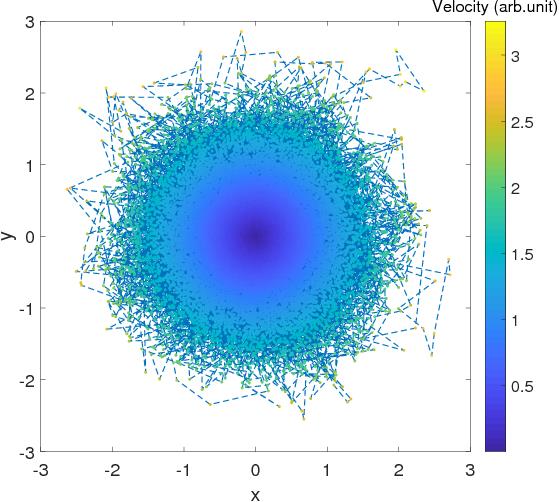}  
\captionof{figure}{\small Left: A point-particle (the dot near the center) subject to the osmotic velocity field $-2a\alpha\,(x(t),y(t))$, due to the ground state of the 2D harmonic oscillator at time $t$; \\ { Right: Color plot of the velocities along a trajectory for the evolution under Nelson dynamics, for the ground state of the 2D harmonic oscillator.}  The simulation (for $a=0.5$ and $\alpha=1$) started from the initial position $(-2,1)$ and was sampled up to $t=1\,000$ with step $\Delta t=0.01$.}\label{gaussplane}
\end{figure}
\end{center}

The same can be said, in fact,  for the 2-dimensional oscillator \raw{(and even for the 2D corral as can be seen in Ref. \cite{v3})} which will be the main topic of section \ref{s6}. Some results of a simulation of a single trajectory under the Nelson dynamics for the ground state of this system are shown in Figure \ref{gaussplane}, in which the red dot in the plot on the left-hand side indicates the (final) position of the particle at time $t$. The probability distribution obtained by sampling the trajectory, clearly decreases with the distance to the origin along concentric circles. 

{\section{Nelson dynamics: a phenomenological dynamical model for walkers?}\label{s6}}\subsection{2D harmonic oscillator \label{s6bis}}

Experimentally, it has proven possible to study the dynamics of bouncing droplets under the influence of an effective harmonic potential in two dimensions, thanks to a well-chosen electro-magnetic configuration and magnetic droplets \cite{Labousse2016}. It {is therefore interesting} to compare predictions that we, on our side, can make in the framework of Nelson dynamics, with actual experimental observations of droplets dynamics\footnote{See \cite{Grossing2010} for a pioneering work very similar to ours in the case of the double slit experiment.}. To our mind, one important comparison to make concerns the convergence to equilibrium.

For example, if the initial distribution of positions projected along a reference axis, say $X$,  fits a mixture of the ground state and the $n$th Fock  \raw{(\cite{v3}, appendix 2A)} state ($n=1,2\cdots$) for the 2D harmonic oscillator (conveniently weighted in order to respect the ineluctable constraint of positivity), \rw{our Nelson-like} model predicts that the typical time of convergence to equilibrium will scale like the inverse of the eigenvalue of the nth Fock state, i.e. as $1/n$, which constitutes a very precise quantitative prediction. This follows from the representation \eqref{pn}, when $\sqrt{P_{st}(x)}$ is the Gaussian ground state of the 1D harmonic oscillator and where the eigenfunctions $g_k$ are the \raw{energy eigenstates (Fock states) of the harmonic oscillator}  {(this{, of course,} because of the separability of the Schr\"odinger equation and of \rw{ our Nelson dynamics} along $X$ and $Y$ in the case of an isotropic 2D oscillator)}. 

A possible way to measure this characteristic time would be to record {the projections along $X$ of} trajectories that correspond to an equally spaced grid of initial positions, weighted such as to fit a mixture of the ground state with the $n$th Fock state ($n=1,2\cdots$), and to compare the histogram constructed in this way at different times with theoretical predictions derived from (\ref{pn}). 

Another precise quantitative (theoretical) prediction, which is even simpler to verify, is that if we prepare {a droplet many times at exactly the same initial position,} the position obtained after averaging over all trajectories will (1) decrease exponentially in time and (2) be characterised by a decay time which scales like $1/a\alpha$, by virtue of the discussion in section \ref{ss54} and in particular equation 
(\ref{meanNels}).  \rw{It has been suggested that droplet trajectories might be characterized by a quantum-like Zitterbewegung, which can be seen in relativistic quantum dynamics as an intrinsic brownian motion at the Compton scale \cite{hestenes,cowi} and various proposals have been formulated in order to express the amplitude and frequency of this Zitterbewegung \cite{Bush2015,Gilet} in terms of the parameters characterizing droplet dynamics \raw{(these are e.g. the viscosity of the fluid, the mass of the droplets, the ratio between the amplitude of the vibrations imposed on the bath and the Faraday threshold, the oil temperature and so on)}.  Exploring these analogies in depth lies beyond the scope of this paper, but the aforementioned attempts, \raw{(Ref. \cite{Gilet} in particular)}, pave the way for \raw{introducing a brownian component in the description of} droplet trajectories.}

\subsection{{Presence of zeros in the interference pattern}}\label{corralsec}
One of our first motivations, when we decided to incorporate a Brownian component in the dBB theory in order to simulate the dynamics of droplets, was the pioneering paper \cite{Harris2013} reporting on observations of a walker trapped in a spherical 2D cavity (corral), for which the histogram of positions occupied over time by a single droplet trajectory faithfully reproduces the Bessel function $J_0$.\footnote{Which is also related to the Green function of the Helmholtz equation, with  a typical length equal to the Faraday wave length of the vibrating bath over which droplets propagate \cite{Dubertrand2016}.} These observations reveal, in a telling way, the presence of a pilot-wave that guides the dynamics of the particles, and also raise the question of ergodicity. 

{If {we try the approach we used} for the 2D harmonic oscillator in the case of the corral (effectively replacing the Gaussian ground state of the 2D harmonic oscillator by the zero order Bessel function), we {are immediately confronted} with problems caused by the presence of zeros in the Bessel function.}  These problems are briefly explained in appendix, where we show that certain formal methods aimed at solving the Fokker-Planck equation (such as those introduced in section \ref{ss54}) are only relevant when the pilot wave possesses no zeros. In particular, the eigenvalues $-\lambda_{k}$ of the Fokker-Planck operator (\ref{schgk}) are not always negative when zeros are present, which of course would menace the stability of the {relaxation} process. 
However, as we already indicated in section \ref{Nelsonrelax}, although the effect of zeros of the pilot wave in \rw{our} Nelson dynamics is by no means trivial, there are several observations that indicate that this problem is not really crucial.

First of all, as mentioned in section \ref{Nelsonrelax}, the Wiener process makes it in principle possible to ``jump'' over the zeros of the equilibrium distribution. This has actually been confirmed in numerical simulations for the case of the 1D harmonic oscillator, where we imposed that the equilibrium distribution $P_{st}$ is the square modulus of the first excited (Fock) state \raw{(\cite{v3}, appendix 2A)}, with amplitude: 
 \begin{equation}
 P_{st} = |\Psi_{st}|^2=|\Psi_{1}(x,t)|^2 =\left(\frac{2 a}{\pi }\right)^{\frac{1}{2}} \left( a \, x^2\right)e^{-2a x^2}.
 \label{psi1}
 \end{equation}
 
\begin{figure}[!t]\centering
\includegraphics[scale=0.4]{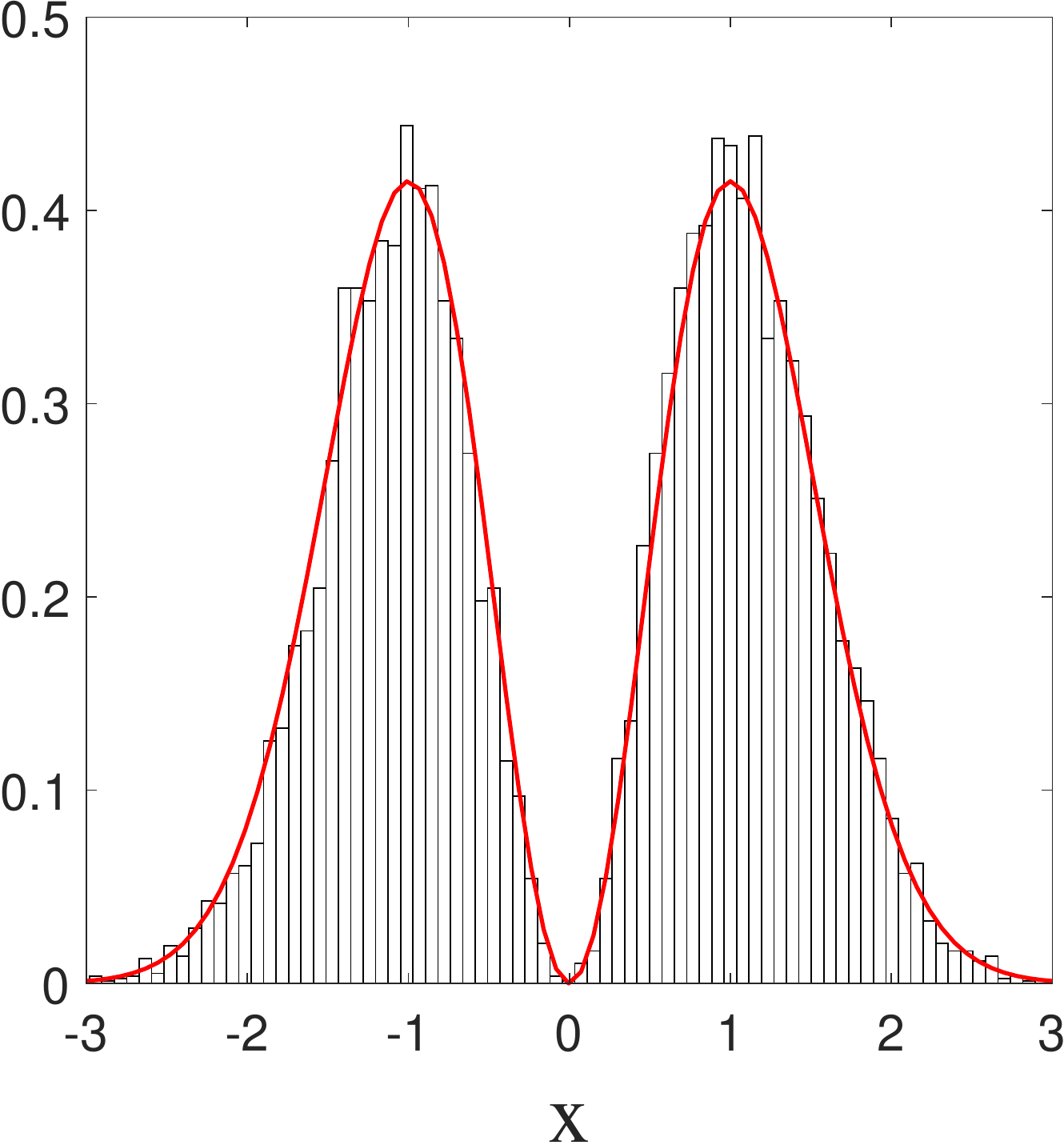}  
\caption{\small Histogram of the positions in $x$ of a single particle, in the case of the first Fock state \eqref{psi1}. The full curve (red) corresponds to the quantum probability $\vert\Psi_{1}\vert^{2}$. Here $a=0.5$ and $\alpha=1$. The total simulation time t is $t=1000$ and the sampling time step is $\Delta t=0.01$. The initial position is $x_{i}=1$ and the number of bins $N_{b}=75$, each with width $\Delta x=  0.08$.}
\label{fn1}
\end{figure}

\begin{figure}[!t]\centering
\includegraphics[scale=0.35]{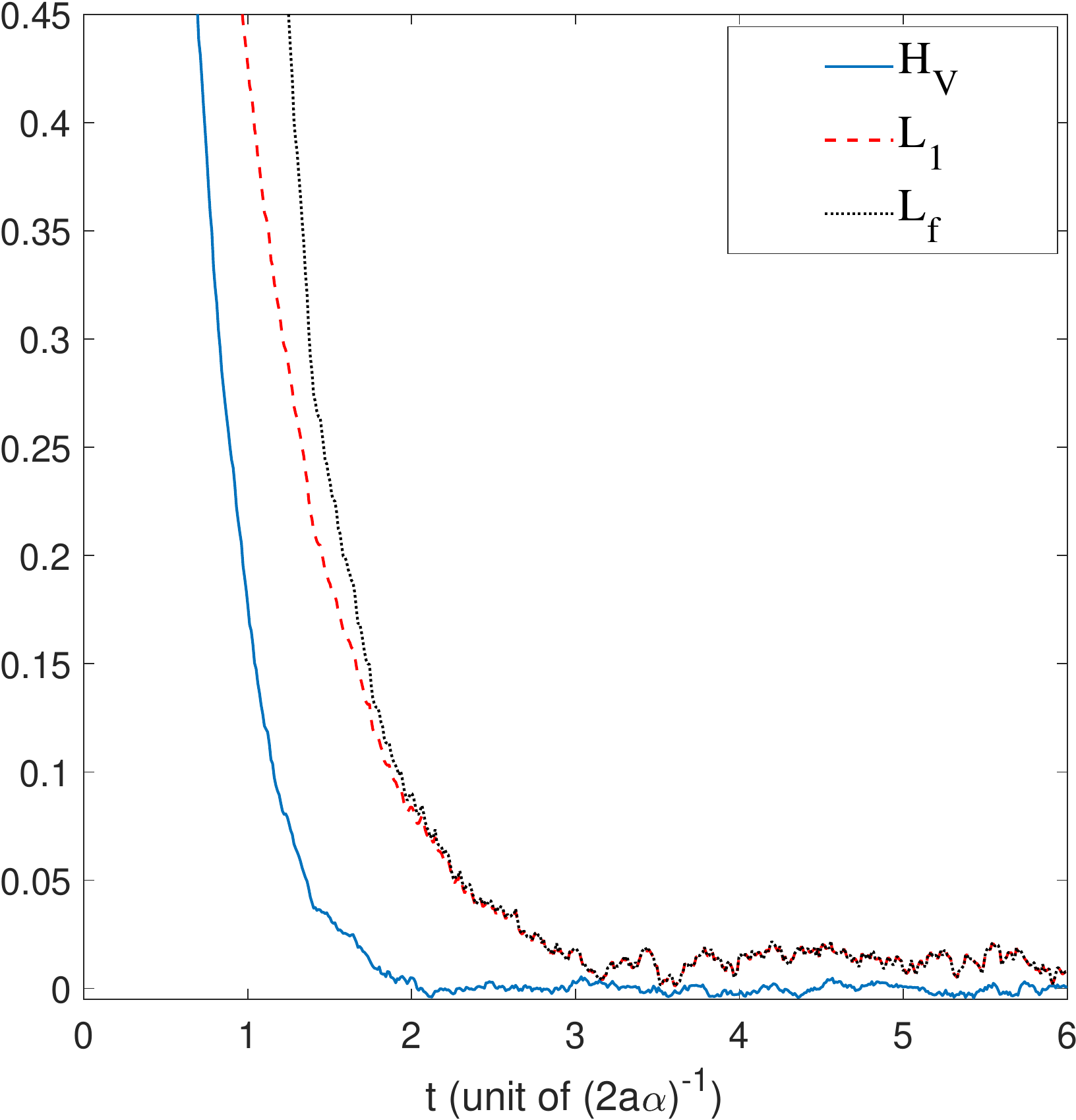}  
\caption{{\small Evolution in time of  $H_V$ \eqref{hv}, $L_f$ \eqref{Lf} and of the $L_1$ \eqref{L1} norm, for a uniform initial probability distribution, showing the relaxation towards the distribution of the first excited state $\vert\Psi_{1}\vert^{2}$ \eqref{psi1}. The simulation is performed for $\alpha=1$, $a=0.5$, $\Delta t=0.01$ and from $20\,000$ uniformly distributed initial conditions.}}
\label{hlfUnitopsi1}
\end{figure}

Indeed, as can be clearly seen from Figure \ref{fn1}, the particle will, from time to time,  jump over the zero in the middle (with the same probability from left to right as in the opposite direction), in such a way that finally the trajectory covers the full real axis, while the histogram of positions faithfully reproduces the quantum prediction $P_{st}=|\Psi_{st}|^2=|\Psi_{1}(x,t)|^2$. This indicates that even in the presence of a zero in the equilibrium distribution, the relaxation process is still ergodic.
The relaxation of a uniform initial distribution to this quantum equilibrium is shown in Figure \ref{hlfUnitopsi1}, for the quantities $H_V, L_f$ and $L_1$.

\begin{figure}[!h]
 \centering
  \includegraphics[width=0.365\linewidth]{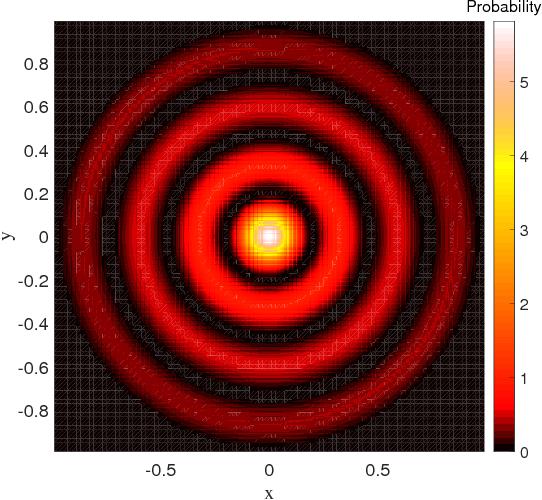}
  \includegraphics[width=0.39\linewidth]{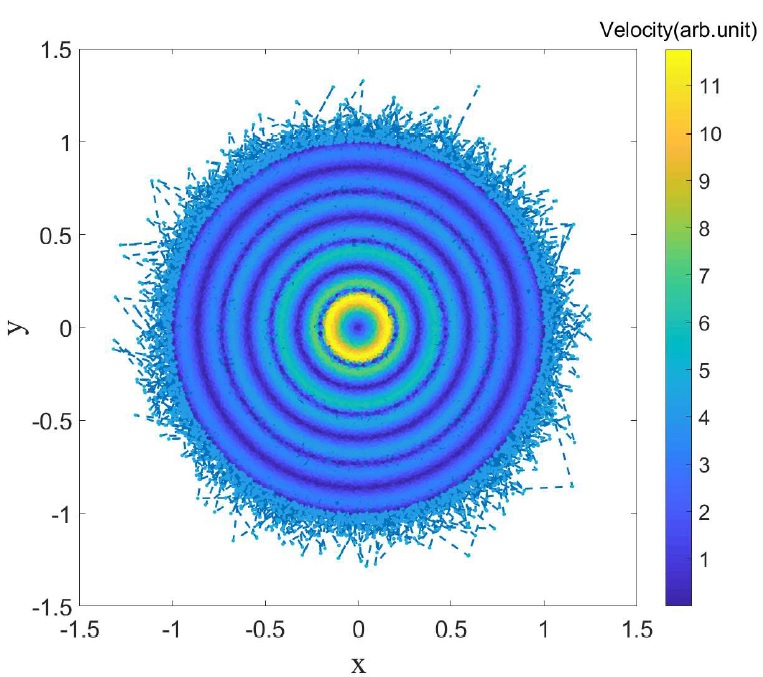}
\captionof{figure}{\small Left: The quantum probability associated to the Bessel function of the first kind $J_{0}$. {Right: Color plot of the velocities reached along the trajectory for an evolution corresponding to \eqref{vj0}.} The initial position was $(1,1)$, the simulation time $t=5\,000$ and the sampling time step $\Delta t=0.005$. We chose $\alpha=0.1$, $\epsilon=0.2$ and the size of the domain is $L=2$. On the boundary we impose a harmonic field force of the form: $-2a\alpha\,\textbf{r}$.}
\label{bessel}
\end{figure}

A second indication that the problem posed by the presence of zeros is not so serious, stems in fact from the experimental observations.  Indeed, if we study the observations reported in \cite{Harris2013} for the case of a corral, it is clear that the minima of the histogram expressing the  distribution of positions of the droplet are in fact not zeros. This, undoubtedly, due to the presence of a non-negligible residual {background}. Actually, without this {background}, the droplet would never pass between regions separated by zeros: due to the rotational symmetry of the corral, the zeros form circles centered at the origin and the position histogram obtained from a trajectory would remain confined to a torus comprising the initial position. This, however, is clearly not the case.
Which then suggests the following strategy: to simulate Nelson dynamics with a static distribution $P_{st}=\vert\Psi_{st}|^2$ given by the Bessel function $J_0$ but supplemented with a constant {positive background ${\epsilon}$},
\begin{equation}
d\textbf{x}(t)= \frac{\alpha}{2}\frac{\boldsymbol{\nabla}J_{0}(r)^{2}}{J_{0}(r)^{2}+\epsilon}\,dt + \sqrt{\alpha}d\boldsymbol{W}(t) .
\label{vj0}
\end{equation}
In this case, the singularities of the Fokker-Planck equation automatically disappear and, despite the fact that we have no analytic expression for the solutions as in the case of the ground state of the harmonic oscillator, we are able to numerically simulate Nelson dynamics without difficulty. The results of these simulations are shown in Figure \ref{bessel}. The osmotic velocity in the Nelson dynamics clearly tends to bring back the particle to regions where $\vert\Psi\vert^{2}$ has extrema and the resemblance with the plot on the left is striking. The fact that this result again does not depend on the choice of initial condition strongly suggests that the relaxation process to quantum equilibrium is also ergodic in this case.

{\section{ Relaxation to quantum equilibrium with dBB and Nelson dynamics: non-static case \label{s7}}}

\subsection{Nelson dynamics and asymptotic coherent states}\label{CSrelax}
Up to now, we have developed analytic and numerical tools aimed at {studying} the onset of equilibrium when the asymptotic {equilibrium} distribution is static. Actually, as the H-theorem of section \ref{Nelsonrelax} is also valid for non-stationary processes, one of course expects relaxation to take place even if the asymptotic state is not static, for instance when it is a Gaussian distribution the center of which periodically oscillates at the classical frequency $\omega$  of the oscillator without deformation (typical for coherent states). In fact, our numerical simulations not only show that equilibrium is reached even in this case, but also that this relaxation is ergodic.

More precisely, we considered a wave function in the coherent state
\begin{equation}
\Psi(x,t)=\left(\frac{2a}{\pi}\right)^{\frac{1}{4}}\,e^{-a\,\left(x-\bar{x}_t\right)^{2}+i\frac{\bar{p}_{t}\,x}{\hbar}+i\varphi(t)},
\label{cs}
\end{equation}
where $\varphi$ is a global phase containing the energy and $\bar{x}_{t}$ ($\bar{p}_{t}$) is the mean position (momentum) of a classical oscillator at time t:
\begin{equation}
\bar{x}_{t} = \bar{x}_{0}\,cos\left(\omega t\right)
   \quad\text{and}\quad 
\bar{p}_{t} =-m \bar{x}_{0}\,sin\left(\omega t\right),
\end{equation}
with $\omega= 2 a \alpha$ ($\alpha=\hbar/m$).
For this ansatz we solved the Ito equation \eqref{Nelsondyn1} numerically for a collection of initial conditions.
\begin{figure}[!t]
\begin{center}
    \includegraphics[scale=.35]{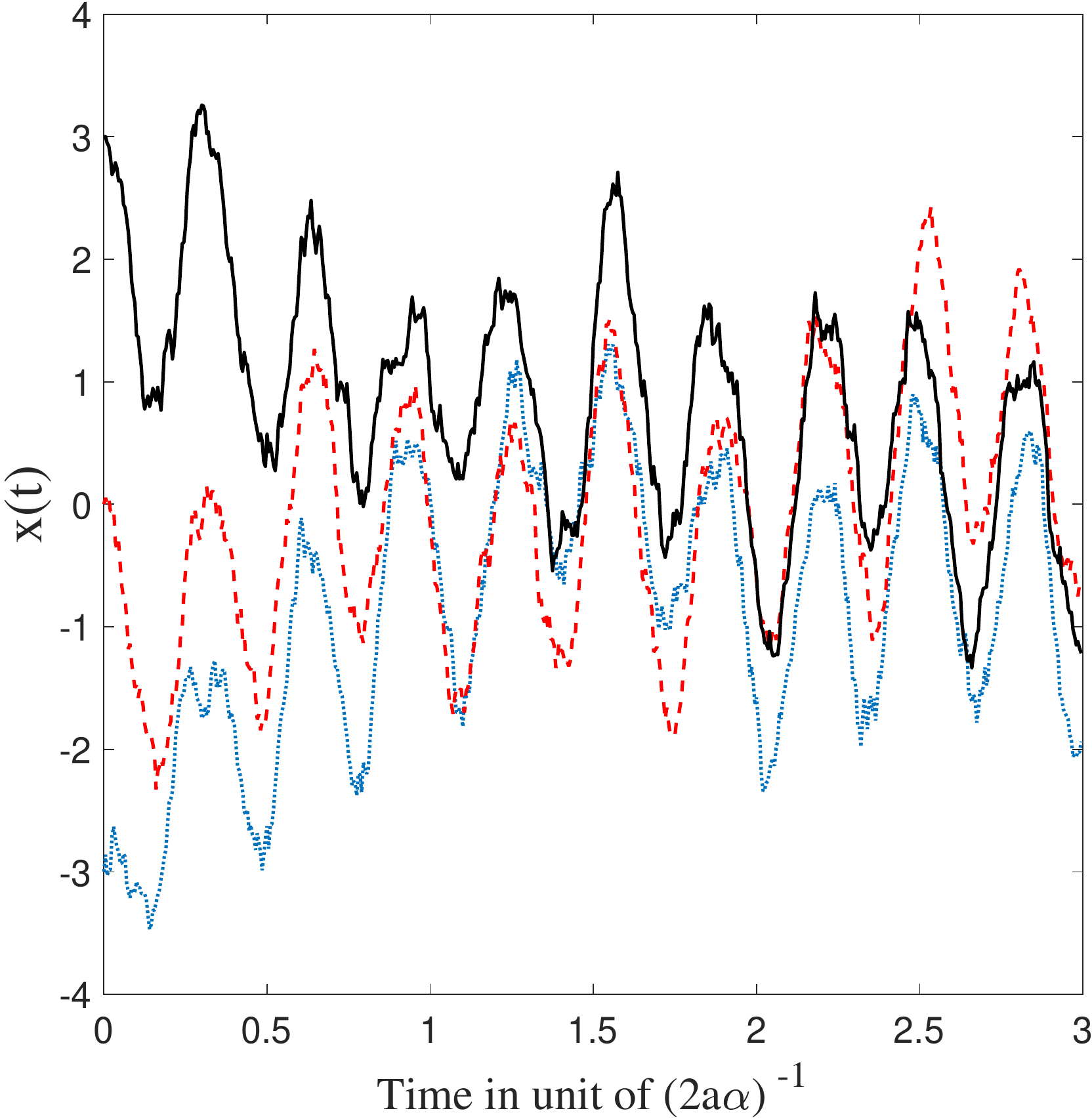}
\caption{\small Numerical solutions of the Ito stochastic differential equation~\eqref{Nelsondyn1} corresponding to the coherent state \eqref{cs}, for three different initial conditions. We used $\bar{x}_{0}=1$, $a=0.5$, $\alpha=1$ and expressed the results in natural units.}
\label{3rea}
\end{center}
\end{figure}

\begin{figure}[!t]
\centering
\begin{subfigure}[b]{0.3\textwidth}
\textbf{a)}\,t=0
\includegraphics[width=\textwidth]{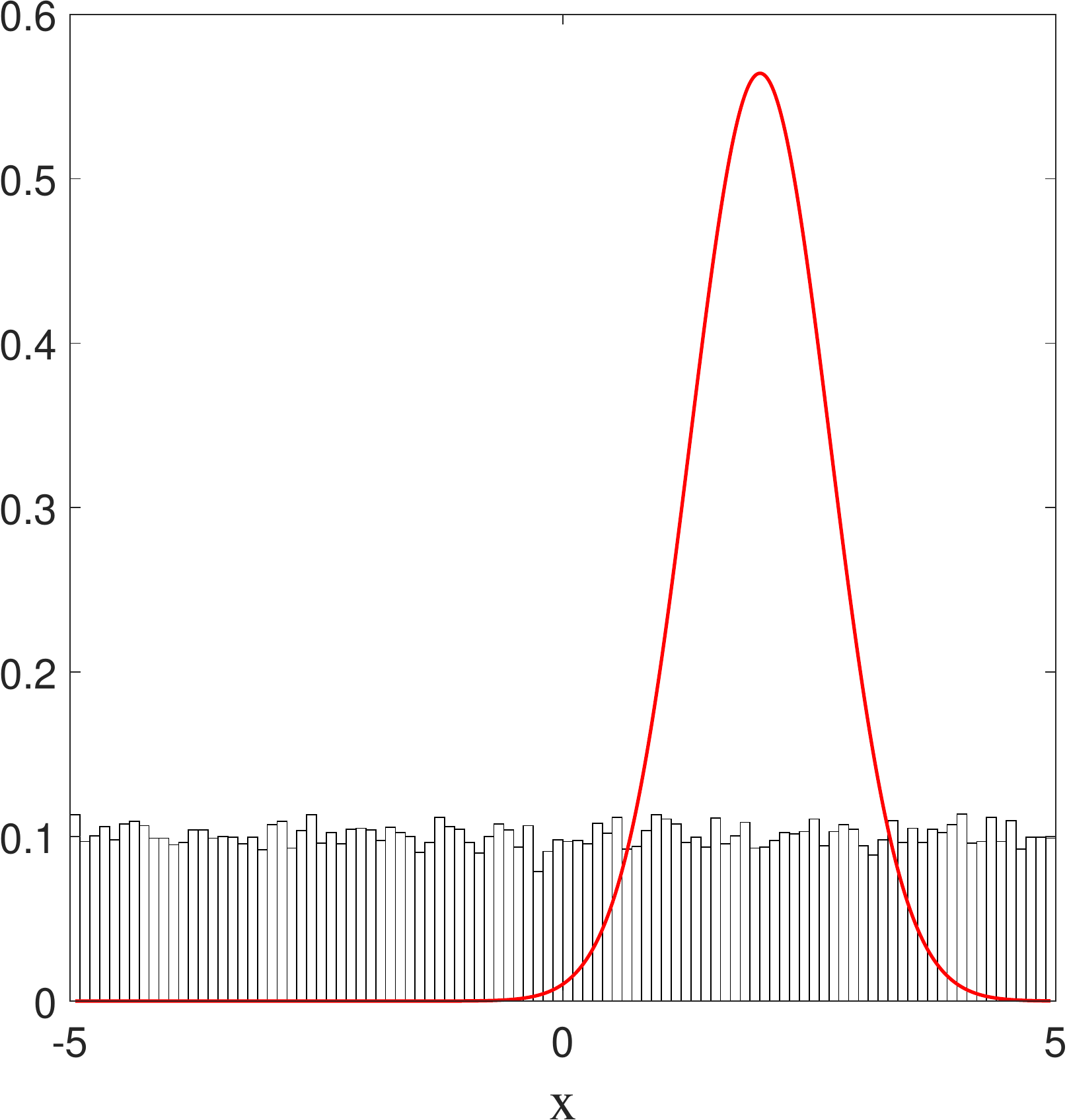}
\end{subfigure}
\begin{subfigure}[b]{0.3\textwidth}
\textbf{b)}\,t=1.2
\includegraphics[width=\textwidth]{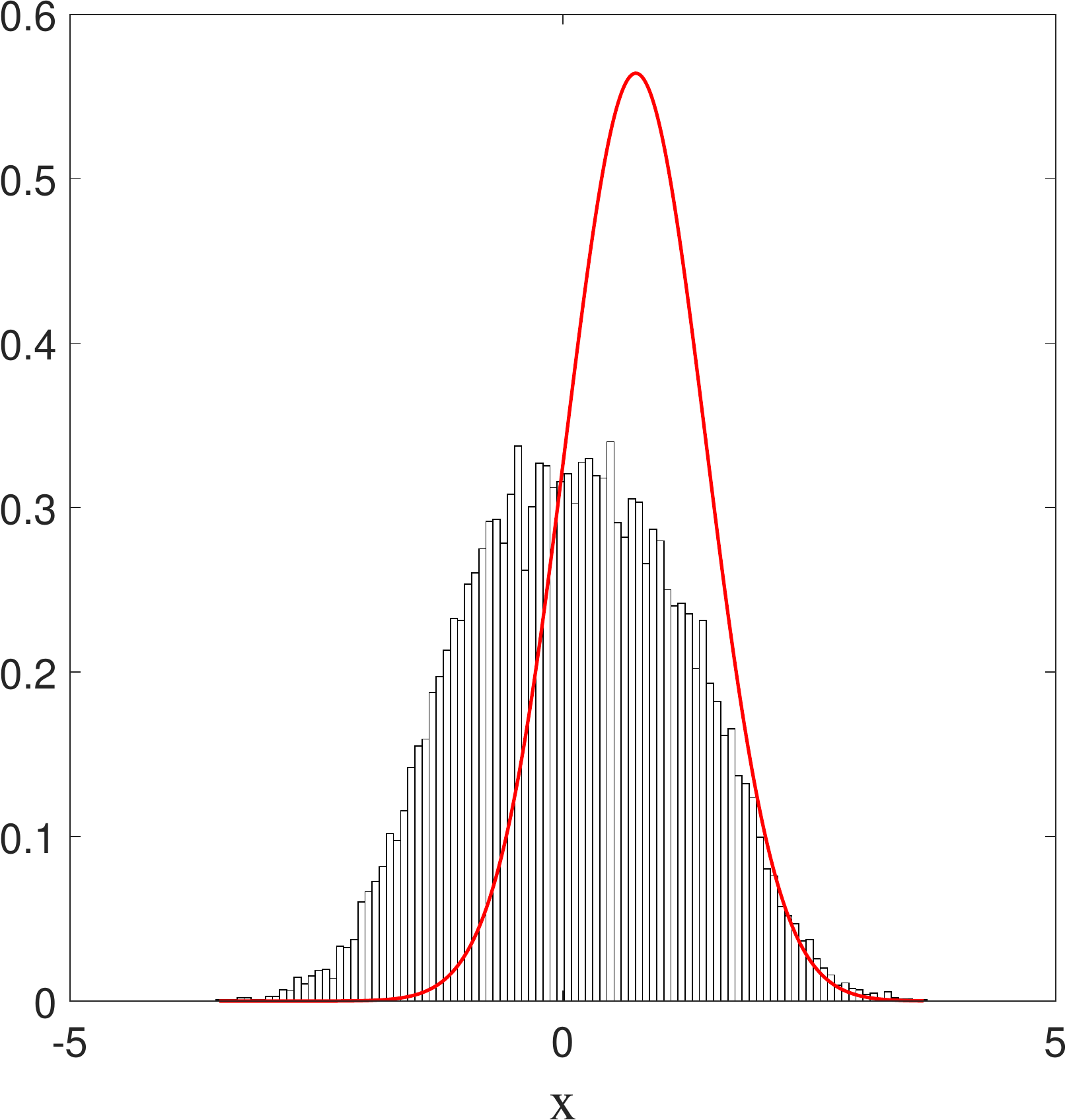}
\end{subfigure}
\begin{subfigure}[b]{0.3\textwidth}
\textbf{c)}\,t=2.4
\includegraphics[width=\textwidth]{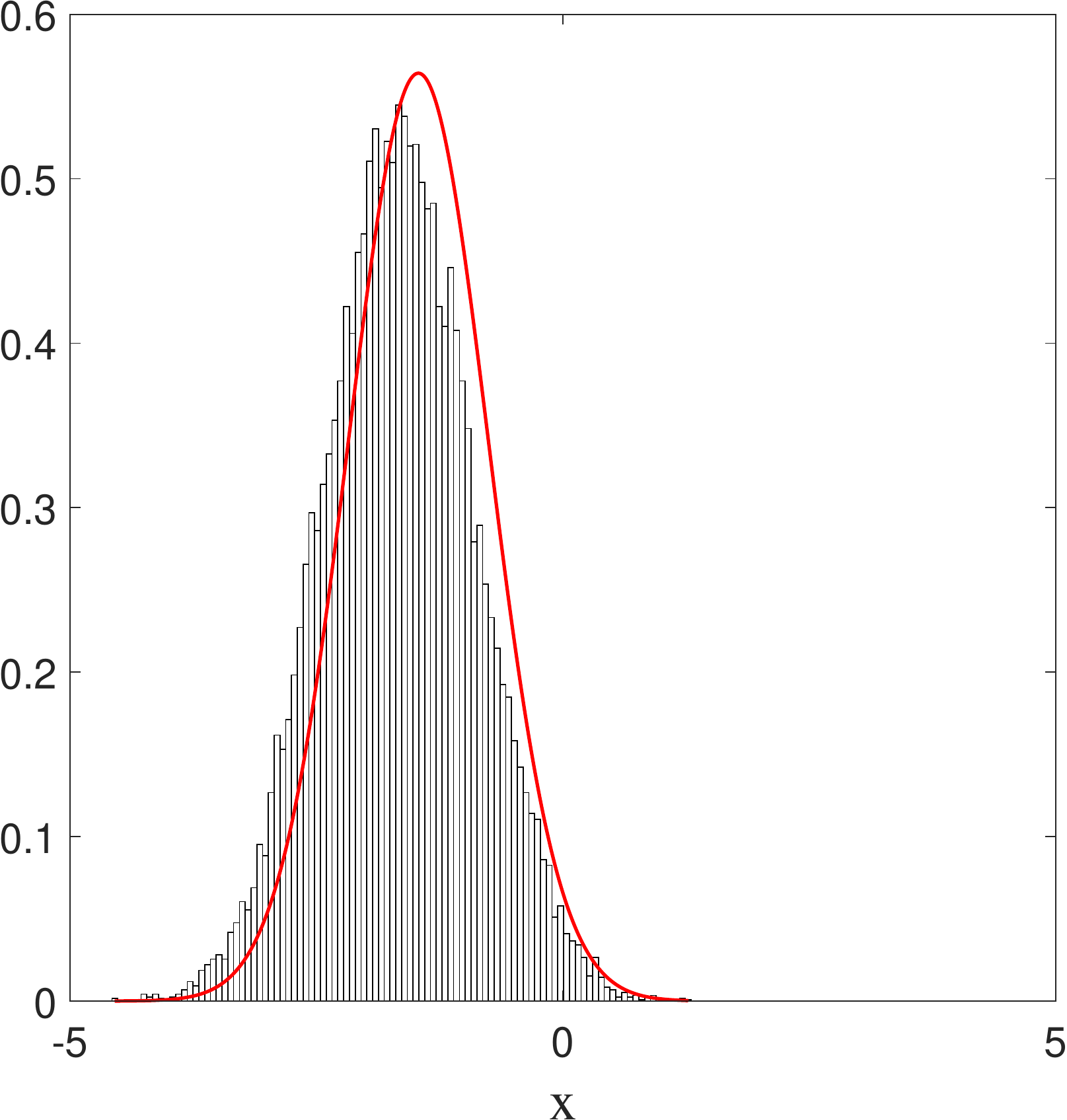}
\end{subfigure}
\begin{subfigure}[b]{0.3\textwidth}
\textbf{d)}\,t=3.6
\includegraphics[width=\textwidth]{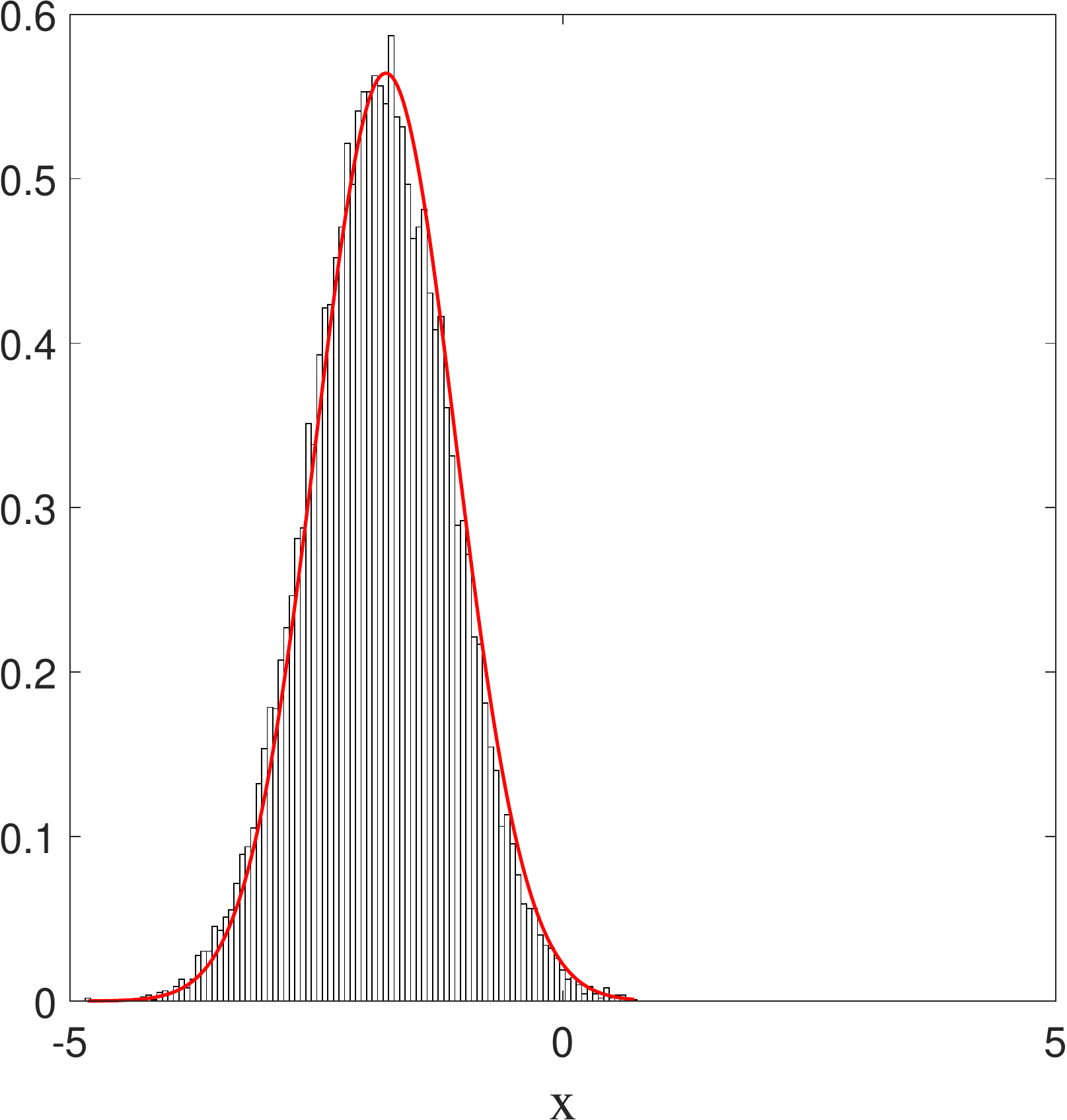}
\end{subfigure}
\begin{subfigure}[b]{0.3\textwidth}
\textbf{e)}\,t=4.8
\includegraphics[width=\textwidth]{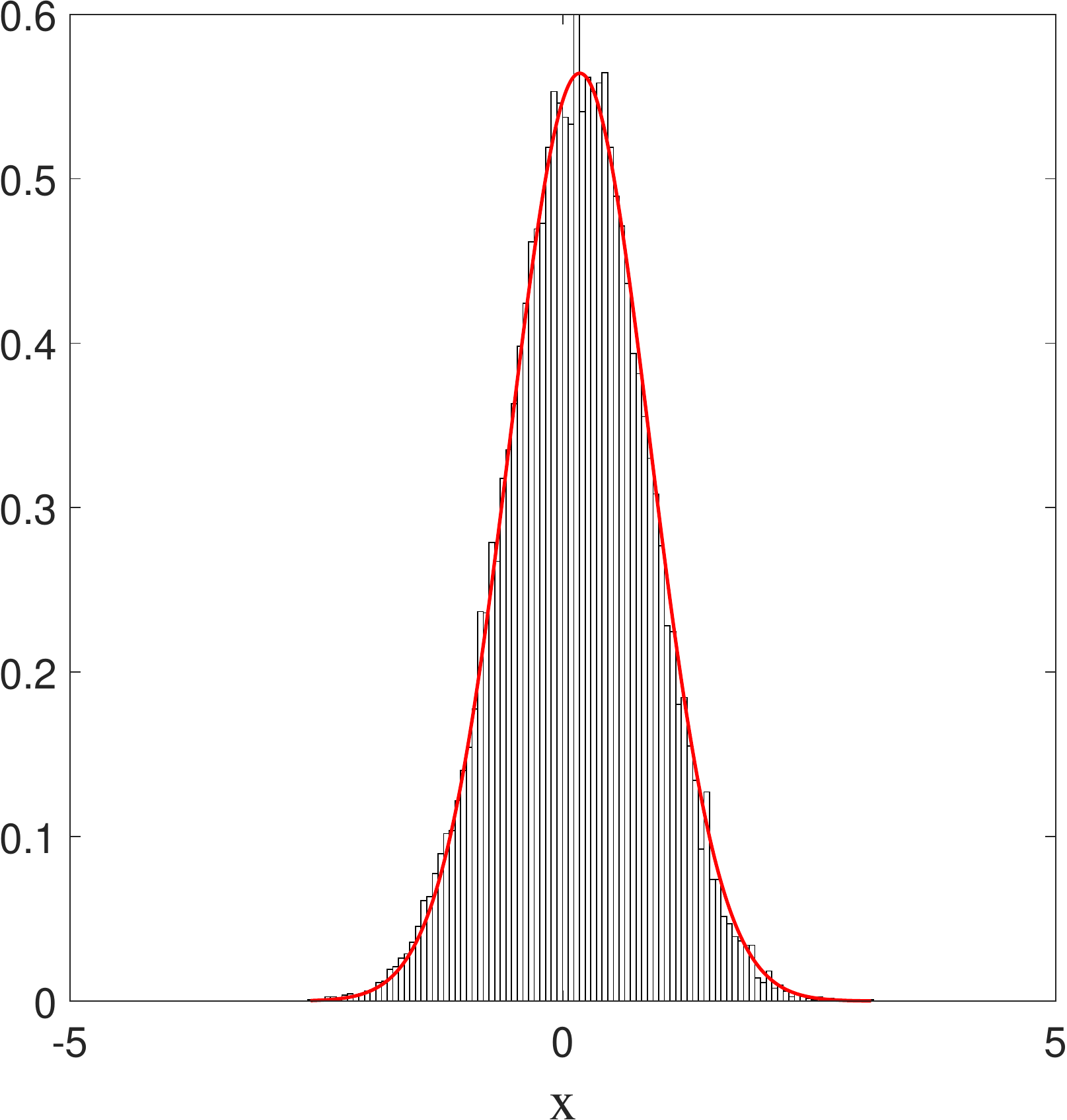}
\end{subfigure}
\begin{subfigure}[b]{0.3\textwidth}
\textbf{f)}\,t=6
\includegraphics[width=\textwidth]{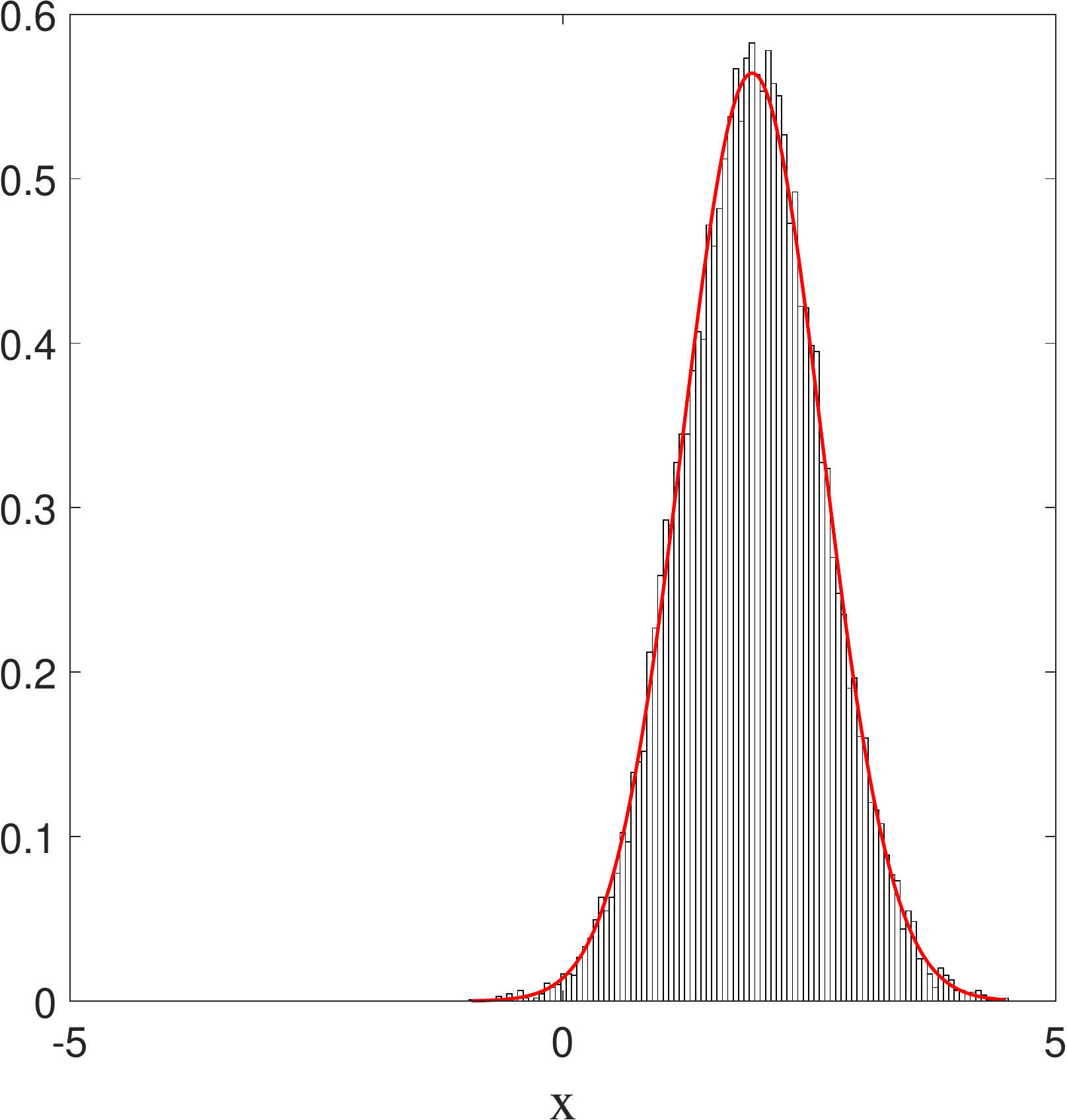}
\end{subfigure}
\caption{\small The time evolution of a non-equilibrium  ensemble, illustrated with position histograms at six different times. The continuous curve is the squared modulus $\vert\Psi\vert^{2}$ for the coherent state (\ref{cs}). As can be seen from figures (d,e,f), once equilibrium is reached, the distribution clings to the coherent state and follows its oscillation faithfully. The center of the wave packet moves between $-2$ and $2$ with a period $2\pi$. We started from a uniform distribution of initial conditions and chose $a=0.5$, $\alpha=1$ and $x_{0}=2$. The sampling time step is $\Delta t=0.01$ and the number of bins is $N_{b}=50$, each with width $\Delta x= 0.0461$.}
\label{rel}
\end{figure}

As can be seen on Figure \ref{3rea}, the trajectories are affected by the stochastic evolution but keep oscillating at the same period because of the deterministic part of the Ito process. 
Notice however that the trajectories seem to be getting closer to classical trajectories that only differ from each other by a simple shift. 
This can be explained as follows:  at equilibrium (cf. Figure \ref{rel}), the Brownian motion is balanced by the osmotic velocity and the dBB velocity is recovered ``on average''. 
Now,  the center of the Gaussian distribution moves at a classical velocity by virtue of Ehrenfest's theorem and, moreover, in the present case the dBB velocities can only depend on time and not on space as the envelope of a coherent state moves without deformation. Hence, the dBB trajectories obtained at equilibrium are, in fact, classical trajectories that only differ by a mere shift in space (the magnitude of which however may change over time).

\begin{figure}[!t]\centering
    \includegraphics[scale=0.33]{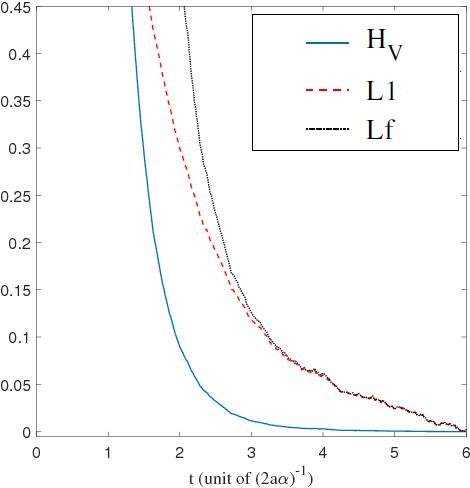}
\captionof{figure}{{\small Time evolution of $H_V$ \eqref{hv}, $L_f$ \eqref{Lf} and $L_1$ \eqref{L1}, for a uniform initial probability distribution, showing the relaxation towards the distribution $\vert\Psi\vert^{2}$ of the coherent state \eqref{cs}. \\ The simulation is performed for $\alpha=1$, $a=0.5$, $\Delta t=0.01$ and from $20\,000$ uniformly distributed initial conditions.}}
\label{fig-entropiesCS}
\end{figure}

\begin{figure}[!t]\centering
\includegraphics[scale=0.33]{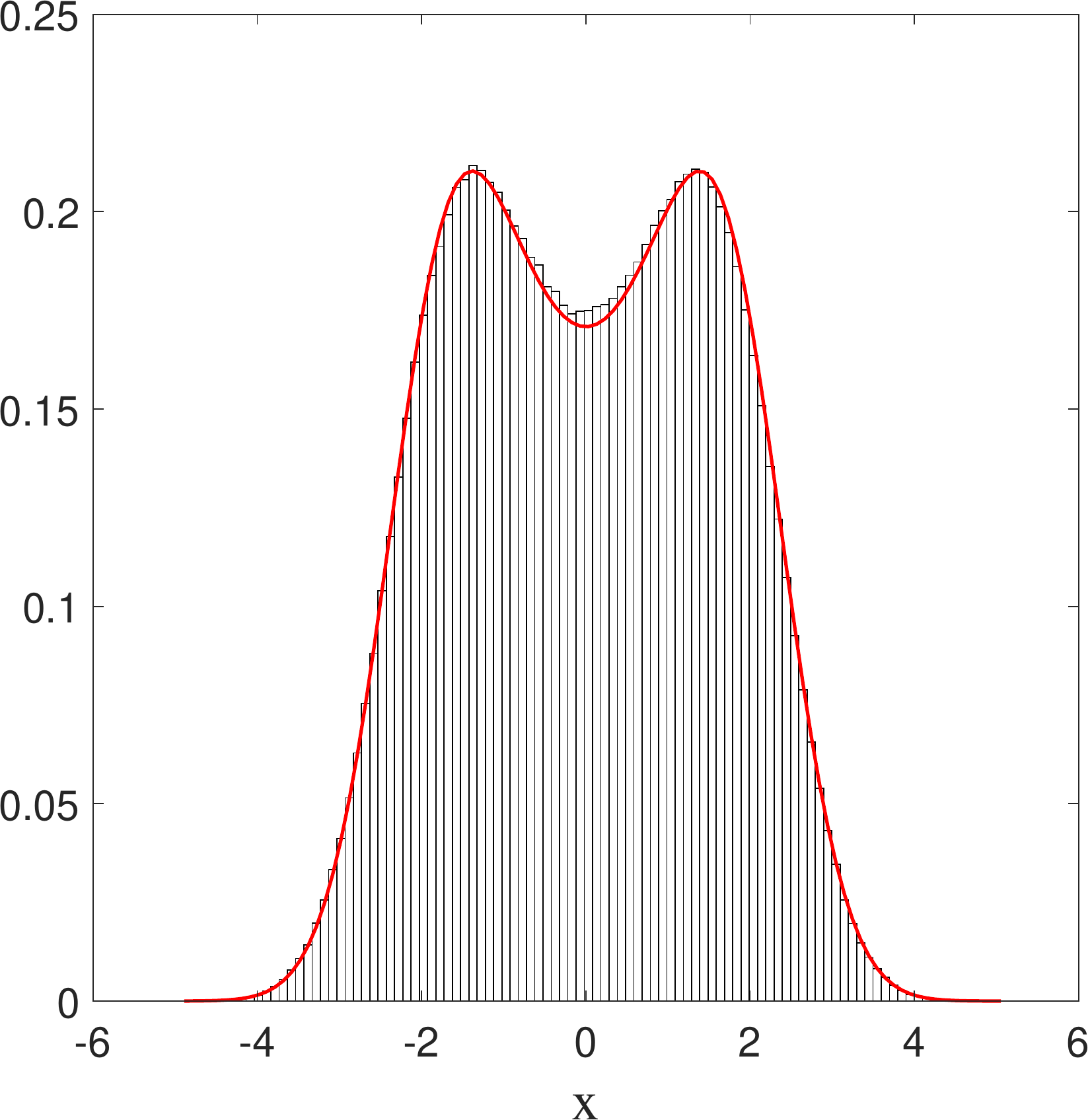}  
\captionof{figure}{\small Histogram of the positions for a unique trajectory satisfying the Ito equation \eqref{Nelsondyn1} for \eqref{cs}. The full curve corresponds to the integration of $\vert\Psi\vert^{2}$ over one period. The center of the wave packet moves between $-2$ and $2$ with a period $2\pi$. Here $a=0.5$ and $\alpha=1$. Total simulation time t is $t=30\,000$ and and the samping time step is $\Delta t=0.01$. The initial position is $x_{i}=1$ and the number of bins $N_{b}=100$, each with width $\Delta x=0.1$.}
\label{histo-cs}
\end{figure}

Secondly, as can be clearly seen on Figure \ref{rel}, even for a uniform initial probability distribution, the convergence to the quantum equilibrium is remarkably fast and the converged distribution faithfully follows the oscillating motion of the non-stationary equilibrium distribution. The remarkable speed of the convergence to quantum equilibrium is corroborated by the decay of the functions $H_V$ and $L_f$ and of the $L_1$ norm shown in Figure \ref{fig-entropiesCS}.
Moreover, Figure \ref{histo-cs} depicts the sampling time average (as defined in section \ref{ergodicity}) of a single trajectory for this non-stationary stochastic process. The convergence of the sampling distribution to a static distribution $\Phi(x)$, described by the integral of $\vert\Psi(x,t)\vert^2$ as given by \eqref{cs}, over a period of the oscillation
\begin{equation}
\Phi(x) = \frac{\omega}{2\pi} \int_0^{2\pi/\omega} \vert\Psi(x,t)\vert^2 dt,
\end{equation}
is striking. As the asymptotic distribution $\Phi(x)$ does not depend on the choice of initial condition, we conclude that the relaxation to equilibrium for the non-stationary stochastic process associated with Nelson dynamics for the coherent state \eqref{cs} is ergodic (in the sense explained in section \ref{ergodicity}).

\subsection{Onset of equilibrium with a dynamical attractor in dBB dynamics and Nelson dynamics\label{VIIIB}}
If one wants to investigate the onset of equilibrium in dBB dynamics, one obviously has to consider non-static asymptotic distributions since in static cases the dBB dynamics freezes the trajectories (as the phase of the wave function is then position independent). {Even in the case of a coherent state (see section \ref{CSrelax}) the distribution of dBB positions would never reach equilibrium because it moves {as a whole (as the shape of a coherent state remains the same throughout time). In a sense coherent states behave as solitary waves.} \rw{Moreover, the absence of zeroes in the wave function might explain why mixing does not occur. }In Figure (\ref{threedbb}) we show the result of simulations of dBB trajectories in the case of a 2D harmonic oscillator for a quantum state consisting of a superposition of equally weighted products of states along $X$ and $Y$, chosen among $M$ energy (Fock) states  (\cite{v3}, appendix 2A), with randomly-chosen initial phases $\theta_{n_{x},n_{y}}$:
\begin{equation}
\Psi(x,y,t)=\frac{1}{\sqrt{M}}\sum_{n_{x}=0}^{\sqrt{M}-1}\sum_{n_{y}=0}^{\sqrt{M}-1}\,e^{i\,\theta_{n_{x},n_{y}}-i\omega\left(n_{x}+n_{y}+1\right)\,t}\,\psi_{n_{x}}\left(x\right)\psi_{n_{y}}\left(y\right).
\label{supoh}
\end{equation}
\begin{figure}[!t]
\centering
\begin{subfigure}[b]{0.31\textwidth}
\includegraphics[width=\textwidth]{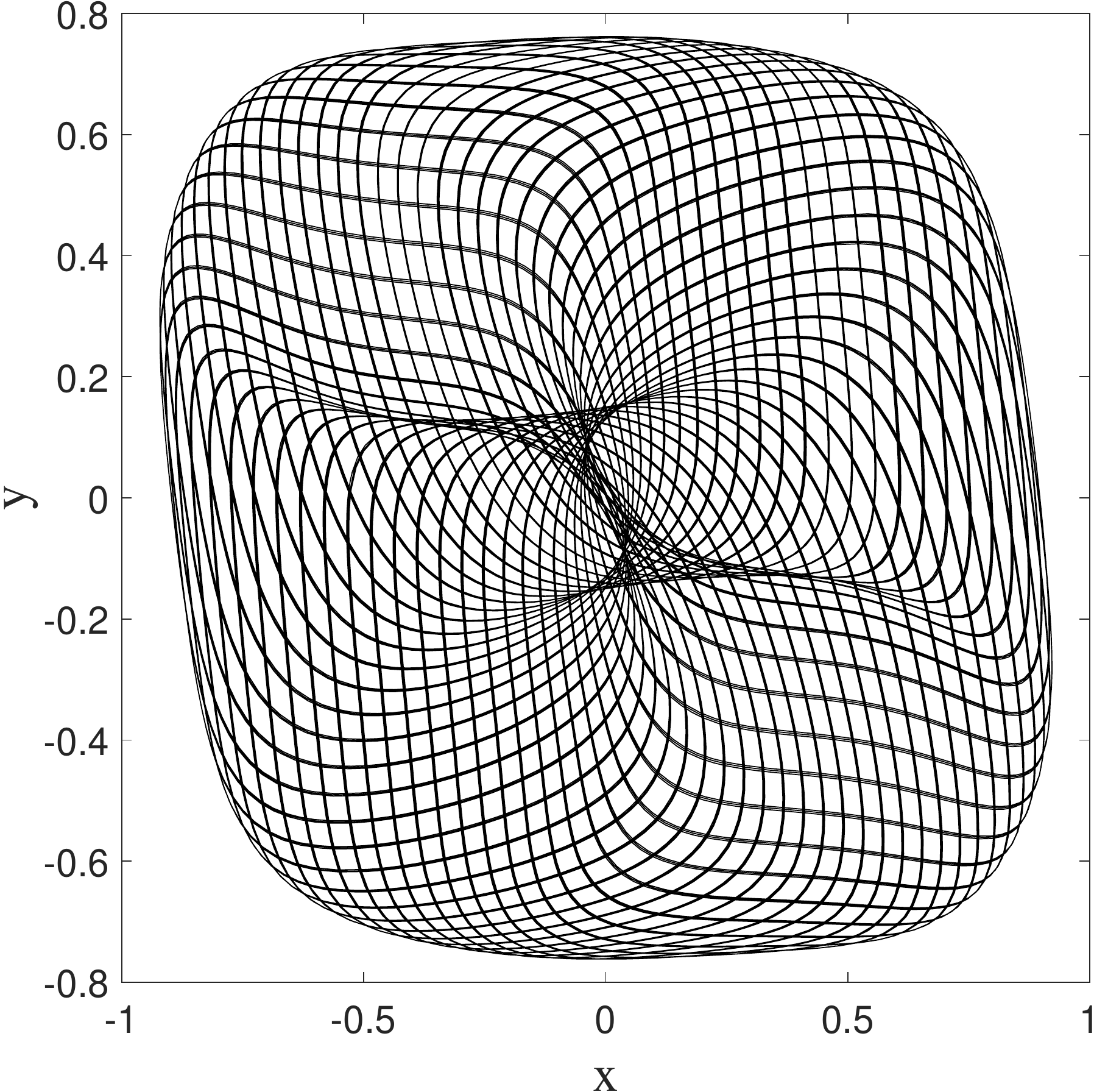}
\end{subfigure}
\begin{subfigure}[b]{0.31\textwidth}
\includegraphics[width=\textwidth]{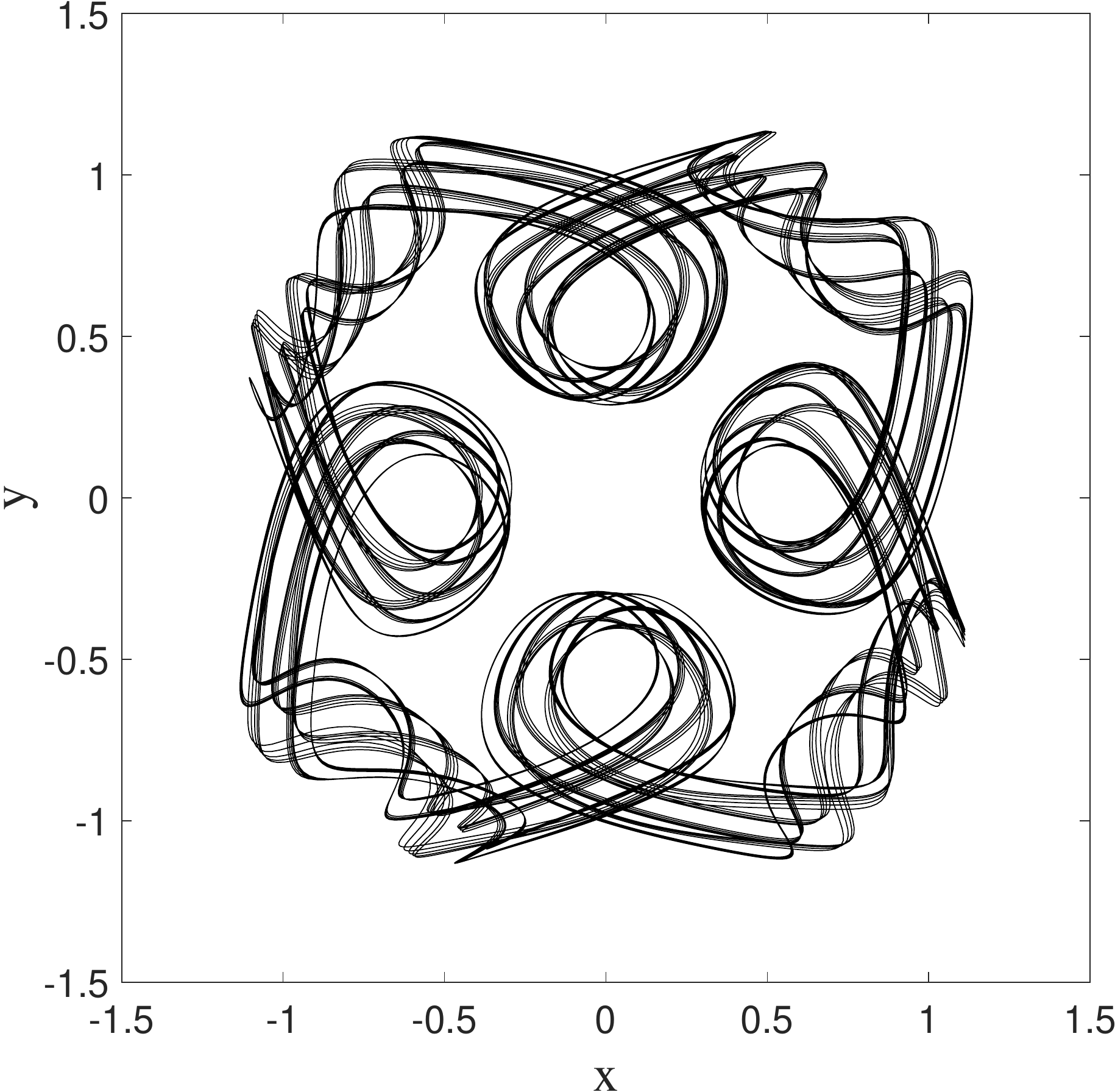}
\end{subfigure}
\begin{subfigure}[b]{0.31\textwidth}
\includegraphics[width=\textwidth]{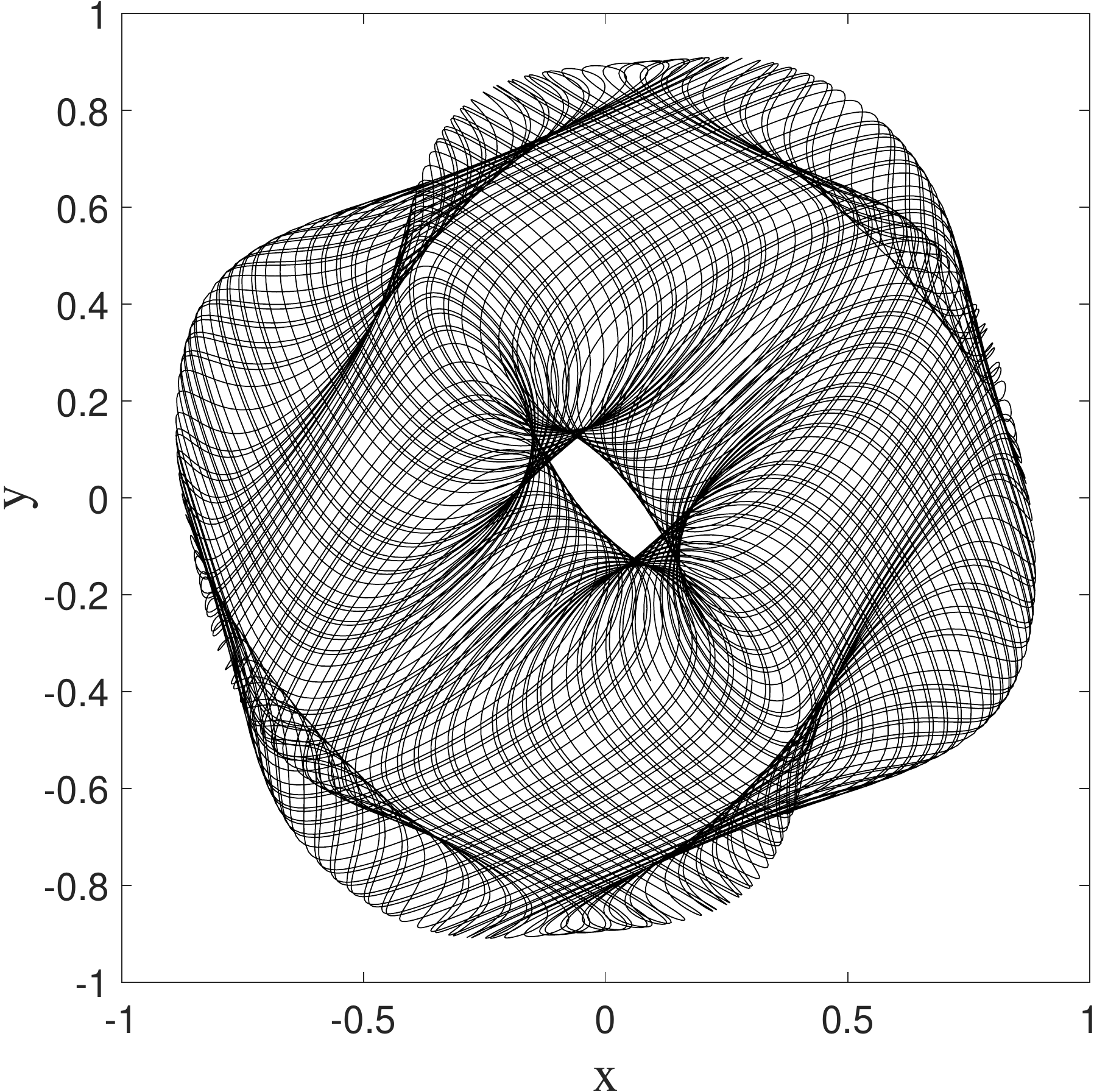}
\end{subfigure}
\caption{\small {Plots showing three possible dBB trajectories for a single point particle in the case of \eqref{supoh} with $M=2^2=4$. Each plot is associated to different initial random phases and different initial positions.}}
\label{threedbb}
\end{figure}

We then compared the relaxation process for dBB with the quantum thermostat given by Nelson dynamics for $M=4^2=16$ energy states. {The results are shown in Figure (\ref{figh}) in which the two H-functions $H_{V}$ (for the dBB and for the Nelson dynamics), as well as $L_{1}$ (for both the dBB and Nelson dynamics) are plotted at the (same) coarse-grained level.} We started from a uniform distribution of positions; we took $\alpha = 0.1$. In both cases, the position distributions $\cal{P}$ and $P$ converge to $\vert\Psi\vert^{2}$. Moreover, we recover an exponential decay for $\overline{H}_{V}$, as already observed in \cite{valentini042}, {even} in the absence of stochastic (Brownian) noise {\it \`a la} Nelson. However, we observe that the convergence to equilibrium occurs faster  in the presence of the quantum thermostat.
\begin{figure}[!t]\centering
\includegraphics[scale=0.32]{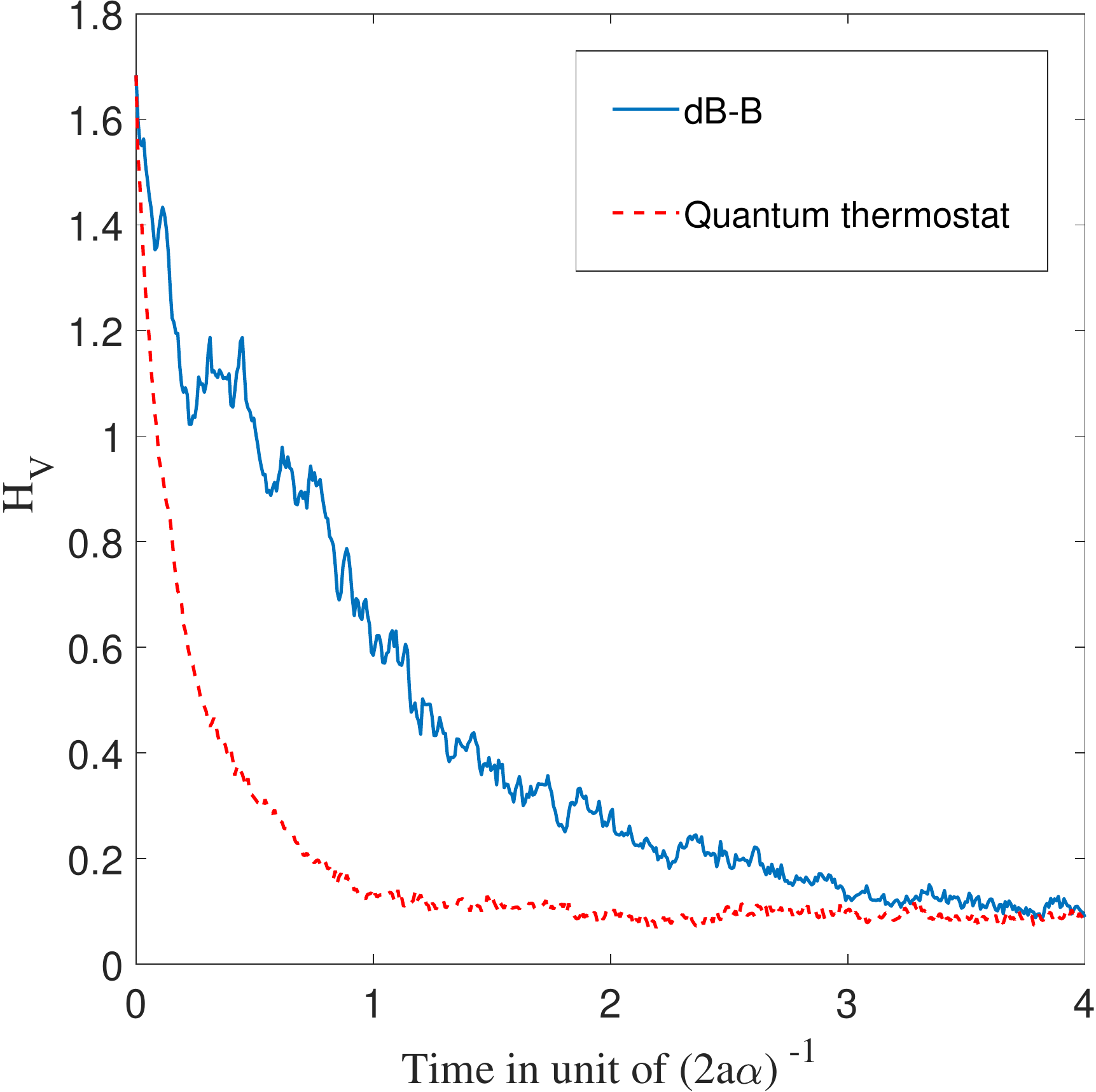}  
\includegraphics[scale=0.32]{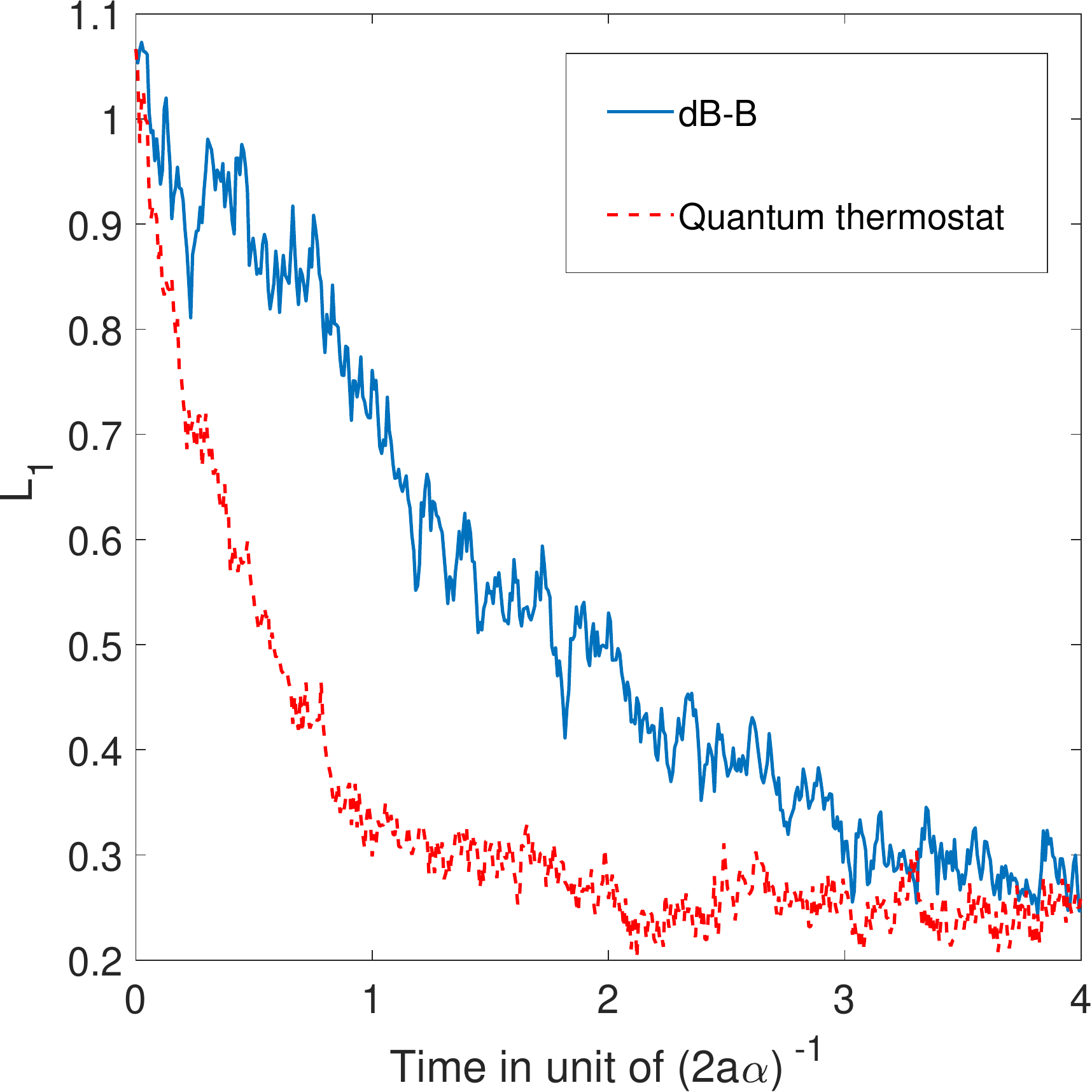}  
\captionof{figure}{\small {Plots of the evolution in time of the coarse-grained H-functions $H_{V}$ (left) and $L_{1}$ (right) for the Nelson and dBB dynamics. The full line corresponds to the dBB dynamics and the dashed line corresponds to the quantum thermostat. We started from $10\,000$ initial positions uniformly distributed in a a box of size 10x10; we chose $a=0.5$, $\alpha=0.1$ and $M=4^2=16$ energy states. }}
\label{figh}
\end{figure}

\section{Dynamical model for droplets and double quantization of the 2-D harmonic oscillator \label{doublequant}\label{s9}}

In this section we shall focus on the description of droplets dynamics as described in \cite{Perrard2014,Labousse2016}, for a magnetised droplet moving in an isotropic 2-D harmonic potential. We shall show that dBB dynamics 
  allows us to {reproduce} some of the main features of the experimental observations. In\cite{Perrard2014,Labousse2016}, it is reported that stable structures appear in the droplets dynamics whenever a double quantisation condition is satisfied.  {The Hamiltonian of the isotropic 2-D harmonic oscillator being invariant under rotations, we may indeed impose a double quantisation constraint, requiring that the energy states of the $2$D quantum harmonic oscillator are also eigenstates of the angular momentum.} 
{In polar coordinates, these states (which are parameterized by two quantum numbers, the energy number $n$ and the magnetic number $m$) are expressed as follows\cite{cohen}:}
\begin{equation}
\psi_{n,m}\left(r,\theta,t\right)=\sqrt{\frac{a}{\pi}\,\frac{k!}{\left(k+\vert m \vert\,\right)!}}\,e^{-\frac{a r^2}{2}}\left(\sqrt{a}\,r\right)^{\left| m\right| } \mathcal{L}_{k}^{\left| m\right|
   }\left[a\,r^2\right] e^{-i \omega(n+1) t+i  m\theta}
   \label{psirth}
\end{equation}
where $\mathcal{L}_{k}^{\left| m\right|}$ are the generalized Laguerre polynomials and $k=\frac{n-\left| m\right| }{2}$. Note that these solutions are linear combinations of the product of Fock states {in $x$ and $y$}. 

A first experimental result reported in\cite{Perrard2014} is the following: trajectories are chaotic and nearly unpredictable unless the spring constant of the harmonic potential takes quantized values which are strongly reminiscent of energy quantization (under the condition that, during the experiment, the size of the orbits is fixed once and for all).  {For quantized energies -- in our case given by $E_n=(n+1) \hbar \omega$, for some `effective' value of $\hbar$ to be determined from actual experiments -- stable orbits appear to which one can attribute yet another quantum number, this time for the angular momentum, which is strongly reminiscent of the magnetic number (the eigenvalue of the orbital momentum, perpendicular to the surface of the vessel, is given by the product of $\hbar$ and $m$).} In \cite{Perrard2014} it is shown, for instance, that for the first excitation ($n$=1, $m=\pm 1$) droplet orbits are circular or oval, turning clockwise or anti-clockwise depending on the sign of $m$. At the second energy level ($n$=2, $m=-2,0,+2$), ovals appear again for $m=\pm 2$ and lemniscates for {an average value angular momentum} $<\!m\!>=0$. At the fourth energy level ($n$=4, $m=-4,-2,0,2,4$) trefoils  appear (for $m=\pm 2$). 

 We simulated dBB trajectories, always considering a superposition of one of the aforementioned doubly quantized eigenstates  $\psi_{n,m}$ with the ground state:
\begin{equation}
\Psi\left(r,\theta,t\right)= \xi_0\,e^{-i \varphi_0}\,\psi_{0,0}\left(r,\theta,t\right)+\sum_{j=0}^{n}\xi_{j+1}\,e^{-i \varphi_{j+1}}\,\psi_{n,-n+2j}\left(r,\theta,t\right)
\label{psirth2}
\end{equation}
where $\varphi_j$ and $\xi_{j}$ are real numbers with  {$0<\xi_{0}\ll\xi_{j\neq0}$}. 
Computing the guidance relation \eqref{guid3} for a single eigenstate \eqref{psirth}, one ends up with a value for $\nabla S $ for which  the trajectories are cirles of radius $R$ around the origin, with tangential velocities proportional to $m/R$. In particular, the dynamics is frozen when $m=0$.

\begin{figure}[!t]
\centering
\begin{subfigure}[b]{0.3\textwidth}
\textbf{a)}\,$n=1,\,m=+1$
\includegraphics[width=\textwidth]{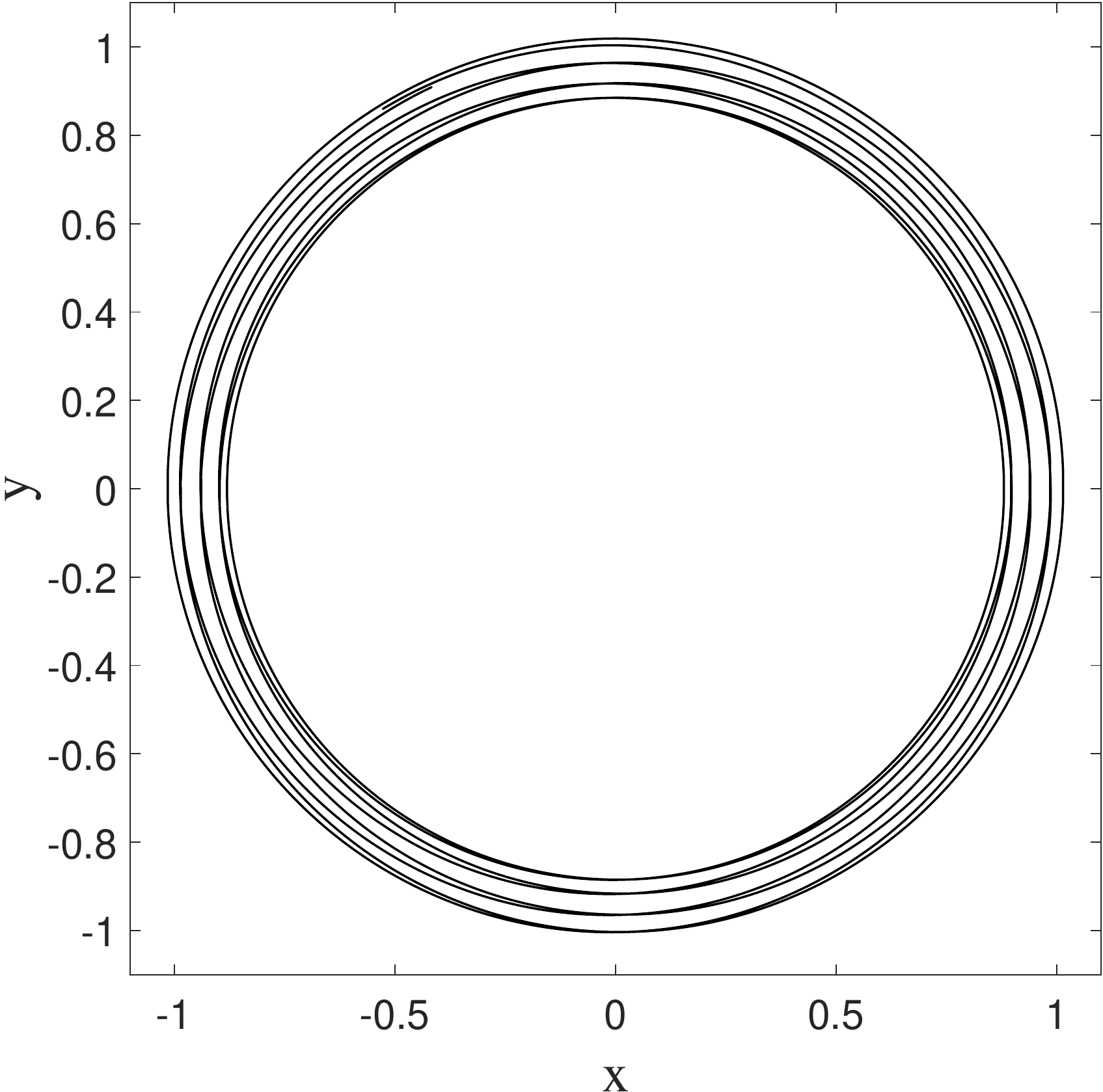}
\end{subfigure}
\begin{subfigure}[b]{0.3\textwidth}
\textbf{b)}\,$n=2,\,m=+2$
\includegraphics[width=\textwidth]{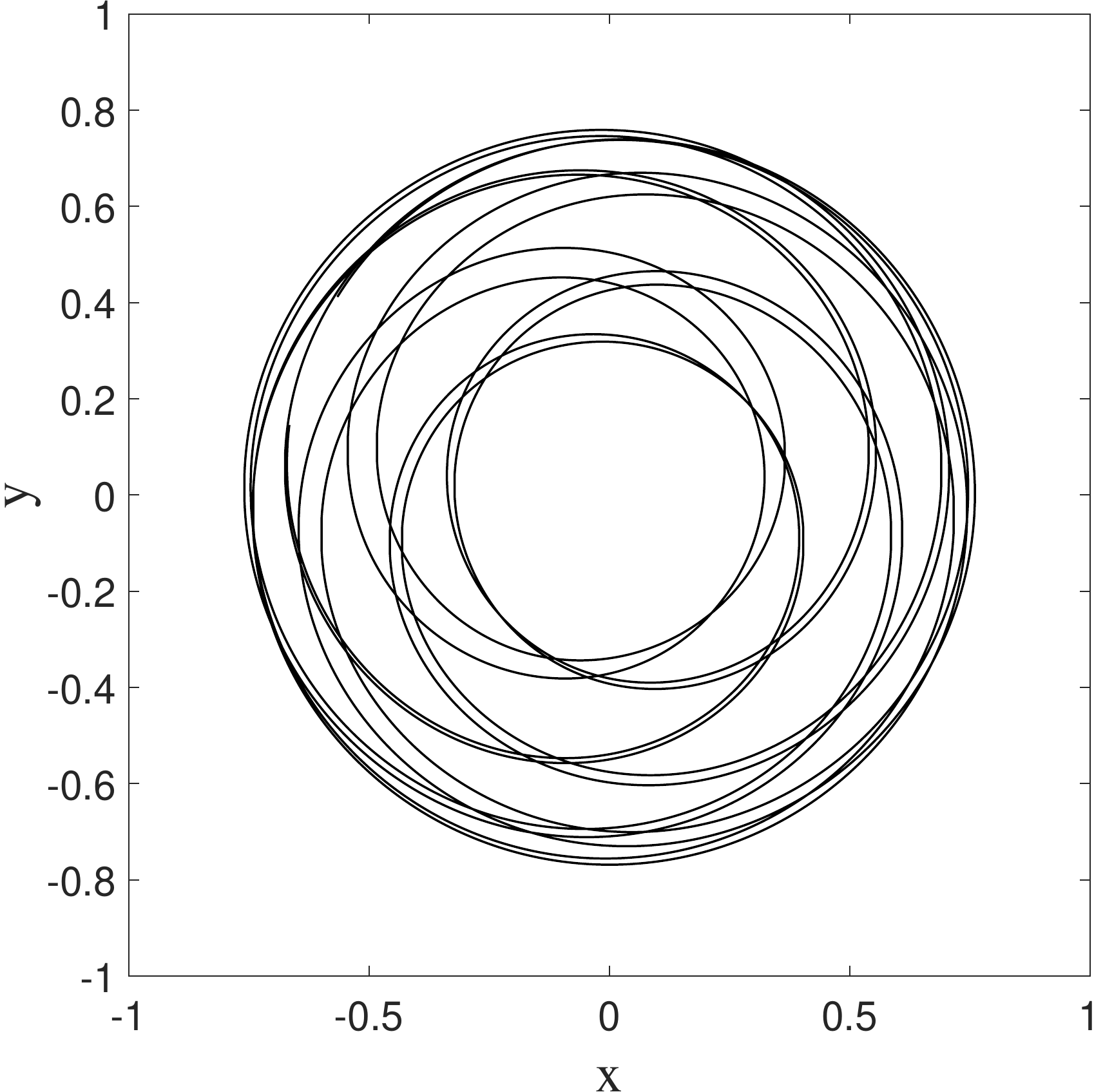}
\end{subfigure}
\begin{subfigure}[b]{0.3\textwidth}
\textbf{c)}\,$n=2,\,m=0,\pm 2$
\includegraphics[width=\textwidth]{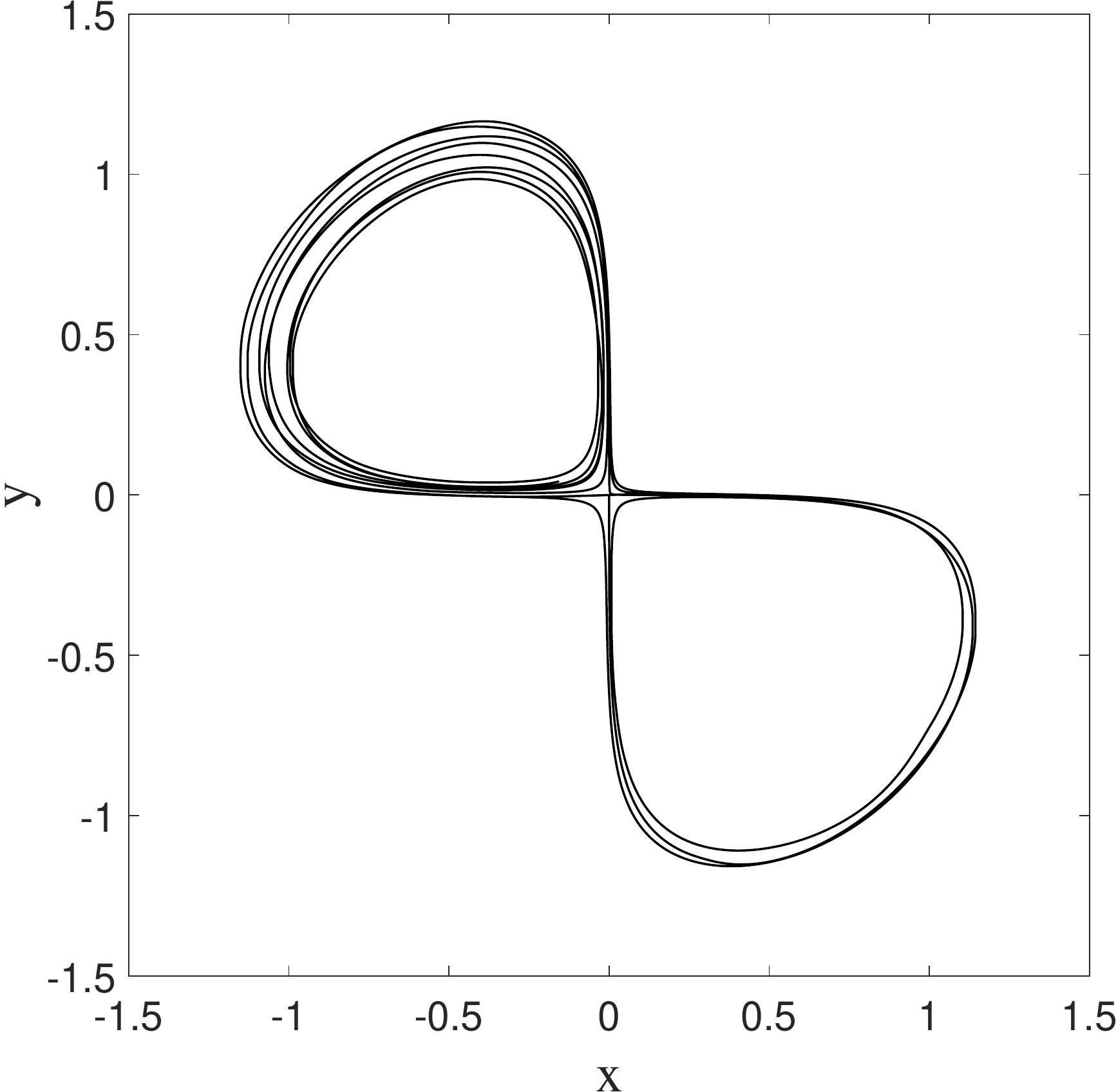}
\end{subfigure}
\caption{\small dB-B trajectories obtained for a single point particle in a superposition of eigenstates \eqref{psirth2}. Each plot is associated to a different combination $(n,m)$, as indicated. \rw{In the graphs \textbf{(a,b)} we imposed $a =1$, and $\omega=1, \frac{ \xi_0}{ \xi_2} = 0.05$ and $\omega=0.5, \frac{ \xi_0}{ \xi_3} = 0.05$ respectively ; for \textbf{(c)} we imposed $a=3, \omega=0.5$, $\frac{ \xi_0}{ \xi_{3}}=0.0708, \frac{ \xi_0}{ \xi_{2}}=0.0456$ and $\frac{ \xi_0}{ \xi_{1}}=0.0773$.   }}
\label{Lz1}
\end{figure}

\begin{figure}[!t]
\centering
\begin{subfigure}[b]{0.3\textwidth}
\textbf{a)}
\includegraphics[width=\textwidth]{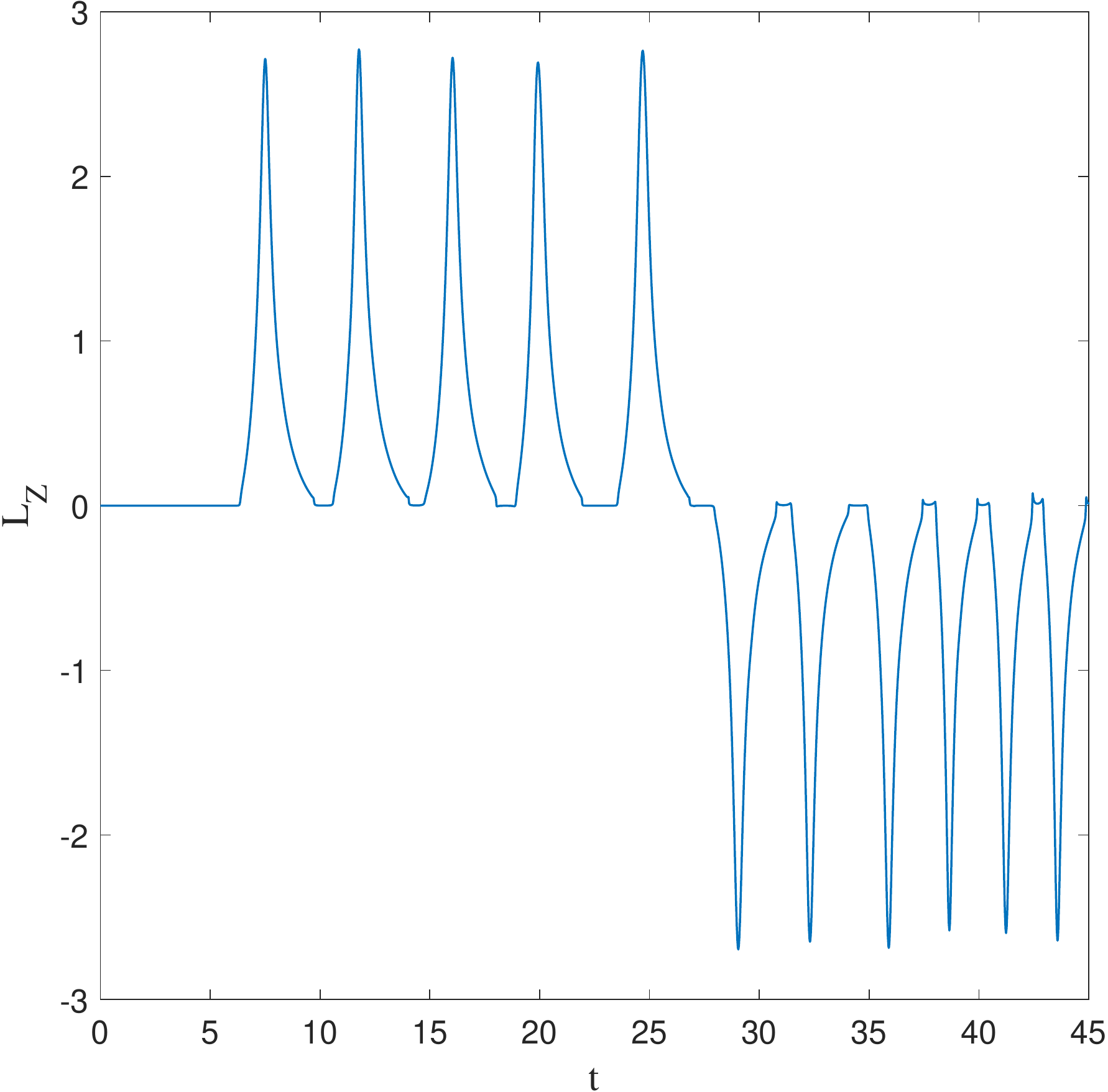}
\end{subfigure}
\begin{subfigure}[b]{0.33\textwidth}
\textbf{b)}
\includegraphics[width=\textwidth]{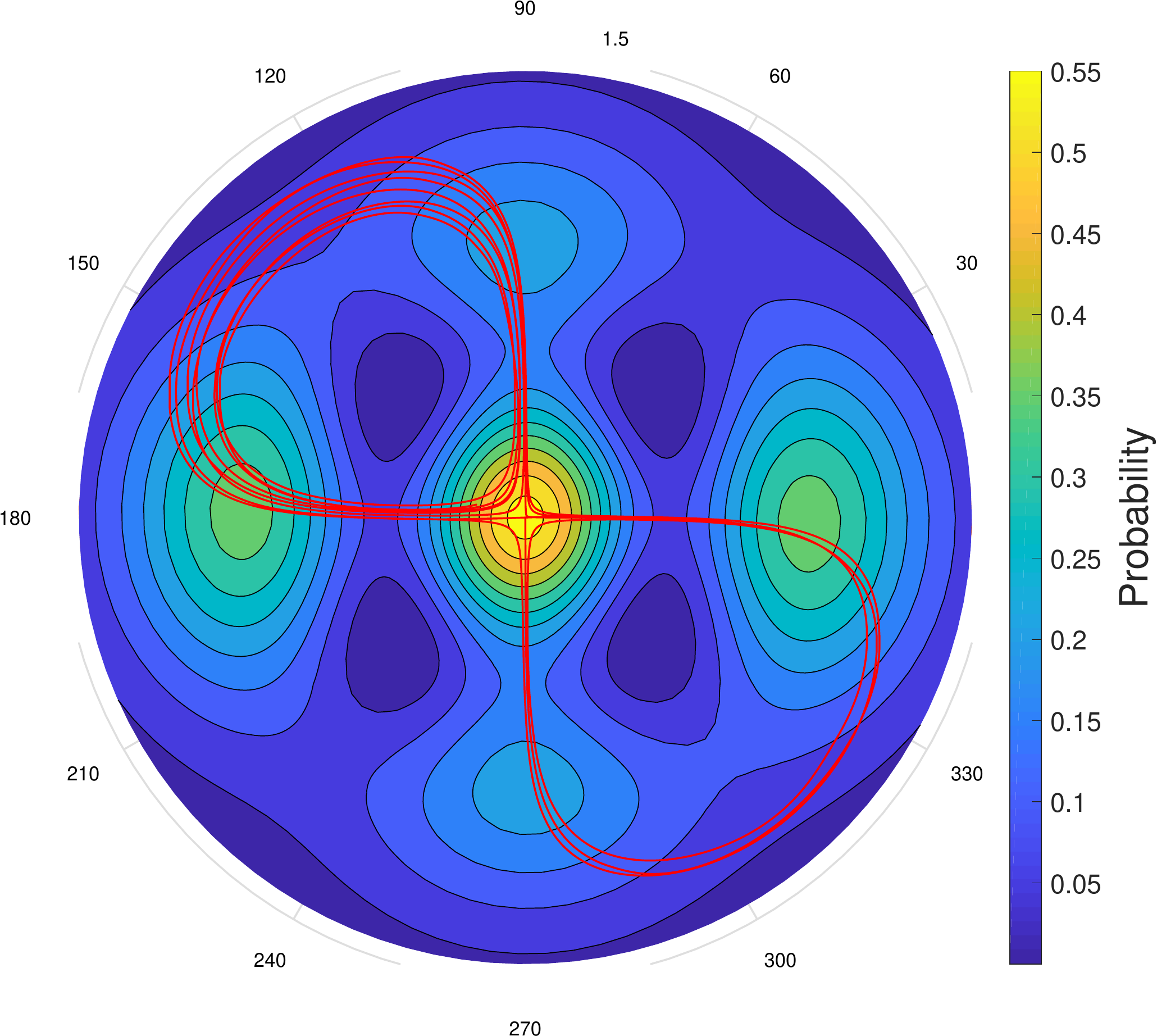}
\end{subfigure}
\begin{subfigure}[b]{0.33\textwidth}
\textbf{c)}
\includegraphics[width=\textwidth]{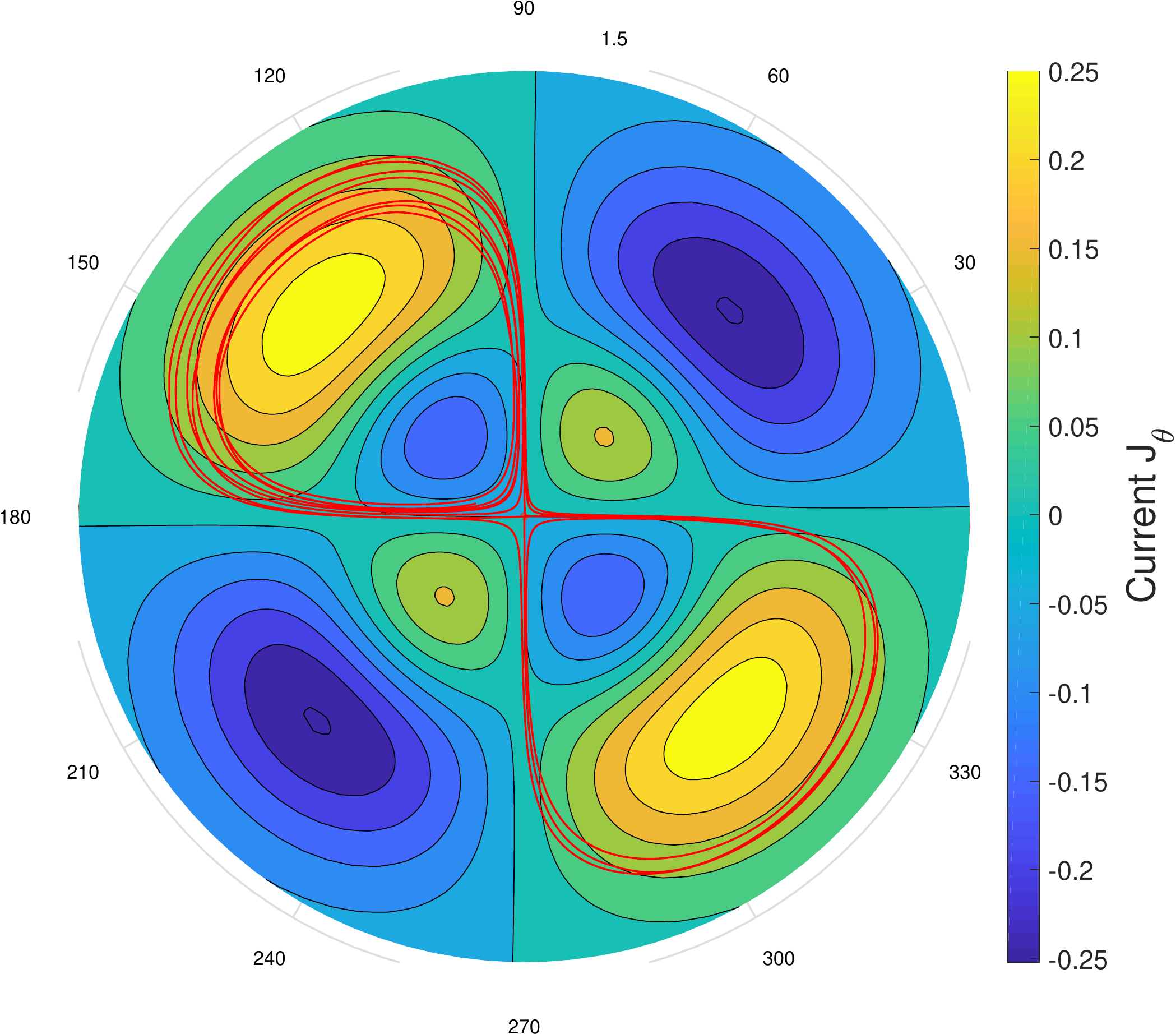}
\end{subfigure}
\caption{\small \rw{ Plots of three quantities associated to the lemniscate in figure \ref{Lz1}$c$.} a) shows the $L_z$-component of the angular momentum and \rw{the polar plots b) and c) show the probability density $\vert\psi\vert^2$ (b) and the $\theta$-component of the probability current \eqref{j1} along the trajectory (c).}}
\label{Lz2}
\end{figure}

 Mixing the wave function with the ground state, however, generates a periodic (in time) component in the dBB velocity field, which turns circular orbits into ovals when $\xi_0$ is small enough, and eventually generates more complex structures like rosaces otherwise. We also tuned the energy difference between the ground state and the excited states such that two timescales characterize the dynamics. These are the ``centrifugal'' period, necessary for drawing a full circle around the origin, which varies as $m/R^2$, and the ``Bohr'' period  which varies like $T/(n+1)$, where $T$ is the classical period of the oscillator.  {Tuning these parameters we were able to simulate dBB trajectories very similar to those reported in \cite{Perrard2014}. For instance, we found circles and ovals (see figures \ref{Lz1} $a,b$) for $(n,m)=(1,1)$ and $(n,m)=(2,2)$. Note that the lemniscate cannot be obtained with a superposition of the ground state and the $(n,m)=(2,0)$ state for which dBB velocities are necessarily purely radial, contrary to the suggestion made in\cite{Perrard2014}{, but rather should be generated with \rw{a superposition of the ground state with $(n,m)=(2,+2), (2,-2)$ and $(2,0)$ in which the weights of the $m=+2$ and $-2$ components are slightly different (see figure \ref{Lz1} $c$). Figure \ref{Lz2} shows further detail of the evolution along this trajectory.} Tuning the energy, we were also able to generate a trefoil and a ``rosace'' (see figure \ref{Lz3})}.

\begin{figure}[!t]
\centering
\begin{subfigure}[b]{0.3\textwidth}
\textbf{a)}\,$n=4,\,m=+2$
\includegraphics[width=\textwidth]{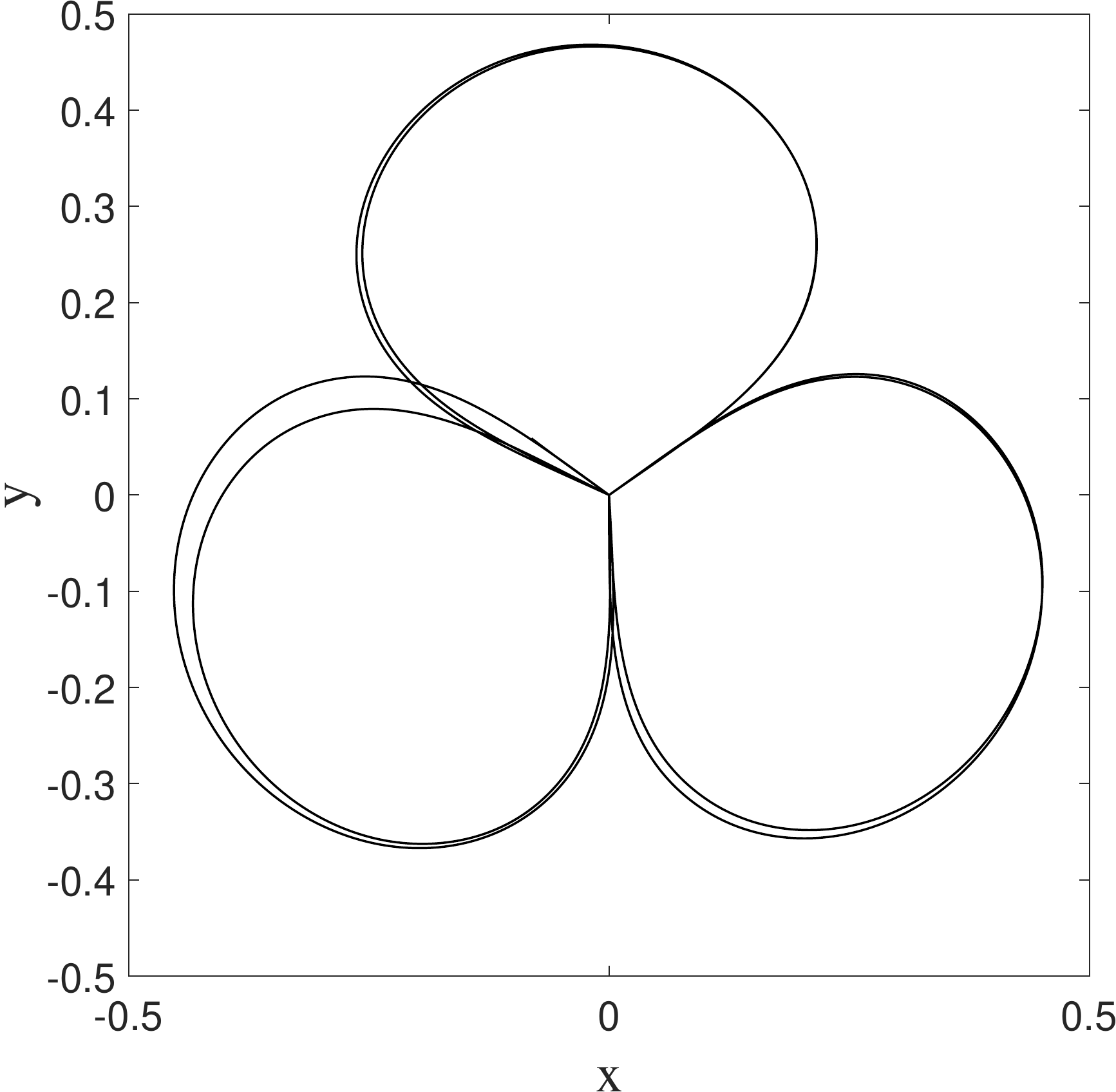}
\end{subfigure}
\begin{subfigure}[b]{0.3\textwidth}
\textbf{b)}\,$n=4,\,m=+2$
\includegraphics[width=\textwidth]{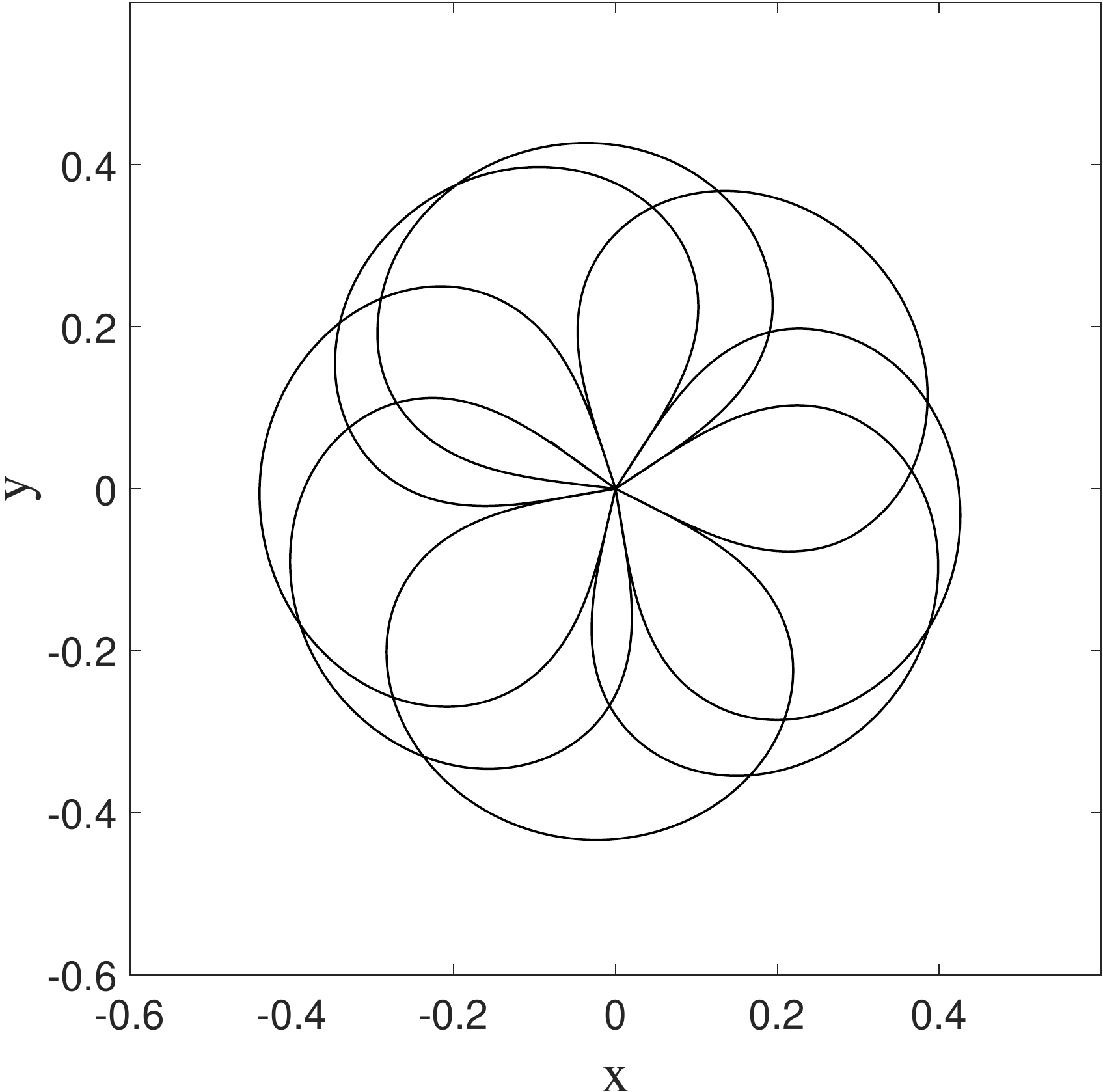}
\end{subfigure}
\begin{subfigure}[b]{0.3\textwidth}
\textbf{c)}\,$n=4,\,m=+2$
\includegraphics[width=\textwidth]{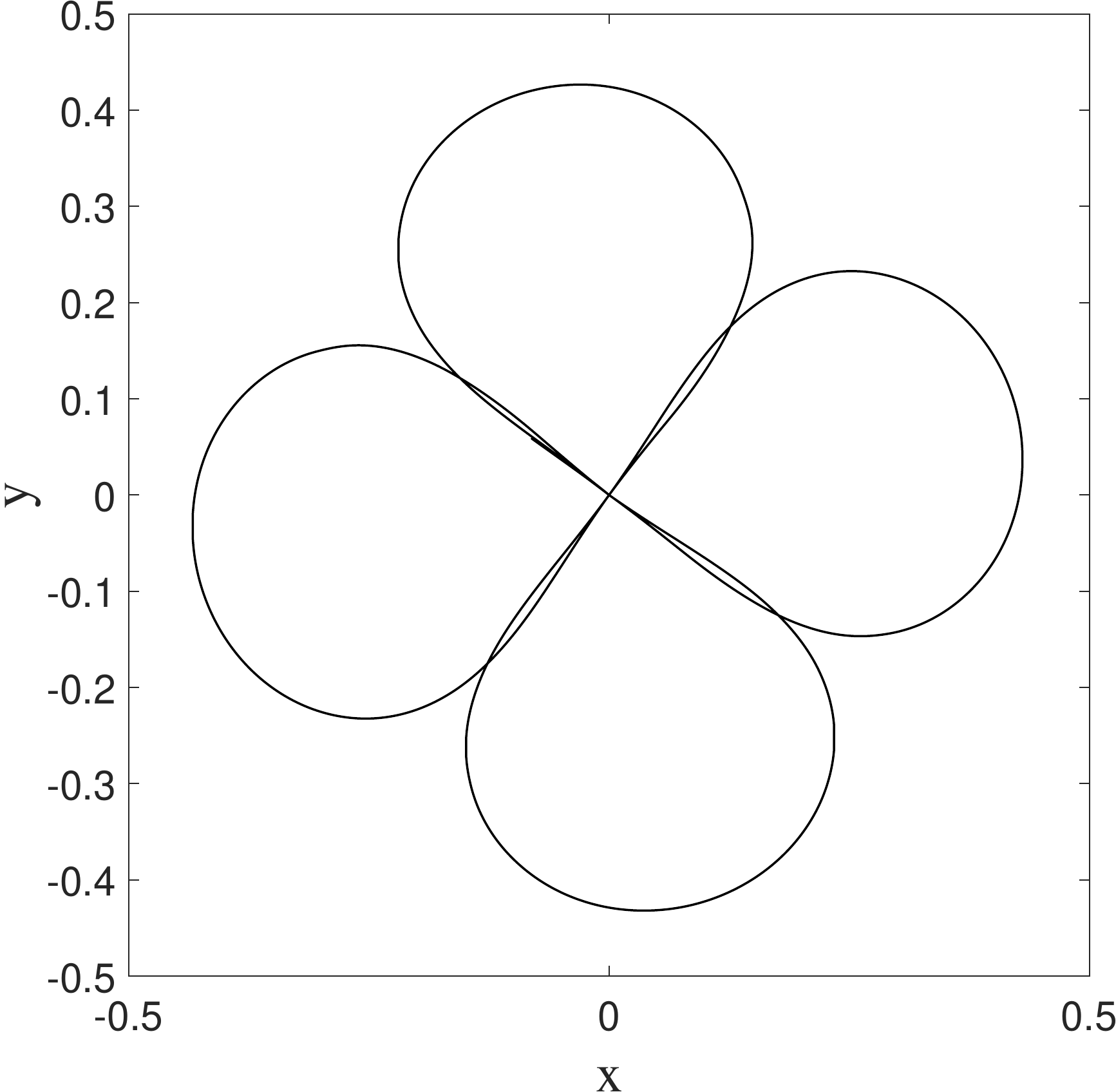}
\end{subfigure}
\caption{\small {dB-B trajectories obtained for a single point particle in a superposition of eigenstates \eqref{psirth2}. Plots $(a,b)$ correspond to $\omega=0,7$ and $\omega=1$ respectively. Case $c$ is obtained after multiplying the amplitude of the $(n,m)=(4,2)$ state by a complex phase ($e^{(0.3 i)}$). We took $a=1$  in all cases.}}
\label{Lz3}
\end{figure}
{It is worth noting however that chaos is omnipresent in the dBB dynamics for this system,  in the sense that the trajectories exhibit an extreme sensitivity to the initial conditions, which explains why these dBB orbits mimicking stable droplets orbits are in general unstable. For instance, \rw{figure \ref{Lz4} shows intermittent transitions between an oval trajectory and a lemniscate (as has {also} been reported in \cite{Perrard2014}), for a superposition of the {ground state with the} $(n,m)=(2,+2), (2,-2)$ and $(2,0)$ states.} Preliminary results furthermore show that the trajectories are also unstable under Nelson dynamics, i.e. in the presence of ``noise'', whenever this noise (parameterized by $\alpha$ in (\ref{Nelsondyn1})) exceeds a critical value. \raw{Note that many experiments involving droplets are characterized by a lack of stability and predictability. For instance, the appearance of interferences similar to those obtained in a double slit experiment\footnote{\raw{See Refs. \cite{Carlos2018} and \cite{v3} for a description {\it \`a la} Nelson of the double slit experiment.}} has been attributed to ``air currents''  in Ref. \cite{Pucci2018}.}
Therefore, although our approach might not explain all details of the double quantization reported in \cite{Perrard2014}, it does reproduce many of its essential features and we believe it would be very interesting to be able to deepen this analogy. For instance, having access to the empirical values of the weights of the ground state, or of the effective values of $\hbar$ and of the mass in the case of droplets  \cite{Gilet} would allow us to test our model in real detail.  

Another experiment, reported in \cite{saenz}, during which both the position of the droplet and the excitation of the bath are monitored, and where a superposition between two distinct modes of the bath is reported, could also provide more insight and might offer some means to test the validity of our model: using exactly the same observation device, but this time in the case where the droplet undergoes a 2-D isotropic potential, would allow one to check whether the modes of the bath are similar to the $(n,m)$ quantum modes  which we associate with the quantized droplets trajectories.

\begin{figure}[!t]
\centering
\begin{subfigure}[b]{0.3\textwidth}
~$n=2,\,m=0,\,m=\pm 2$
\includegraphics[width=\textwidth]{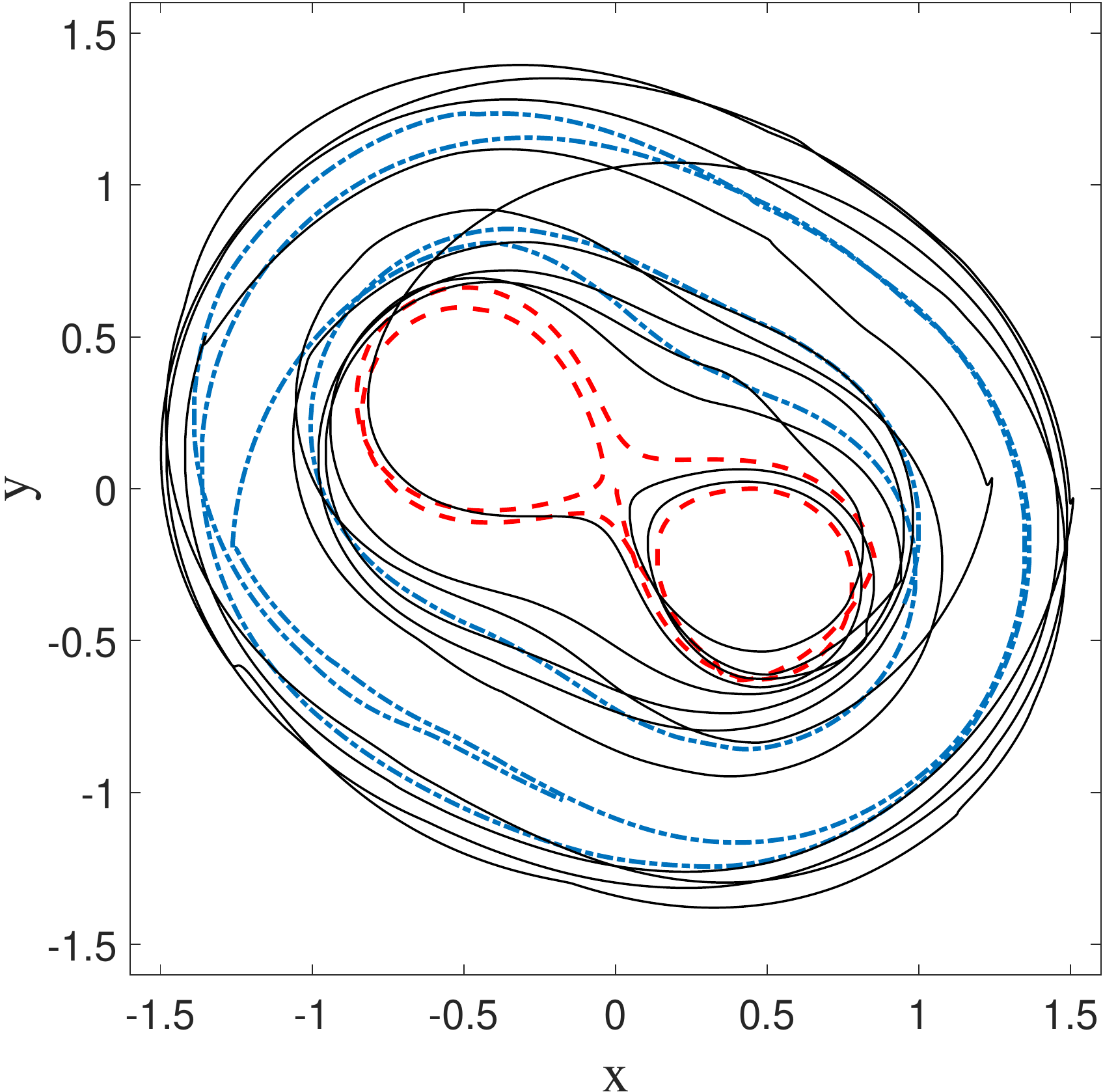}
\end{subfigure}
\caption{\small {dB-B trajectories obtained for a single point particle in a superposition of eigenstates \eqref{psirth2} showing intermittent transitions between two types of trajectories. \rw{The relevant parameter values are $\omega=0.2$ , $a=1$ and $\frac{ \xi_0}{ \xi_{3}}=0.0342, \frac{ \xi_0}{ \xi_{2}}=0.2547$ and $\frac{ \xi_0}{ \xi_{1}}=0.0505$.}}    }
\label{Lz4}
\end{figure}

\section {Conclusions -- open questions}
{In this paper we studied \rw{stochastic, Nelson-like dynamics as well as dBB dynamics}, with the aim of simulating the dynamics of droplets. The \rw{stochastic} approach has the merit that it explicitly takes into account the influence of noise on the dynamics\cite{deBroglieDurt,Carlos2018}. In contrast to certain experiments where noise is considered to be a parameter that should be minimized, here, noise is considered to be a relevant parameter for the dynamics (see also \cite{Grossing2010}). For instance, as we have shown, it plays an essential role {in the relaxation towards equilibrium and in the ergodicity of the dynamics}. In the dBB approach, on the other hand, the main ingredient is the chaotic nature of the dynamics \cite{efthymiopoulos4}. Both models thus shed a different light on the dynamics and could possibly fit diverse set of regimes in droplets dynamics.} {Note that in the limit where the amplitude of the Brownian motion in \rw{our} Nelson dynamics tends to zero, the dynamics approaches dBB dynamics very closely. In sufficiently complex situations (e.g. when the mixing process due to the presence of zeros in the wave function becomes effective \sam{\cite{valentini042,efthymiopoulos}}), we expect {the relaxation to equilibrium to be accompanied by chaotic rather than stochastic dynamics, as one has in Nelson dynamics}\footnote{{Although Nelson dynamics with small but non-zero Brownian motion is hard to distinguish from dBB dynamics, it has the advantage that relaxation is guaranteed to occur, even in the absence of coarse graining (and/or mixing).}}. } }

Ultimately, experiment ought to indicate whether it is relevant, with respect to droplet phenomenology, to formalize  the dynamical influence of noise {\`a la} Nelson {(and/or dBB)} as we did in the present paper. We formulated several proposals in this sense in {sections \ref{s6bis} and \ref{doublequant}}. 
As emphasized throughout the paper however, our models should be seen as a first step in the direction of a dynamical model, which remains to be formulated, combining Nelson's stochastic dynamics (and/or dBB dynamics) and memory effects. {We think that the results of section \ref{doublequant} show that this is a promising program for future research.}

{Finally, it is worth recalling some of the problems that arose when, first de Broglie, and then Bohm and Nelson developed their theories  aimed at deriving quantum dynamics (statistics) as an emergent property, i.e., resulting from an underlying ``hidden'' dynamics. }

{The most severe problem is undoubtedly non-locality, which was recognised by Bohm\cite{bohm521,bohm522} to be an irreducible feature of dBB dynamics (see also\cite{bohmh,nelsonloc} for similar conclusions concerning \rw{Nelson-type} dynamics). Today, under the influence of the work of John Bell  \cite{bell} and his followers, it is widely recognized that quantum theory is irreducibly non-local, which makes it particularly difficult to mimick with classical models.} \raw{Note that entanglement and non-locality (as well as decoherence, which is the corollary of entanglement \cite{Proeminent}) only appear if we consider more than one particle at a time, which explains why we did not address these fundamental questions in the core of the paper, where we were always describing a single droplet.}

{Another problem concerns the fact that the pilot wave is a complex function. This poses still unresolved problems in the case of Nelson dynamics because Nelson's diffusion process does not make it possible \cite{bacciaphase,wallstrom} to fix the phase of the wave function {unequivocally}\footnote{\raw{See Ref. \cite{Derakhshani2015} for an interesting proposal involving a multivalued wave function, also based on Zitterbewegung.}}.}
In our approach, which is mainly of quantum inspiration, complex wave functions and imaginary phases appear spontaneously, but if we wish to scrutinize the link with the empirically observed modes at the surface of oil baths\cite{Perrard2014,Labousse2016,Gilet,saenz}, it will be important to interpret the exact meaning of this complex phase.
\rw{In the framework of his double solution program \cite{bacval,CDWannales} de Broglie, and others, showed how to derive the Schr\"odinger equation from a Klein-Gordon equation in the non-relativistic limit. This is only possible provided the real wave bounces at an extremely high frequency (of the order of $mc^2/h$). A similar approach has been proposed in the context of droplets phenomenology in Ref. \cite{Borghesi2017} where a complex Schr\"odinger equation is derived from the Klein-Gordon equation along these lines. Although such (interesting and promising) alternative studies of droplets solve the problem of the appearance of a complex phase in a classical context, it is worth noting that the phenomenological results outlined in section \ref{s9}, concerning the quantization of droplet orbits in the case of a harmonic potential  \cite{Perrard2014,Labousse2016}, cannot be explained simply in terms of excited modes of the oil bath, because in these experiments only the droplet undergoes the harmonic potential, the oil bath being electromagnetically neutral. This difficulty actually concerns any classical model in which droplet dynamics is formulated in terms of classical modes of the bath only.}

\rw{To conclude, in our view, the programs that aim at simulating droplet dynamics with quantum tools or at describing the emergence of quantum dynamics based on droplet dynamics, are still largely incomplete and raise challenging fundamental questions. This Pandora box is now open and it will not be closed soon, which is however not something to be feared as it offers new and stimulating perspectives for future research in the field.}

\section{Acknowledgements}
The authors gratefully acknowledge funding and support from the John Templeton foundation (grant 60230, Non-Linearity and Quantum Mechanics: Limits of the No-Signaling Condition, 2016-2019) and a FQXi Physics of what happens grant (Quantum Rogue Waves as Emergent Quantum Events, 2015-2017). \rw{RW would also like to thank FQXi for support through a mini-grant (Grant number FQXi-MGA-1819), which enabled him to elucidate crucial aspects of the relaxation to quantum equilibrium. 
SC also thanks the Foundational Questions Institute (fqxi.org) for its support though a mini-grant (Grant number FQXi-MGA-1705).
TD also thanks Aur\'elien Dr\'ezet (CNRS-Grenoble) for drawing our attention to the paper of Kyprianidis and for useful comments on the draft of our paper during TD's personal visit to Grenoble (May 3 2018).}
\section{Appendix 1. Numerical simulations}
Firstly, we discuss the case of the dBB dynamics.
It is assumed that we have an analytical solution of the Schr\"odinger equation $\Psi(t,{\bf x})$. 
We want to compute the evolution of a given initial non-equilibrium density $\pdbb(t_i,{\bf x})$ 
up to a final time $t_f$ and for intermediate time events (we denote all these events by $t_k$, with $t_0=t_i$ and $t_f=t_K$). 
In particular, we are interested in the coarse-grained non-equilibrium density
\begin{equation}
\overline{\pdbb}({\bf x},t_k)=\frac{1}{\epsilon^3}\int_{\textrm{CG cell}\ni{\bf x}} d^3 x  \pdbb({\bf x},t_k)\,,
\end{equation}
which is defined in \eqref{cgquantities1}.

Numerically, we replace that integral by a discrete sum over a finite set of points ${\bf x}^{l}$, which are uniformly distributed over the CG cells. 
In order to obtain the value of each $\pdbb({\bf x}^{l},t_k)$ we use the Liouville relation
\begin{equation}
\frac{\pdbb({\bf x}^{l},t_k)}{|\Psi({\bf x}^{l},t_k)|^2}=\frac{\pdbb({\bf x}^{l}_{i},t_i)}{|\Psi({\bf x}^{l}_{i},t_i)|^2},
\end{equation}
where ${\bf x}^{l}_i$ is the position of the particle which, when evolved according to \eqref{guid3} from $t_i$ up to $t_k$, 
gives ${\bf x}^{l}$.

In order to obtain ${\bf x}^{l}_i$ for each ${\bf x}^{l}$, we consider the time-reversed evolution with wave-function $\Psi^\ast(-t,{\bf x})$ and initial 
condition ${\bf x}^{l}$ at time $-t_k$. The position ${\bf x}^{l}$, if time evolved from $-t_k$ up to $-t_i$ according to \eqref{guid3}, will give the position ${\bf x}^{l}_{i}$.
As there is usually no analytical solution of \eqref{guid3}, we use a Runge-Kutta (RK) algorithm \cite{nr} to obtain a numerical estimate of the position ${\bf x}^{l}_{i}$.
In order to know if we can trust the result of the Runge-Kutta algorithm, we perform two realizations of the algorithm 
with different choices of a so-called tolerance parameter  (the smaller the value of that tolerance parameter, the more precise the computation), say $\gamma$ and 
$10^{-1}\gamma$.
If the distance between the two positions is less than some chosen value $\delta$, the result of the last iteration of the RK algorithm is trusted.
Otherwise, we perform another iteration with $10^{-2}\gamma$ and we compare it to the previous realization of the RK algorithm.
We repeat the procedure until the constraint on the distance between the two successive results of the RK algorithm is satisfied, or until we reach 
some minimal value of the tolerance parameter. 
In that case, the position ${\bf x}^{l}$ is considered as a bad position and it is discarded from the numerical integration of \eqref{cgquantities1}. This method was used in\cite{valentini042}.

That is one possible method but we could also adopt a more brute-force method: Randomly generate a set of $N$ initial positions according to $\pdbb(t_i,{\bf x})$ and let them evolve according to an Euler algorithm (that is, we divide the time-interval in small time-steps of length $\Delta t$ and we increment the position by ${\bf v}(t) \Delta t$ at each time-step). We record the positions of the $N$ particles for each value of $t_k$, we count the number of particles in each CG cell for each time $t_k$ (say $n_{CG}$) and we divide $n_{CG}$ by $N$ in order to define $\overline{\pdbb}({\bf x},t_k)$. The first method turns out to be more efficient in the case of the dBB dynamics but it is not applicable in the presence of stochastic terms.

In the case of Nelson dynamics we used the Euler-Maruyama method for stochastic processes to approximate the solution of the Ito equation \eqref{ito}. In the same way as Euler's method, the time $T$ is divided into $N$ small discrete time steps $\Delta t$. For each time $t_{i}$ we generated a random variable normally distributed $\Delta W_{i}=\sqrt{\Delta t}\,\mathcal{N}\left(0,1\right)$. The integration scheme has the form:
\begin{equation}
x_{i+1}=x_{i} + v(x_{i},i\,\Delta t)\Delta t + \sqrt{\alpha}\,\Delta W_{i}.
\end{equation}
We invite the reader interested in the details to read \cite{Higham2001}. 
The remaining question is how to choose the time step $\Delta t$ so that one can trust the result of the numerical simulations. 
One way to do this is the following. 
We know that the Born distribution remains invariant under Nelson's dynamics (equivariance). 
We therefore start with some value for $\Delta t$ and decrease it until the result of the numerical simulation confirms this theoretical prediction.
We then perform the numerical simulation for the non-equilibrium distribution with the value of $\Delta t$ thus obtained.

\bibliography{RefEntropyV13.bbl}

\begin{thebibliography}{72}%
\makeatletter
\providecommand \@ifxundefined [1]{%
 \@ifx{#1\undefined}
}%
\providecommand \@ifnum [1]{%
 \ifnum #1\expandafter \@firstoftwo
 \else \expandafter \@secondoftwo
 \fi
}%
\providecommand \@ifx [1]{%
 \ifx #1\expandafter \@firstoftwo
 \else \expandafter \@secondoftwo
 \fi
}%
\providecommand \natexlab [1]{#1}%
\providecommand \enquote  [1]{``#1''}%
\providecommand \bibnamefont  [1]{#1}%
\providecommand \bibfnamefont [1]{#1}%
\providecommand \citenamefont [1]{#1}%
\providecommand \href@noop [0]{\@secondoftwo}%
\providecommand \href [0]{\begingroup \@sanitize@url \@href}%
\providecommand \@href[1]{\@@startlink{#1}\@@href}%
\providecommand \@@href[1]{\endgroup#1\@@endlink}%
\providecommand \@sanitize@url [0]{\catcode `\\12\catcode `\$12\catcode
  `\&12\catcode `\#12\catcode `\^12\catcode `\_12\catcode `\%12\relax}%
\providecommand \@@startlink[1]{}%
\providecommand \@@endlink[0]{}%
\providecommand \url  [0]{\begingroup\@sanitize@url \@url }%
\providecommand \@url [1]{\endgroup\@href {#1}{\urlprefix }}%
\providecommand \urlprefix  [0]{URL }%
\providecommand \Eprint [0]{\href }%
\providecommand \doibase [0]{http://dx.doi.org/}%
\providecommand \selectlanguage [0]{\@gobble}%
\providecommand \bibinfo  [0]{\@secondoftwo}%
\providecommand \bibfield  [0]{\@secondoftwo}%
\providecommand \translation [1]{[#1]}%
\providecommand \BibitemOpen [0]{}%
\providecommand \bibitemStop [0]{}%
\providecommand \bibitemNoStop [0]{.\EOS\space}%
\providecommand \EOS [0]{\spacefactor3000\relax}%
\providecommand \BibitemShut  [1]{\csname bibitem#1\endcsname}%
\let\auto@bib@innerbib\@empty
\bibitem [{\citenamefont {Couder}\ and\ \citenamefont {Fort}(2006)}]{couder1}%
  \BibitemOpen
  \bibfield  {author} {\bibinfo {author} {\bibfnamefont {Y.}~\bibnamefont
  {Couder}}\ and\ \bibinfo {author} {\bibfnamefont {E.}~\bibnamefont {Fort}},\
  }\bibfield  {title} {\enquote {\bibinfo {title} {Single-particle diffraction
  and interference at a macroscopic scale.}}\ }\href@noop {} {\bibfield
  {journal} {\bibinfo  {journal} {Phys. Rev. Lett.}\ }\textbf {\bibinfo
  {volume} {97}},\ \bibinfo {pages} {15410} (\bibinfo {year}
  {2006})}\BibitemShut {NoStop}%
\bibitem [{\citenamefont {Couder}\ \emph {et~al.}(2010)\citenamefont {Couder},
  \citenamefont {Boudaoud}, \citenamefont {Proti\`ere},\ and\ \citenamefont
  {Fort}}]{couder2}%
  \BibitemOpen
  \bibfield  {author} {\bibinfo {author} {\bibfnamefont {Y.}~\bibnamefont
  {Couder}}, \bibinfo {author} {\bibfnamefont {A.}~\bibnamefont {Boudaoud}},
  \bibinfo {author} {\bibfnamefont {S.}~\bibnamefont {Proti\`ere}}, \ and\
  \bibinfo {author} {\bibfnamefont {E.}~\bibnamefont {Fort}},\ }\bibfield
  {title} {\enquote {\bibinfo {title} {Walking droplets, a form of
  wave-particle duality at macroscopic scale?}}\ }\href@noop {} {\bibfield
  {journal} {\bibinfo  {journal} {Europhysics News}\ }\textbf {\bibinfo
  {volume} {41}},\ \bibinfo {pages} {14--18} (\bibinfo {year}
  {2010})}\BibitemShut {NoStop}%
\bibitem [{\citenamefont {Couder}\ \emph {et~al.}(2005)\citenamefont {Couder},
  \citenamefont {Proti\`ere}, \citenamefont {Fort},\ and\ \citenamefont
  {Boudaoud}}]{couder3}%
  \BibitemOpen
  \bibfield  {author} {\bibinfo {author} {\bibfnamefont {Y.}~\bibnamefont
  {Couder}}, \bibinfo {author} {\bibfnamefont {S.}~\bibnamefont {Proti\`ere}},
  \bibinfo {author} {\bibfnamefont {E.}~\bibnamefont {Fort}}, \ and\ \bibinfo
  {author} {\bibfnamefont {A.}~\bibnamefont {Boudaoud}},\ }\bibfield  {title}
  {\enquote {\bibinfo {title} {Dynamical phenomena: walking and orbiting
  droplets.}}\ }\href@noop {} {\bibfield  {journal} {\bibinfo  {journal}
  {Nature}\ }\textbf {\bibinfo {volume} {437}},\ \bibinfo {pages} {208}
  (\bibinfo {year} {2005})}\BibitemShut {NoStop}%
\bibitem [{\citenamefont {Harris}\ \emph {et~al.}(2013)\citenamefont {Harris},
  \citenamefont {Moukhtar}, \citenamefont {Fort}, \citenamefont {Couder},\ and\
  \citenamefont {Bush}}]{Harris2013}%
  \BibitemOpen
  \bibfield  {author} {\bibinfo {author} {\bibfnamefont {D.~M.}\ \bibnamefont
  {Harris}}, \bibinfo {author} {\bibfnamefont {J.}~\bibnamefont {Moukhtar}},
  \bibinfo {author} {\bibfnamefont {E.}~\bibnamefont {Fort}}, \bibinfo {author}
  {\bibfnamefont {Y.}~\bibnamefont {Couder}}, \ and\ \bibinfo {author}
  {\bibfnamefont {J.~W.}\ \bibnamefont {Bush}},\ }\bibfield  {title} {\enquote
  {\bibinfo {title} {Wavelike statistics from pilot-wave dynamics in a circular
  corral},}\ }\href@noop {} {\bibfield  {journal} {\bibinfo  {journal} {Phys.
  Rev. E.}\ }\textbf {\bibinfo {volume} {88}},\ \bibinfo {pages} {011001}
  (\bibinfo {year} {2013})}\BibitemShut {NoStop}%
\bibitem [{\citenamefont {Bush}(2015{\natexlab{a}})}]{Bush2015}%
  \BibitemOpen
  \bibfield  {author} {\bibinfo {author} {\bibfnamefont {J.~W.~M.}\
  \bibnamefont {Bush}},\ }\bibfield  {title} {\enquote {\bibinfo {title}
  {{Pilot-wave hydrodynamics}},}\ }\href@noop {} {\bibfield  {journal}
  {\bibinfo  {journal} {Annu. Rev. Fluid Mech.}\ }\textbf {\bibinfo {volume}
  {47}},\ \bibinfo {pages} {269--292} (\bibinfo {year}
  {2015}{\natexlab{a}})}\BibitemShut {NoStop}%
\bibitem [{\citenamefont {Couder}\ and\ \citenamefont
  {Fort}(2012)}]{Couder2012}%
  \BibitemOpen
  \bibfield  {author} {\bibinfo {author} {\bibfnamefont {Y.}~\bibnamefont
  {Couder}}\ and\ \bibinfo {author} {\bibfnamefont {E.}~\bibnamefont {Fort}},\
  }\bibfield  {title} {\enquote {\bibinfo {title} {Probabilities and
  trajectories in a classical wave-particle duality},}\ }\href@noop {}
  {\bibfield  {journal} {\bibinfo  {journal} {J. Phys.: Conf. Ser.}\ }\textbf
  {\bibinfo {volume} {361}},\ \bibinfo {pages} {012001} (\bibinfo {year}
  {2012})}\BibitemShut {NoStop}%
\bibitem [{\citenamefont {Bush}(2015{\natexlab{b}})}]{Bush2015a}%
  \BibitemOpen
  \bibfield  {author} {\bibinfo {author} {\bibfnamefont {J.~W.~M.}\
  \bibnamefont {Bush}},\ }\bibfield  {title} {\enquote {\bibinfo {title} {The
  new wave of pilot-wave theory},}\ }\href@noop {} {\bibfield  {journal}
  {\bibinfo  {journal} {Physics Today}\ }\textbf {\bibinfo {volume} {68}},\
  \bibinfo {pages} {47} (\bibinfo {year} {2015}{\natexlab{b}})}\BibitemShut
  {NoStop}%
\bibitem [{\citenamefont {Eddi}\ \emph {et~al.}(2011)\citenamefont {Eddi},
  \citenamefont {Sultan}, \citenamefont {Moukhtar}, \citenamefont {Fort},
  \citenamefont {Rossi},\ and\ \citenamefont {Couder}}]{eddi2011}%
  \BibitemOpen
  \bibfield  {author} {\bibinfo {author} {\bibfnamefont {A.}~\bibnamefont
  {Eddi}}, \bibinfo {author} {\bibfnamefont {E.}~\bibnamefont {Sultan}},
  \bibinfo {author} {\bibfnamefont {J.}~\bibnamefont {Moukhtar}}, \bibinfo
  {author} {\bibfnamefont {E.}~\bibnamefont {Fort}}, \bibinfo {author}
  {\bibfnamefont {M.}~\bibnamefont {Rossi}}, \ and\ \bibinfo {author}
  {\bibfnamefont {Y.}~\bibnamefont {Couder}},\ }\bibfield  {title} {\enquote
  {\bibinfo {title} {{Information stored in Faraday waves: the origin of a path
  memory}},}\ }\href@noop {} {\bibfield  {journal} {\bibinfo  {journal}
  {Journal of Fluid Mechanics}\ }\textbf {\bibinfo {volume} {674}},\ \bibinfo
  {pages} {433--463} (\bibinfo {year} {2011})}\BibitemShut {NoStop}%
\bibitem [{\citenamefont {Perrard}\ \emph {et~al.}(2014)\citenamefont
  {Perrard}, \citenamefont {Labousse}, \citenamefont {Miskin}, \citenamefont
  {Fort},\ and\ \citenamefont {Couder}}]{Perrard2014}%
  \BibitemOpen
  \bibfield  {author} {\bibinfo {author} {\bibfnamefont {S.}~\bibnamefont
  {Perrard}}, \bibinfo {author} {\bibfnamefont {M.}~\bibnamefont {Labousse}},
  \bibinfo {author} {\bibfnamefont {M.}~\bibnamefont {Miskin}}, \bibinfo
  {author} {\bibfnamefont {E.}~\bibnamefont {Fort}}, \ and\ \bibinfo {author}
  {\bibfnamefont {Y.}~\bibnamefont {Couder}},\ }\bibfield  {title} {\enquote
  {\bibinfo {title} {Self-organization into quantized eigenstates of a
  classical wave-driven particle},}\ }\href@noop {} {\bibfield  {journal}
  {\bibinfo  {journal} {Nature communications}\ }\textbf {\bibinfo {volume}
  {5}},\ \bibinfo {pages} {3219} (\bibinfo {year} {2014})}\BibitemShut
  {NoStop}%
\bibitem [{\citenamefont {Durt}(1999)}]{Durt1999}%
  \BibitemOpen
  \bibfield  {author} {\bibinfo {author} {\bibfnamefont {T.}~\bibnamefont
  {Durt}},\ }\bibfield  {title} {\enquote {\bibinfo {title} {Do dice
  remember?}}\ }\href@noop {} {\bibfield  {journal} {\bibinfo  {journal}
  {International journal of theoretical physics}\ }\textbf {\bibinfo {volume}
  {38}},\ \bibinfo {pages} {457--473} (\bibinfo {year} {1999})}\BibitemShut
  {NoStop}%
\bibitem [{\citenamefont {Bohm}(1952{\natexlab{a}})}]{bohm521}%
  \BibitemOpen
  \bibfield  {author} {\bibinfo {author} {\bibfnamefont {D.}~\bibnamefont
  {Bohm}},\ }\bibfield  {title} {\enquote {\bibinfo {title} {{A Suggested
  Interpretation of the Quantum Theory in Terms of ``Hidden'' Variables. I}},}\
  }\href@noop {} {\bibfield  {journal} {\bibinfo  {journal} {Phys. Rev.}\
  }\textbf {\bibinfo {volume} {85}},\ \bibinfo {pages} {166--179} (\bibinfo
  {year} {1952}{\natexlab{a}})}\BibitemShut {NoStop}%
\bibitem [{\citenamefont {Bohm}(1952{\natexlab{b}})}]{bohm522}%
  \BibitemOpen
  \bibfield  {author} {\bibinfo {author} {\bibfnamefont {D.}~\bibnamefont
  {Bohm}},\ }\bibfield  {title} {\enquote {\bibinfo {title} {{A Suggested
  Interpretation of the Quantum Theory in Terms of ``Hidden" Variables. II}},}\
  }\href@noop {} {\bibfield  {journal} {\bibinfo  {journal} {Phys. Rev.}\
  }\textbf {\bibinfo {volume} {85}},\ \bibinfo {pages} {180--193} (\bibinfo
  {year} {1952}{\natexlab{b}})}\BibitemShut {NoStop}%
\bibitem [{\citenamefont {Nelson}(1967)}]{Nelson1967}%
  \BibitemOpen
  \bibfield  {author} {\bibinfo {author} {\bibfnamefont {E.}~\bibnamefont
  {Nelson}},\ }\bibfield  {title} {\enquote {\bibinfo {title} {{Dynamical
  Theories of Brownian Motion}},}\ }\href@noop {} {\bibfield  {journal}
  {\bibinfo  {journal} {Mathematical Notes}\ }\textbf {\bibinfo {volume}
  {131}},\ \bibinfo {pages} {2381--2396} (\bibinfo {year} {1967})}\BibitemShut
  {NoStop}%
\bibitem [{\citenamefont {Labousse}(2014)}]{Labousse2014}%
  \BibitemOpen
  \bibfield  {author} {\bibinfo {author} {\bibfnamefont {M.}~\bibnamefont
  {Labousse}},\ }\emph {\bibinfo {title} {{Etude d'une dynamique \`a m\'emoire
  de chemin: une exp\'erimentation th\'eorique}}},\ \href@noop {} {Ph.D.
  thesis},\ \bibinfo  {school} {{Universit\'e Pierre et Marie Curie UPMC Paris
  VI}} (\bibinfo {year} {2014})\BibitemShut {NoStop}%
\bibitem [{\citenamefont {Fort}\ \emph {et~al.}(2010)\citenamefont {Fort},
  \citenamefont {Eddi}, \citenamefont {Boudaoud}, \citenamefont {Moukhtar},\
  and\ \citenamefont {Couder}}]{Fort2010}%
  \BibitemOpen
  \bibfield  {author} {\bibinfo {author} {\bibfnamefont {E.}~\bibnamefont
  {Fort}}, \bibinfo {author} {\bibfnamefont {A.}~\bibnamefont {Eddi}}, \bibinfo
  {author} {\bibfnamefont {A.}~\bibnamefont {Boudaoud}}, \bibinfo {author}
  {\bibfnamefont {J.}~\bibnamefont {Moukhtar}}, \ and\ \bibinfo {author}
  {\bibfnamefont {Y.}~\bibnamefont {Couder}},\ }\bibfield  {title} {\enquote
  {\bibinfo {title} {Path-memory induced quantization of classical orbits},}\
  }\href@noop {} {\bibfield  {journal} {\bibinfo  {journal} {Proceedings of the
  National Academy of Sciences}\ }\textbf {\bibinfo {volume} {107}},\ \bibinfo
  {pages} {17515--17520} (\bibinfo {year} {2010})}\BibitemShut {NoStop}%
\bibitem [{\citenamefont {Dubertrand}\ \emph {et~al.}(2016)\citenamefont
  {Dubertrand}, \citenamefont {Hubert}, \citenamefont {Schlagheck},
  \citenamefont {Vandewalle}, \citenamefont {Bastin},\ and\ \citenamefont
  {Martin}}]{Dubertrand2016}%
  \BibitemOpen
  \bibfield  {author} {\bibinfo {author} {\bibfnamefont {R.}~\bibnamefont
  {Dubertrand}}, \bibinfo {author} {\bibfnamefont {M.}~\bibnamefont {Hubert}},
  \bibinfo {author} {\bibfnamefont {P.}~\bibnamefont {Schlagheck}}, \bibinfo
  {author} {\bibfnamefont {N.}~\bibnamefont {Vandewalle}}, \bibinfo {author}
  {\bibfnamefont {T.}~\bibnamefont {Bastin}}, \ and\ \bibinfo {author}
  {\bibfnamefont {J.}~\bibnamefont {Martin}},\ }\bibfield  {title} {\enquote
  {\bibinfo {title} {{Scattering theory of walking droplets in the presence of
  obstacles}},}\ }\href@noop {} {\bibfield  {journal} {\bibinfo  {journal} {New
  Journal of Physics}\ }\textbf {\bibinfo {volume} {18}},\ \bibinfo {pages}
  {113037} (\bibinfo {year} {2016})}\BibitemShut {NoStop}%
\bibitem [{\citenamefont {Tadrist}\ \emph {et~al.}(2017)\citenamefont
  {Tadrist}, \citenamefont {Shim}, \citenamefont {Gilet},\ and\ \citenamefont
  {Schlagheck}}]{Tadrist2017}%
  \BibitemOpen
  \bibfield  {author} {\bibinfo {author} {\bibfnamefont {L.}~\bibnamefont
  {Tadrist}}, \bibinfo {author} {\bibfnamefont {J.-B.}\ \bibnamefont {Shim}},
  \bibinfo {author} {\bibfnamefont {T.}~\bibnamefont {Gilet}}, \ and\ \bibinfo
  {author} {\bibfnamefont {P.}~\bibnamefont {Schlagheck}},\ }\bibfield  {title}
  {\enquote {\bibinfo {title} {Faraday instability and subthreshold faraday
  waves: surface waves emitted by walkers},}\ }\href@noop {} {\  (\bibinfo
  {year} {2017})},\ \Eprint {http://arxiv.org/abs/1711.06791}
  {arXiv:1711.06791} \BibitemShut {NoStop}%
\bibitem [{\citenamefont {Bohm}\ and\ \citenamefont
  {Vigier}(1954)}]{bohm-vigier}%
  \BibitemOpen
  \bibfield  {author} {\bibinfo {author} {\bibfnamefont {D.}~\bibnamefont
  {Bohm}}\ and\ \bibinfo {author} {\bibfnamefont {J.-P.}\ \bibnamefont
  {Vigier}},\ }\bibfield  {title} {\enquote {\bibinfo {title} {{Model of the
  causal interpretation of quantum theory in terms of a fluid with irregular
  fluctuations}},}\ }\href@noop {} {\bibfield  {journal} {\bibinfo  {journal}
  {Phys. Rev.}\ }\textbf {\bibinfo {volume} {96}},\ \bibinfo {pages} {208}
  (\bibinfo {year} {1954})}\BibitemShut {NoStop}%
\bibitem [{\citenamefont {Bohm}\ and\ \citenamefont {Hiley}(1989)}]{bohmh}%
  \BibitemOpen
  \bibfield  {author} {\bibinfo {author} {\bibfnamefont {D.}~\bibnamefont
  {Bohm}}\ and\ \bibinfo {author} {\bibfnamefont {B.}~\bibnamefont {Hiley}},\
  }\bibfield  {title} {\enquote {\bibinfo {title} {Non-locality and locality in
  the stochastic interpretation of quantum mechanics.}}\ }\href@noop {}
  {\bibfield  {journal} {\bibinfo  {journal} {Physics Reports}\ }\textbf
  {\bibinfo {volume} {172}},\ \bibinfo {pages} {93--122} (\bibinfo {year}
  {1989})}\BibitemShut {NoStop}%
\bibitem [{\citenamefont {Kyprianidis}(1992)}]{kyprianidis}%
  \BibitemOpen
  \bibfield  {author} {\bibinfo {author} {\bibfnamefont {P.}~\bibnamefont
  {Kyprianidis}},\ }\bibfield  {title} {\enquote {\bibinfo {title} {{The
  Principles of a Stochastic Formulation of Quantum Theory}},}\ }\href@noop {}
  {\bibfield  {journal} {\bibinfo  {journal} {Found. Phys.}\ }\textbf {\bibinfo
  {volume} {22}},\ \bibinfo {pages} {1449--1483} (\bibinfo {year}
  {1992})}\BibitemShut {NoStop}%
\bibitem [{\citenamefont {Valentini}(1991{\natexlab{a}})}]{valentini91a}%
  \BibitemOpen
  \bibfield  {author} {\bibinfo {author} {\bibfnamefont {A.}~\bibnamefont
  {Valentini}},\ }\bibfield  {title} {\enquote {\bibinfo {title} {{Signal
  locality, uncertainty and the subquantum H-theorem. I}},}\ }\href@noop {}
  {\bibfield  {journal} {\bibinfo  {journal} {Phys.\ Lett.\ A}\ }\textbf
  {\bibinfo {volume} {156}},\ \bibinfo {pages} {5--11} (\bibinfo {year}
  {1991}{\natexlab{a}})}\BibitemShut {NoStop}%
\bibitem [{\citenamefont {Petroni}(1995)}]{Petroni1994}%
  \BibitemOpen
  \bibfield  {author} {\bibinfo {author} {\bibfnamefont {N.~C.}\ \bibnamefont
  {Petroni}},\ }\bibfield  {title} {\enquote {\bibinfo {title} {{Asymptotic
  behaviour of densities for Nelson processes}},}\ }in\ \href@noop {} {\emph
  {\bibinfo {booktitle} {Quantum Communications and Measurement}}}\ (\bibinfo
  {publisher} {Springer},\ \bibinfo {year} {1995})\ pp.\ \bibinfo {pages}
  {43--52}\BibitemShut {NoStop}%
\bibitem [{\citenamefont {Petroni}\ and\ \citenamefont
  {Guerra}(1995)}]{Petroni1995}%
  \BibitemOpen
  \bibfield  {author} {\bibinfo {author} {\bibfnamefont {N.~C.}\ \bibnamefont
  {Petroni}}\ and\ \bibinfo {author} {\bibfnamefont {F.}~\bibnamefont
  {Guerra}},\ }\bibfield  {title} {\enquote {\bibinfo {title} {{Quantum
  Mechanical States as Attractors for Nelson Processes}},}\ }\href@noop {}
  {\bibfield  {journal} {\bibinfo  {journal} {Found. Phys.}\ }\textbf {\bibinfo
  {volume} {25}},\ \bibinfo {pages} {297--315} (\bibinfo {year}
  {1995})}\BibitemShut {NoStop}%
\bibitem [{\citenamefont {Guerra}(1995)}]{guerra1995}%
  \BibitemOpen
  \bibfield  {author} {\bibinfo {author} {\bibfnamefont {F.}~\bibnamefont
  {Guerra}},\ }\bibfield  {title} {\enquote {\bibinfo {title} {{Introduction to
  Nelson stochastic mechanics as a model for quantum mechanics}},}\ }in\
  \href@noop {} {\emph {\bibinfo {booktitle} {The Foundations of Quantum
  Mechanics}}}\ (\bibinfo  {publisher} {Kluwer Academic Publishers},\ \bibinfo
  {year} {1995})\ pp.\ \bibinfo {pages} {339--355}\BibitemShut {NoStop}%
\bibitem [{\citenamefont {Efthymiopoulos}, \citenamefont {Contopoulos},\ and\
  \citenamefont {Tzemos}(2017)}]{efthymiopoulos4}%
  \BibitemOpen
  \bibfield  {author} {\bibinfo {author} {\bibfnamefont {C.}~\bibnamefont
  {Efthymiopoulos}}, \bibinfo {author} {\bibfnamefont {G.}~\bibnamefont
  {Contopoulos}}, \ and\ \bibinfo {author} {\bibfnamefont {A.~C.}\ \bibnamefont
  {Tzemos}},\ }\bibfield  {title} {\enquote {\bibinfo {title} {{Chaos in de
  Broglie - Bohm quantum mechanics and the dynamics of quantum relaxation}},}\
  }\href@noop {} {\bibfield  {journal} {\bibinfo  {journal} {Ann. Fond. de
  Broglie}\ }\textbf {\bibinfo {volume} {42}},\ \bibinfo {pages} {133}
  (\bibinfo {year} {2017})}\BibitemShut {NoStop}%
\bibitem [{\citenamefont {Valentini}\ and\ \citenamefont
  {Westman}(2005)}]{valentini042}%
  \BibitemOpen
  \bibfield  {author} {\bibinfo {author} {\bibfnamefont {A.}~\bibnamefont
  {Valentini}}\ and\ \bibinfo {author} {\bibfnamefont {H.}~\bibnamefont
  {Westman}},\ }\bibfield  {title} {\enquote {\bibinfo {title} {Dynamical
  origin of quantum probabilities},}\ }\href@noop {} {\bibfield  {journal}
  {\bibinfo  {journal} {Proc. R. Soc. A}\ }\textbf {\bibinfo {volume} {461}},\
  \bibinfo {pages} {253--272} (\bibinfo {year} {2005})}\BibitemShut {NoStop}%
\bibitem [{\citenamefont {Colin}\ and\ \citenamefont {Struyve}(2010)}]{cost10}%
  \BibitemOpen
  \bibfield  {author} {\bibinfo {author} {\bibfnamefont {S.}~\bibnamefont
  {Colin}}\ and\ \bibinfo {author} {\bibfnamefont {W.}~\bibnamefont
  {Struyve}},\ }\bibfield  {title} {\enquote {\bibinfo {title} {{Quantum
  non-equilibrium and relaxation to quantum equilibrium for a class of de
  Broglie-Bohm-type theories}},}\ }\href@noop {} {\bibfield  {journal}
  {\bibinfo  {journal} {New\ J.\ Phys.}\ }\textbf {\bibinfo {volume} {12}},\
  \bibinfo {pages} {043008} (\bibinfo {year} {2010})}\BibitemShut {NoStop}%
\bibitem [{\citenamefont {Towler}, \citenamefont {Russell},\ and\ \citenamefont
  {Valentini}(2011)}]{toruva}%
  \BibitemOpen
  \bibfield  {author} {\bibinfo {author} {\bibfnamefont {M.}~\bibnamefont
  {Towler}}, \bibinfo {author} {\bibfnamefont {N.~J.}\ \bibnamefont {Russell}},
  \ and\ \bibinfo {author} {\bibfnamefont {A.}~\bibnamefont {Valentini}},\
  }\bibfield  {title} {\enquote {\bibinfo {title} {{Time scales for dynamical
  relaxation to the Born rule}},}\ }\href@noop {} {\bibfield  {journal}
  {\bibinfo  {journal} {Proc. R. Soc. A}\ }\textbf {\bibinfo {volume} {468}},\
  \bibinfo {pages} {990--1013} (\bibinfo {year} {2011})}\BibitemShut {NoStop}%
\bibitem [{\citenamefont {Colin}(2012)}]{colin2012}%
  \BibitemOpen
  \bibfield  {author} {\bibinfo {author} {\bibfnamefont {S.}~\bibnamefont
  {Colin}},\ }\bibfield  {title} {\enquote {\bibinfo {title} {{Relaxation to
  quantum equilibrium for Dirac fermions in the de Broglie-Bohm pilot-wave
  theory}},}\ }\href@noop {} {\bibfield  {journal} {\bibinfo  {journal} {Proc.
  R. Soc. A}\ }\textbf {\bibinfo {volume} {468}},\ \bibinfo {pages}
  {1116--1135} (\bibinfo {year} {2012})}\BibitemShut {NoStop}%
\bibitem [{\citenamefont {Contopoulos}, \citenamefont {Delis},\ and\
  \citenamefont {Efthymiopoulos}(2012)}]{efthymiopoulos3}%
  \BibitemOpen
  \bibfield  {author} {\bibinfo {author} {\bibfnamefont {G.}~\bibnamefont
  {Contopoulos}}, \bibinfo {author} {\bibfnamefont {N.}~\bibnamefont {Delis}},
  \ and\ \bibinfo {author} {\bibfnamefont {C.}~\bibnamefont {Efthymiopoulos}},\
  }\bibfield  {title} {\enquote {\bibinfo {title} {{Order in de Broglie - Bohm
  quantum mechanics}},}\ }\href@noop {} {\bibfield  {journal} {\bibinfo
  {journal} {J. Phys. A: Math. Theor.}\ }\textbf {\bibinfo {volume} {45}},\
  \bibinfo {pages} {165301} (\bibinfo {year} {2012})}\BibitemShut {NoStop}%
\bibitem [{\citenamefont {Abraham}, \citenamefont {Colin},\ and\ \citenamefont
  {Valentini}(2014)}]{abcova}%
  \BibitemOpen
  \bibfield  {author} {\bibinfo {author} {\bibfnamefont {E.}~\bibnamefont
  {Abraham}}, \bibinfo {author} {\bibfnamefont {S.}~\bibnamefont {Colin}}, \
  and\ \bibinfo {author} {\bibfnamefont {A.}~\bibnamefont {Valentini}},\
  }\bibfield  {title} {\enquote {\bibinfo {title} {Long-time relaxation in the
  pilot-wave theory},}\ }\href@noop {} {\bibfield  {journal} {\bibinfo
  {journal} {J. Phys. A: Math. Theor.}\ }\textbf {\bibinfo {volume} {47}},\
  \bibinfo {pages} {395306} (\bibinfo {year} {2014})}\BibitemShut {NoStop}%
\bibitem [{\citenamefont {{de La Pe{\~{n}}a}}\ and\ \citenamefont
  {Cetto}(2013)}]{DelaPena2013}%
  \BibitemOpen
  \bibfield  {author} {\bibinfo {author} {\bibfnamefont {L.}~\bibnamefont {{de
  La Pe{\~{n}}a}}}\ and\ \bibinfo {author} {\bibfnamefont {A.~M.}\ \bibnamefont
  {Cetto}},\ }\href@noop {} {\emph {\bibinfo {title} {{The quantum dice: an
  introduction to stochastic electrodynamics}}}}\ (\bibinfo  {publisher}
  {Springer Science {\&} Business Media},\ \bibinfo {year} {2013})\BibitemShut
  {NoStop}%
\bibitem [{\citenamefont {de~la Pe{\~{n}}a}, \citenamefont {Cetto},\ and\
  \citenamefont {{Vald{\'{e}}s Hern{\'{a}}ndez}}(2015)}]{DelaPena2015}%
  \BibitemOpen
  \bibfield  {author} {\bibinfo {author} {\bibfnamefont {L.}~\bibnamefont
  {de~la Pe{\~{n}}a}}, \bibinfo {author} {\bibfnamefont {A.~M.}\ \bibnamefont
  {Cetto}}, \ and\ \bibinfo {author} {\bibfnamefont {A.}~\bibnamefont
  {{Vald{\'{e}}s Hern{\'{a}}ndez}}},\ }\href@noop {} {\emph {\bibinfo {title}
  {{The Emerging Quantum}}}}\ (\bibinfo  {publisher} {Springer},\ \bibinfo
  {year} {2015})\BibitemShut {NoStop}%
\bibitem [{\citenamefont {Deotto}\ and\ \citenamefont
  {Ghirardi}(1998)}]{deotto}%
  \BibitemOpen
  \bibfield  {author} {\bibinfo {author} {\bibfnamefont {E.}~\bibnamefont
  {Deotto}}\ and\ \bibinfo {author} {\bibfnamefont {G.-C.}\ \bibnamefont
  {Ghirardi}},\ }\bibfield  {title} {\enquote {\bibinfo {title} {Bohmian
  mechanics revisited.}}\ }\href@noop {} {\bibfield  {journal} {\bibinfo
  {journal} {Found. Phys.}\ }\textbf {\bibinfo {volume} {28}},\ \bibinfo
  {pages} {1--30} (\bibinfo {year} {1998})}\BibitemShut {NoStop}%
\bibitem [{\citenamefont {de~Broglie}(1987)}]{deBroglieend}%
  \BibitemOpen
  \bibfield  {author} {\bibinfo {author} {\bibfnamefont {L.}~\bibnamefont
  {de~Broglie}},\ }\bibfield  {title} {\enquote {\bibinfo {title}
  {Interpretation of quantum mechanics by the double solution theory},}\
  }\href@noop {} {\bibfield  {journal} {\bibinfo  {journal} {Annales de la
  Fondation Louis de Broglie}\ }\textbf {\bibinfo {volume} {12}},\ \bibinfo
  {pages} {1--23} (\bibinfo {year} {1987})}\BibitemShut {NoStop}%
\bibitem [{\citenamefont {Bacciagaluppi}(1999)}]{Bacciagaluppi1999}%
  \BibitemOpen
  \bibfield  {author} {\bibinfo {author} {\bibfnamefont {G.}~\bibnamefont
  {Bacciagaluppi}},\ }\bibfield  {title} {\enquote {\bibinfo {title}
  {{Nelsonian mechanics revisited}},}\ }\href@noop {} {\bibfield  {journal}
  {\bibinfo  {journal} {Foundations of Physics Letters}\ }\textbf {\bibinfo
  {volume} {12}},\ \bibinfo {pages} {1--16} (\bibinfo {year}
  {1999})}\BibitemShut {NoStop}%
\bibitem [{\citenamefont {deBroglie}(1927)}]{deBroglie27}%
  \BibitemOpen
  \bibfield  {author} {\bibinfo {author} {\bibfnamefont {L.}~\bibnamefont
  {deBroglie}},\ }\bibfield  {title} {\enquote {\bibinfo {title} {La
  m\'ecanique ondulatoire et la structure atomique de la mati\`ere et du
  rayonnement},}\ }\href@noop {} {\bibfield  {journal} {\bibinfo  {journal} {J.
  Phys. Radium}\ }\textbf {\bibinfo {volume} {8}},\ \bibinfo {pages} {225--241}
  (\bibinfo {year} {1927})}\BibitemShut {NoStop}%
\bibitem [{\citenamefont {Valentini}(1992)}]{valentini-phd}%
  \BibitemOpen
  \bibfield  {author} {\bibinfo {author} {\bibfnamefont {A.}~\bibnamefont
  {Valentini}},\ }\emph {\bibinfo {title} {On the pilot-wave theory of
  classical, quantum and subquantum physics.}},\ \href@noop {} {Ph.D. thesis},\
  \bibinfo  {school} {SISSA} (\bibinfo {year} {1992})\BibitemShut {NoStop}%
\bibitem [{\citenamefont {Valentini}(1991{\natexlab{b}})}]{valentini91b}%
  \BibitemOpen
  \bibfield  {author} {\bibinfo {author} {\bibfnamefont {A.}~\bibnamefont
  {Valentini}},\ }\bibfield  {title} {\enquote {\bibinfo {title} {{Signal
  locality, uncertainty and the subquantum H-theorem. II}},}\ }\href@noop {}
  {\bibfield  {journal} {\bibinfo  {journal} {Phys.\ Lett.\ A}\ }\textbf
  {\bibinfo {volume} {158}},\ \bibinfo {pages} {1--8} (\bibinfo {year}
  {1991}{\natexlab{b}})}\BibitemShut {NoStop}%
\bibitem [{\citenamefont {Efthymiopoulos}, \citenamefont {Kalapotharakos},\
  and\ \citenamefont {Contopoulos}(2009)}]{efthymiopoulos}%
  \BibitemOpen
  \bibfield  {author} {\bibinfo {author} {\bibfnamefont {C.}~\bibnamefont
  {Efthymiopoulos}}, \bibinfo {author} {\bibfnamefont {C.}~\bibnamefont
  {Kalapotharakos}}, \ and\ \bibinfo {author} {\bibfnamefont {G.}~\bibnamefont
  {Contopoulos}},\ }\bibfield  {title} {\enquote {\bibinfo {title} {Origin of
  chaos near critical points of quantum flow},}\ }\href@noop {} {\bibfield
  {journal} {\bibinfo  {journal} {Phys. Rev. E}\ }\textbf {\bibinfo {volume}
  {79}},\ \bibinfo {pages} {036203} (\bibinfo {year} {2009})}\BibitemShut
  {NoStop}%
\bibitem [{\citenamefont {Valentini}(2010)}]{valentini2010}%
  \BibitemOpen
  \bibfield  {author} {\bibinfo {author} {\bibfnamefont {A.}~\bibnamefont
  {Valentini}},\ }\bibfield  {title} {\enquote {\bibinfo {title} {Inflationary
  cosmology as a probe of primordial quantum mechanics},}\ }\href@noop {}
  {\bibfield  {journal} {\bibinfo  {journal} {Phys. Rev. D}\ }\textbf {\bibinfo
  {volume} {82}},\ \bibinfo {pages} {063513} (\bibinfo {year}
  {2010})}\BibitemShut {NoStop}%
\bibitem [{\citenamefont {Underwood}\ and\ \citenamefont
  {Valentini}(2015)}]{underwood}%
  \BibitemOpen
  \bibfield  {author} {\bibinfo {author} {\bibfnamefont {N.~G.}\ \bibnamefont
  {Underwood}}\ and\ \bibinfo {author} {\bibfnamefont {A.}~\bibnamefont
  {Valentini}},\ }\bibfield  {title} {\enquote {\bibinfo {title} {Quantum field
  theory of relic nonequilibrium systems},}\ }\href@noop {} {\bibfield
  {journal} {\bibinfo  {journal} {Phys. Rev. D}\ }\textbf {\bibinfo {volume}
  {92}},\ \bibinfo {pages} {063531} (\bibinfo {year} {2015})}\BibitemShut
  {NoStop}%
\bibitem [{\citenamefont {Bricmont}(2001)}]{bricmont}%
  \BibitemOpen
  \bibfield  {author} {\bibinfo {author} {\bibfnamefont {J.}~\bibnamefont
  {Bricmont}},\ }\bibfield  {title} {\enquote {\bibinfo {title} {{ Bayes,
  Boltzmann and Bohm: Probabilities in Physics}},}\ }in\ \href@noop {} {\emph
  {\bibinfo {booktitle} {Chances in Physics}}}\ (\bibinfo  {publisher}
  {Bricmont J., Ghirardi G., D\"urr D., Petruccione F., Galavotti M.C., Zanghi
  N. (eds), Springer},\ \bibinfo {year} {2001})\ pp.\ \bibinfo {pages}
  {3--21}\BibitemShut {NoStop}%
\bibitem [{\citenamefont {J\"ungel}(2016)}]{jungel}%
  \BibitemOpen
  \bibfield  {author} {\bibinfo {author} {\bibfnamefont {A.}~\bibnamefont
  {J\"ungel}},\ }\href@noop {} {\emph {\bibinfo {title} {Entropy Methods for
  Partial Differential Equations}}}\ (\bibinfo  {publisher} {Springer briefs
  for Mathematics, Springer},\ \bibinfo {year} {2016})\BibitemShut {NoStop}%
\bibitem [{\citenamefont {Gardiner}(1985)}]{Gardiner1985}%
  \BibitemOpen
  \bibfield  {author} {\bibinfo {author} {\bibfnamefont {C.~W.}\ \bibnamefont
  {Gardiner}},\ }\href@noop {} {\emph {\bibinfo {title} {{Handbook of
  stochastic processes}}}}\ (\bibinfo  {publisher} {Springer-Verlag, New
  York},\ \bibinfo {year} {1985})\BibitemShut {NoStop}%
\bibitem [{\citenamefont {Risken}(1996)}]{Risken1996}%
  \BibitemOpen
  \bibfield  {author} {\bibinfo {author} {\bibfnamefont {H.}~\bibnamefont
  {Risken}},\ }\bibfield  {title} {\enquote {\bibinfo {title} {{Fokker-planck
  equation}},}\ }in\ \href@noop {} {\emph {\bibinfo {booktitle} {The
  Fokker-Planck Equation}}}\ (\bibinfo  {publisher} {Springer},\ \bibinfo
  {year} {1996})\ pp.\ \bibinfo {pages} {63--95}\BibitemShut {NoStop}%
\bibitem [{\citenamefont {Petroni}, \citenamefont {{De Martino}},\ and\
  \citenamefont {{De Siena}}(1998)}]{Petroni1998}%
  \BibitemOpen
  \bibfield  {author} {\bibinfo {author} {\bibfnamefont {N.~C.}\ \bibnamefont
  {Petroni}}, \bibinfo {author} {\bibfnamefont {S.}~\bibnamefont {{De
  Martino}}}, \ and\ \bibinfo {author} {\bibfnamefont {S.}~\bibnamefont {{De
  Siena}}},\ }\bibfield  {title} {\enquote {\bibinfo {title} {{Exact solutions
  of Fokker-Planck equations associated to quantum wave functions}},}\
  }\href@noop {} {\bibfield  {journal} {\bibinfo  {journal} {Physics Letters
  A}\ }\textbf {\bibinfo {volume} {245}},\ \bibinfo {pages} {1--10} (\bibinfo
  {year} {1998})}\BibitemShut {NoStop}%
\bibitem [{\citenamefont {Brics}, \citenamefont {Kaupuzs},\ and\ \citenamefont
  {Mahnke}(2013)}]{brics2013solve}%
  \BibitemOpen
  \bibfield  {author} {\bibinfo {author} {\bibfnamefont {M.}~\bibnamefont
  {Brics}}, \bibinfo {author} {\bibfnamefont {J.}~\bibnamefont {Kaupuzs}}, \
  and\ \bibinfo {author} {\bibfnamefont {R.}~\bibnamefont {Mahnke}},\
  }\bibfield  {title} {\enquote {\bibinfo {title} {How to solve fokker-planck
  equation treating mixed eigenvalue spectrum?}}\ }\href@noop {} {\bibfield
  {journal} {\bibinfo  {journal} {Cond. Matt. Phys.}\ }\textbf {\bibinfo
  {volume} {16}},\ \bibinfo {pages} {13002} (\bibinfo {year}
  {2013})}\BibitemShut {NoStop}%
\bibitem [{\citenamefont {Hatifi}\ \emph {et~al.}(2018)\citenamefont {Hatifi},
  \citenamefont {Willox}, \citenamefont {Colin},\ and\ \citenamefont
  {Durt}}]{v3}%
  \BibitemOpen
  \bibfield  {author} {\bibinfo {author} {\bibfnamefont {M.}~\bibnamefont
  {Hatifi}}, \bibinfo {author} {\bibfnamefont {R.}~\bibnamefont {Willox}},
  \bibinfo {author} {\bibfnamefont {S.}~\bibnamefont {Colin}}, \ and\ \bibinfo
  {author} {\bibfnamefont {T.}~\bibnamefont {Durt}},\ }\bibfield  {title}
  {\enquote {\bibinfo {title} {{Bouncing oil droplets, de Broglie’s quantum
  thermostat and convergence to equilibrium}},}\ }\href@noop {} {\  (\bibinfo
  {year} {2018})},\ \Eprint {http://arxiv.org/abs/1807.00569v3}
  {arXiv:1807.00569v3 [quant-ph]} \BibitemShut {NoStop}%
\bibitem [{\citenamefont {Gray}(2009)}]{ergogray}%
  \BibitemOpen
  \bibfield  {author} {\bibinfo {author} {\bibfnamefont {R.~M.}\ \bibnamefont
  {Gray}},\ }\href@noop {} {\emph {\bibinfo {title} {Probability, random
  processes, and ergodic properties}}}\ (\bibinfo  {publisher} {Springer,
  Dordrecht; Heidelberg},\ \bibinfo {year} {2009})\BibitemShut {NoStop}%
\bibitem [{\citenamefont {Arnold}\ and\ \citenamefont {Avez}(1967)}]{arnavez}%
  \BibitemOpen
  \bibfield  {author} {\bibinfo {author} {\bibfnamefont {V.}~\bibnamefont
  {Arnold}}\ and\ \bibinfo {author} {\bibfnamefont {A.}~\bibnamefont {Avez}},\
  }\href@noop {} {\emph {\bibinfo {title} {Probl\`emes ergodiques de la
  m\'echanique classique}}}\ (\bibinfo  {publisher} {Gauthier-Villars, Paris},\
  \bibinfo {year} {1967})\BibitemShut {NoStop}%
\bibitem [{\citenamefont {Labousse}\ \emph {et~al.}(2016)\citenamefont
  {Labousse}, \citenamefont {Oza}, \citenamefont {Perrard},\ and\ \citenamefont
  {Bush}}]{Labousse2016}%
  \BibitemOpen
  \bibfield  {author} {\bibinfo {author} {\bibfnamefont {M.}~\bibnamefont
  {Labousse}}, \bibinfo {author} {\bibfnamefont {A.~U.}\ \bibnamefont {Oza}},
  \bibinfo {author} {\bibfnamefont {S.}~\bibnamefont {Perrard}}, \ and\
  \bibinfo {author} {\bibfnamefont {J.~W.}\ \bibnamefont {Bush}},\ }\bibfield
  {title} {\enquote {\bibinfo {title} {{Pilot-wave dynamics in a harmonic
  potential: Quantization and stability of circular orbits}},}\ }\href@noop {}
  {\bibfield  {journal} {\bibinfo  {journal} {Physical Review E}\ }\textbf
  {\bibinfo {volume} {93}},\ \bibinfo {pages} {033122} (\bibinfo {year}
  {2016})}\BibitemShut {NoStop}%
\bibitem [{\citenamefont {Gr{\"{o}}ssing}(2010)}]{Grossing2010}%
  \BibitemOpen
  \bibfield  {author} {\bibinfo {author} {\bibfnamefont {G.}~\bibnamefont
  {Gr{\"{o}}ssing}},\ }\bibfield  {title} {\enquote {\bibinfo {title}
  {{Sub-quantum thermodynamics as a basis of emergent quantum mechanics}},}\
  }\href@noop {} {\bibfield  {journal} {\bibinfo  {journal} {Entropy}\ }\textbf
  {\bibinfo {volume} {12}},\ \bibinfo {pages} {1975--2044} (\bibinfo {year}
  {2010})}\BibitemShut {NoStop}%
\bibitem [{\citenamefont {Hestenes}(1990)}]{hestenes}%
  \BibitemOpen
  \bibfield  {author} {\bibinfo {author} {\bibfnamefont {D.}~\bibnamefont
  {Hestenes}},\ }\bibfield  {title} {\enquote {\bibinfo {title} {The
  zitterbewegung interpretation of quantum mechanics.}}\ }\href@noop {}
  {\bibfield  {journal} {\bibinfo  {journal} {Founds. of Phys.}\ }\textbf
  {\bibinfo {volume} {20}},\ \bibinfo {pages} {10} (\bibinfo {year}
  {1990})}\BibitemShut {NoStop}%
\bibitem [{\citenamefont {Colin}\ and\ \citenamefont {Wiseman}(2011)}]{cowi}%
  \BibitemOpen
  \bibfield  {author} {\bibinfo {author} {\bibfnamefont {S.}~\bibnamefont
  {Colin}}\ and\ \bibinfo {author} {\bibfnamefont {H.~M.}\ \bibnamefont
  {Wiseman}},\ }\bibfield  {title} {\enquote {\bibinfo {title} {The zig-zag
  road to reality},}\ }\href@noop {} {\bibfield  {journal} {\bibinfo  {journal}
  {J. Phys. A: Math. Theor.}\ }\textbf {\bibinfo {volume} {44}},\ \bibinfo
  {pages} {345304} (\bibinfo {year} {2011})}\BibitemShut {NoStop}%
\bibitem [{\citenamefont {Gilet}(2016)}]{Gilet}%
  \BibitemOpen
  \bibfield  {author} {\bibinfo {author} {\bibfnamefont {T.}~\bibnamefont
  {Gilet}},\ }\bibfield  {title} {\enquote {\bibinfo {title} {Quantumlike
  statistics of deterministic wave-particle interactions in a circular
  cavity.}}\ }\href@noop {} {\bibfield  {journal} {\bibinfo  {journal} {Phys.
  Rev. E}\ }\textbf {\bibinfo {volume} {93}},\ \bibinfo {pages} {042202}
  (\bibinfo {year} {2016})}\BibitemShut {NoStop}%
\bibitem [{\citenamefont {Cohen-Tannoudji}\ and\ \citenamefont
  {Laloe}(1977)}]{cohen}%
  \BibitemOpen
  \bibfield  {author} {\bibinfo {author} {\bibfnamefont {D.~B.}\ \bibnamefont
  {Cohen-Tannoudji}, \bibfnamefont {C.}}\ and\ \bibinfo {author} {\bibfnamefont
  {F.}~\bibnamefont {Laloe}},\ }\href@noop {} {\emph {\bibinfo {title} {Quantum
  Mechanics}}}\ (\bibinfo  {publisher} {Wiley},\ \bibinfo {year}
  {1977})\BibitemShut {NoStop}%
\bibitem [{\citenamefont {Hatifi}, \citenamefont {Lopez-Fortin},\ and\
  \citenamefont {Durt}(2018)}]{Carlos2018}%
  \BibitemOpen
  \bibfield  {author} {\bibinfo {author} {\bibfnamefont {M.}~\bibnamefont
  {Hatifi}}, \bibinfo {author} {\bibfnamefont {C.}~\bibnamefont
  {Lopez-Fortin}}, \ and\ \bibinfo {author} {\bibfnamefont {T.}~\bibnamefont
  {Durt}},\ }\bibfield  {title} {\enquote {\bibinfo {title} {{De Broglie's
  double solution: limitations of the self-gravity approach}},}\ }\href@noop {}
  {\bibfield  {journal} {\bibinfo  {journal} {Annales de la Fondation Louis de
  Broglie}\ }\textbf {\bibinfo {volume} {43}},\ \bibinfo {pages} {63} (\bibinfo
  {year} {2018})}\BibitemShut {NoStop}%
\bibitem [{\citenamefont {Pucci}\ \emph {et~al.}(2018)\citenamefont {Pucci},
  \citenamefont {Harris}, \citenamefont {Faria},\ and\ \citenamefont
  {Bush}}]{Pucci2018}%
  \BibitemOpen
  \bibfield  {author} {\bibinfo {author} {\bibfnamefont {G.}~\bibnamefont
  {Pucci}}, \bibinfo {author} {\bibfnamefont {D.~M.}\ \bibnamefont {Harris}},
  \bibinfo {author} {\bibfnamefont {L.~M.}\ \bibnamefont {Faria}}, \ and\
  \bibinfo {author} {\bibfnamefont {J.~W.~M.}\ \bibnamefont {Bush}},\
  }\bibfield  {title} {\enquote {\bibinfo {title} {{Walking droplets
  interacting with single and double slits}},}\ }\href@noop {} {\bibfield
  {journal} {\bibinfo  {journal} {Journal of Fluid Mechanics}\ }\textbf
  {\bibinfo {volume} {835}},\ \bibinfo {pages} {1136--1156} (\bibinfo {year}
  {2018})}\BibitemShut {NoStop}%
\bibitem [{\citenamefont {S\'aenz}\ and\ \citenamefont {Bush}(2017)}]{saenz}%
  \BibitemOpen
  \bibfield  {author} {\bibinfo {author} {\bibfnamefont {C.-P.~T.}\
  \bibnamefont {S\'aenz}, \bibfnamefont {P.J.}}\ and\ \bibinfo {author}
  {\bibfnamefont {J.}~\bibnamefont {Bush}},\ }\bibfield  {title} {\enquote
  {\bibinfo {title} {Statistical projection effects in a hydrodynamic
  pilot-wave system.}}\ }\href@noop {} {\bibfield  {journal} {\bibinfo
  {journal} {Nature}\ }\textbf {\bibinfo {volume} {14}},\ \bibinfo {pages} {3}
  (\bibinfo {year} {2017})}\BibitemShut {NoStop}%
\bibitem [{\citenamefont {Durt}(2017)}]{deBroglieDurt}%
  \BibitemOpen
  \bibfield  {author} {\bibinfo {author} {\bibfnamefont {T.}~\bibnamefont
  {Durt}},\ }\bibfield  {title} {\enquote {\bibinfo {title} {{ L. de Broglie's
  double solution and gravitation}},}\ }\href@noop {} {\bibfield  {journal}
  {\bibinfo  {journal} {Annales de la Fondation Louis de Broglie}\ }\textbf
  {\bibinfo {volume} {42}},\ \bibinfo {pages} {73} (\bibinfo {year}
  {2017})}\BibitemShut {NoStop}%
\bibitem [{\citenamefont {Nelson}(2012)}]{nelsonloc}%
  \BibitemOpen
  \bibfield  {author} {\bibinfo {author} {\bibfnamefont {E.}~\bibnamefont
  {Nelson}},\ }\bibfield  {title} {\enquote {\bibinfo {title} {{Review of
  stochastic mechanics}},}\ }\href@noop {} {\bibfield  {journal} {\bibinfo
  {journal} {J. of Phys.: Conf. Ser.}\ }\textbf {\bibinfo {volume} {361}},\
  \bibinfo {pages} {012011} (\bibinfo {year} {2012})}\BibitemShut {NoStop}%
\bibitem [{\citenamefont {Bell}(1964)}]{bell}%
  \BibitemOpen
  \bibfield  {author} {\bibinfo {author} {\bibfnamefont {J.}~\bibnamefont
  {Bell}},\ }\bibfield  {title} {\enquote {\bibinfo {title} {{On the EPR
  paradox}},}\ }\href@noop {} {\bibfield  {journal} {\bibinfo  {journal}
  {Physics}\ }\textbf {\bibinfo {volume} {1}},\ \bibinfo {pages} {165}
  (\bibinfo {year} {1964})}\BibitemShut {NoStop}%
\bibitem [{\citenamefont {Durt}(2004)}]{Proeminent}%
  \BibitemOpen
  \bibfield  {author} {\bibinfo {author} {\bibfnamefont {T.}~\bibnamefont
  {Durt}},\ }\bibfield  {title} {\enquote {\bibinfo {title} {Characterisation
  of an entanglement-free evolution.}}\ }\href@noop {} {\bibfield  {journal}
  {\bibinfo  {journal} {Zeits. fur Nat.}\ }\textbf {\bibinfo {volume} {59A}},\
  \bibinfo {pages} {425} (\bibinfo {year} {2004})}\BibitemShut {NoStop}%
\bibitem [{\citenamefont {Bacciagaluppi}(2005)}]{bacciaphase}%
  \BibitemOpen
  \bibfield  {author} {\bibinfo {author} {\bibfnamefont {G.}~\bibnamefont
  {Bacciagaluppi}},\ }\bibfield  {title} {\enquote {\bibinfo {title} {{ A
  Conceptual Introduction to Nelson's Mechanics}},}\ }in\ \href@noop {} {\emph
  {\bibinfo {booktitle} {Endophysics, time, quantum and the subjective}}}\
  (\bibinfo  {publisher} {R. Buccheri, M. Saniga, A. Elitzur (eds), World
  Scientific},\ \bibinfo {year} {2005})\ pp.\ \bibinfo {pages}
  {367--388}\BibitemShut {NoStop}%
\bibitem [{\citenamefont {Wallstrom}(1994)}]{wallstrom}%
  \BibitemOpen
  \bibfield  {author} {\bibinfo {author} {\bibfnamefont {T.}~\bibnamefont
  {Wallstrom}},\ }\bibfield  {title} {\enquote {\bibinfo {title}
  {{Inequivalence between the Schr\"odinger equation and the Madelung
  hydrodynamic equations}},}\ }\href@noop {} {\bibfield  {journal} {\bibinfo
  {journal} {Phys. Rev. A}\ }\textbf {\bibinfo {volume} {49}},\ \bibinfo
  {pages} {1613} (\bibinfo {year} {1994})}\BibitemShut {NoStop}%
\bibitem [{\citenamefont {Derakhshani}(2015)}]{Derakhshani2015}%
  \BibitemOpen
  \bibfield  {author} {\bibinfo {author} {\bibfnamefont {M.}~\bibnamefont
  {Derakhshani}},\ }\bibfield  {title} {\enquote {\bibinfo {title} {{A
  Suggested Answer To Wallstrom's Criticism (I): Zitterbewegung Stochastic
  Mechanics}},}\ }\href@noop {} {\  (\bibinfo {year} {2015})},\ \Eprint
  {http://arxiv.org/abs/1510.06391} {arXiv:1510.06391 [quant-ph]} \BibitemShut
  {NoStop}%
\bibitem [{\citenamefont {Bacciagaluppi}\ and\ \citenamefont
  {Valentini}(2010)}]{bacval}%
  \BibitemOpen
  \bibfield  {author} {\bibinfo {author} {\bibfnamefont {G.}~\bibnamefont
  {Bacciagaluppi}}\ and\ \bibinfo {author} {\bibfnamefont {A.}~\bibnamefont
  {Valentini}},\ }\href@noop {} {\emph {\bibinfo {title} {Quantum Theory at the
  Crossroads: Reconsidering the 1927 Solvay Conference}}}\ (\bibinfo
  {publisher} {Cambridge University Press, Cambridge},\ \bibinfo {year}
  {2010})\BibitemShut {NoStop}%
\bibitem [{\citenamefont {Colin}, \citenamefont {Durt},\ and\ \citenamefont
  {Willox}(2017)}]{CDWannales}%
  \BibitemOpen
  \bibfield  {author} {\bibinfo {author} {\bibfnamefont {S.}~\bibnamefont
  {Colin}}, \bibinfo {author} {\bibfnamefont {T.}~\bibnamefont {Durt}}, \ and\
  \bibinfo {author} {\bibfnamefont {R.}~\bibnamefont {Willox}},\ }\bibfield
  {title} {\enquote {\bibinfo {title} {{ L. de Broglie's double solution
  program: 90 years later}},}\ }\href@noop {} {\bibfield  {journal} {\bibinfo
  {journal} {Annales de la Fondation Louis de Broglie}\ }\textbf {\bibinfo
  {volume} {42}},\ \bibinfo {pages} {19} (\bibinfo {year} {2017})}\BibitemShut
  {NoStop}%
\bibitem [{\citenamefont {Borghesi}(2017)}]{Borghesi2017}%
  \BibitemOpen
  \bibfield  {author} {\bibinfo {author} {\bibfnamefont {C.}~\bibnamefont
  {Borghesi}},\ }\bibfield  {title} {\enquote {\bibinfo {title} {{Equivalent
  quantum equations with effective gravity in a system inspired by bouncing
  droplets experiments}},}\ }\href@noop {} {\bibfield  {journal} {\bibinfo
  {journal} {arXiv preprint arXiv:1706.05640}\ } (\bibinfo {year}
  {2017})}\BibitemShut {NoStop}%
\bibitem [{\citenamefont {Press}\ \emph {et~al.}(2007)\citenamefont {Press},
  \citenamefont {Teukolsky}, \citenamefont {Vetterling},\ and\ \citenamefont
  {Flannery}}]{nr}%
  \BibitemOpen
  \bibfield  {author} {\bibinfo {author} {\bibfnamefont {W.~H.}\ \bibnamefont
  {Press}}, \bibinfo {author} {\bibfnamefont {S.~A.}\ \bibnamefont
  {Teukolsky}}, \bibinfo {author} {\bibfnamefont {W.~T.}\ \bibnamefont
  {Vetterling}}, \ and\ \bibinfo {author} {\bibfnamefont {B.~P.}\ \bibnamefont
  {Flannery}},\ }\href@noop {} {\emph {\bibinfo {title} {Numerical Recipes 3rd
  Edition: The Art of Scientific Computing}}},\ \bibinfo {edition} {3rd}\ ed.\
  (\bibinfo  {publisher} {Cambridge University Press},\ \bibinfo {address} {New
  York, NY, USA},\ \bibinfo {year} {2007})\BibitemShut {NoStop}%
\bibitem [{\citenamefont {Higham}(2001)}]{Higham2001}%
  \BibitemOpen
  \bibfield  {author} {\bibinfo {author} {\bibfnamefont {D.~J.}\ \bibnamefont
  {Higham}},\ }\bibfield  {title} {\enquote {\bibinfo {title} {An algorithmic
  introduction to numerical simulation of stochastic differential equations},}\
  }\href@noop {} {\bibfield  {journal} {\bibinfo  {journal} {SIAM review}\
  }\textbf {\bibinfo {volume} {43}},\ \bibinfo {pages} {525--546} (\bibinfo
  {year} {2001})}\BibitemShut {NoStop}%
\end{thebibliography}%
\end{document}